\documentclass[a4paper,twoside,reqno]{amsart}

\RequirePackage[authoryear]{natbib}
\RequirePackage[colorlinks,citecolor=blue,urlcolor=blue]{hyperref}
\RequirePackage{graphicx}
\usepackage{overpic}
\usepackage{algorithm}
\usepackage{comment}
\usepackage{amssymb, amsthm, amsmath, amsfonts}
\usepackage{enumitem} 
\usepackage{mathtools}
\usepackage{xcolor}
\usepackage[foot]{amsaddr}

\usepackage{subcaption}
\usepackage{multirow}
\newcommand{\E}{\ensuremath{\mathbb{E}}}
\allowdisplaybreaks

\renewcommand{\vec}[1]{\boldsymbol{#1}}

\theoremstyle{plain}%
\newtheorem*{proposition*}{Proposition}%
\newtheorem*{lemma*}{Lemma}%

\theoremstyle{remark}%
\theoremstyle{definition}%

\numberwithin{equation}{section}

\newcommand{\R}{\ensuremath{\mathbb{R}}}

\newcommand\sym[1]{\mathrm{Sym}({#1})}
\newcommand\spd[1]{\mathcal{S}^+({#1})}

\newcommand\diag[1]{\mathrm{diag}\left({#1}\right)}

\newcommand{\Exp}{\ensuremath{\mathrm{Exp}}}
\newcommand{\Log}{\ensuremath{\mathrm{Log}}}
\newcommand{\Logx}{\ensuremath{\Log}_x}
\newcommand{\Expx}{\ensuremath{\Exp}_x}
\newcommand{\ptrans}[2]{\ensuremath{\mathcal{P}_{{#1},{#2}}}}
\newcommand{\mfd}{\mathcal{M}}
\newcommand{\tgt}[1]{T_{#1}\mfd}
\newcommand\norm[1]{\left\lVert#1\right\rVert}

\newcommand{\Id}[1]{\ensuremath{I_{#1}}}

\newcommand{\daff}[2]{d_g^{\mathrm{aff}}\left(#1,#2\right)}
\newcommand{\deuc}[2]{d_g^{\mathrm{euc}}\left(#1,#2\right)}

\newcommand{\gaff}[3]{g_{#3}^{\mathrm{aff}}\left(#1,#2\right)}
\newcommand{\geuc}[3]{g_{#3}^{\mathrm{euc}}\left(#1,#2\right)}

\DeclareMathOperator*{\argmin}{\mbox{\rm argmin}}

\DeclareMathOperator{\trace}{tr}

\makeatletter
\def\subsection{\@startsection{subsection}{2}%
  \z@{.5\linespacing\@plus.7\linespacing}{.3\linespacing}%
  {\normalfont\bfseries}}
\makeatother
\begin{document}

\title[Manifold-valued time series models]{Manifold-valued models for analysis of EEG time series data}

\author{Tao Ding}
\address[Tao Ding and Tom~M.~W.~Nye]{School of Mathematics, Statistics and Physics\\ Newcastle University\\ Newcastle upon Tyne\\ UK}
\email{t.ding2@newcastle.ac.uk}
\author{Tom~M.~W.~Nye}
\email{tom.nye@ncl.ac.uk}
\author{Yujiang Wang}
\address[Yujiang Wang]{School of Computing\\ Newcastle University\\ Newcastle upon Tyne\\ UK}
\email{yujiang.wang@ncl.ac.uk}


\begin{abstract}
We propose a model for time series taking values on a Riemannian manifold and fit it to time series of covariance matrices derived from EEG data for patients suffering from epilepsy. 
The aim of the study is two-fold: to develop a model with interpretable parameters for different possible modes of EEG dynamics, and to explore the extent to which modelling results are affected by the choice of manifold and its associated geometry. 
The model specifies a distribution for the tangent direction vector at any time point, combining an autoregressive term, a mean reverting term and a form of Gaussian noise. 
Parameter inference is carried out by maximum likelihood estimation, and we compare modelling results obtained using the standard Euclidean geometry on covariance matrices and the affine invariant geometry.  
Results distinguish between epileptic seizures and interictal periods between seizures in patients: between seizures the dynamics have a strong mean reverting component and the autoregressive component is missing, while for the majority of seizures there is a significant autoregressive component and the mean reverting effect is weak. 
The fitted models are also used to compare seizures within and between patients. 
The affine invariant geometry is advantageous and it provides a better fit to the data.  
\end{abstract}

\maketitle

\section{Introduction}

EEG (electroencephalogram) is a method for recording electric activity in the brain. 
Electrodes are placed on the subject's scalp, or in some circumstances placed surgically within the cranium, and fluctuations in the electrical activity at the electrode contacts are recorded across time. EEG is crucial in the diagnosis and treatment of epilepsy, as epileptic seizure activity often stands out from normal brain EEG activity in terms of the magnitude and spectral content of the signal. Common quantitative approaches to distinguish seizure EEG from normal brain EEG (between seizures) include a range of features \citep{siddiqui2020review,boonyakitanont2020review}. In this work, we will focus on the covariance of signals between electrode contacts, as a feature that captures `functional networks' in the brain. 
Typically, such functional networks use correlation between normal EEG signals \citep{bastos2016tutorial,abreu2020pushing,chiarion2023connectivity}. 
However, in this work, we believe that the signal variance is an important component to retain, as it varies dramatically between seizure EEG and interictal EEG (i.e.~normal EEG recorded between seizures).

Covariance matrices are symmetric positive semi-definite, but we further restrict to data sets of strictly positive definite matrices in order to avoid redundancies in the experimental data (see Section~\ref{sec:dimred}). 
Data consisting of samples of $p\times p$ positive definite covariance matrices form a subset of the $p\times p$ symmetric matrices. 
We argue that it is important not only to take this restriction into account, but also the curvature of the space when viewed as a Riemannian manifold. 
Analysis of manifold-valued data is attracting increasing research attention in statistics and machine learning due to the availability of increasing volumes of manifold data generated by a range of different experimental techniques. 
An excellent overview of the area and related topics is given by \cite{marron2014overview}. 
Directional data \citep{jupp2009directional} and some shape data \citep{dryden2016shape} lie on Riemannian manifolds, and manifold-based statistics play an important role in medical image analysis \citep{pennec2019riemannian}. 
Analysis of data sets of symmetric positive definite matrices has received particular attention in this context \citep[Chapter 5]{pennec2019riemannian}, due to the tractable nature of the geometry and the range of applications, especially in different forms of medical imaging such as Diffusion Tensor Imaging \citep{lenglet2006}.  
Data sets of covariance matrices arise in a number of different experimental contexts, and \cite{dryden2009non} describe several alternative geometries for the space of $p\times p$ covariance matrices. 
Some of these geometries take rank deficiency into account, but this can make analysis more involved since data then lie on a manifold-stratified space \citep[Section 6]{marron2014overview}. 
Since we assume the data consist of full-rank matrices, it follows that they lie on a manifold, and in addition, for reasons  described in Section~\ref{sec:aigeom}, we adopt the affine invariant geometry as the Riemannian structure on the manifold.
Under this geometry there are closed formulae for the various operations we require in order to specify our model, and moreover the geometry possesses symmetry properties that are advantageous for the application to EEG data. 

A variety of statistical methods have been developed for manifold-valued data. 
The definition of the sample mean in Euclidean space does not generally apply on a manifold, so an intrinsic mean was developed \citep{frechet1948elements}, together with associated theory for asymptotic properties \citep{bhattacharya2003large,bhattacharya2005large}. 
Analogs of principal component analysis \citep{fletcher2004principal,huckemann2006principal} and linear regression \citep{thomas2013geodesic} have been proposed, replacing straight lines in Euclidean space with geodesics in Riemannian manifolds, and using other higher dimensional subspaces to replace linear Euclidean subspaces \citep{jung2012analysis}. 
Manifold-valued Gaussian processes have been developed and applied to functional data on manifolds \citep{mallasto2018wrapped}. 
Similarly, there are existing methods for learning trajectories from manifold-valued longitudinal data \citep{NIPS2015_186a157b}.  
Recently, continuous-time manifold valued Markov processes which are analogs of the Euclidean Ornstein-Uhlenbeck process have been developed \citep{bui2021inference}, and our model for time series data includes a related mean reverting term. 
Autoregressive time series models for brain dynamics which are not manifold-adapted have been applied more widely. 
For example in \cite{costa2017} an autoregressive model is fitted which enables estimation of an underlying time-varying graph of interactions between brain regions. 

\begin{figure}
\begin{center}
\includegraphics[width=0.8\textwidth,trim={0.3cm, 0.3cm, 0.3cm, 0.3cm},clip]{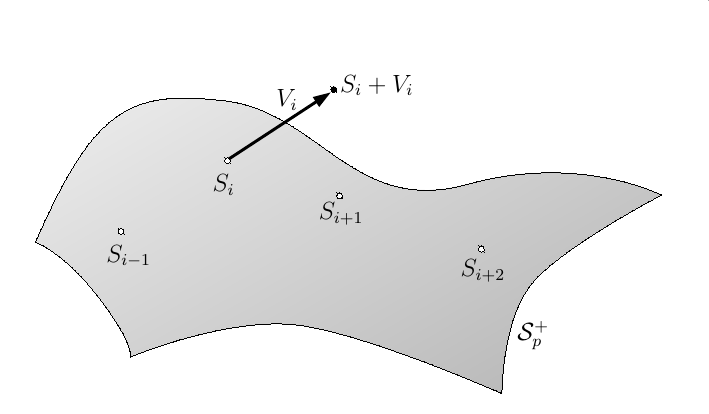}
\begin{caption}{\label{fig:manifold}
Diagram representing a time series $S_i$ of covariance matrices. 
The ambient space is the space of $p\times p$ symmetric matrices $\sym{p}$, while the shaded surface represents the manifold of symmetric positive definite matrices $\spd{p}$ embedded within $\sym{p}$. 
A perturbation of $S_i$ by an arbitrary estimated `velocity' $V_i\in\sym{p}$ does not necessarily lie within the manifold $\spd{p}$.}
\end{caption}
\end{center}
\end{figure}

Let $\sym{p}$ denote the set of (real) symmetric $p\times p$ matrices and let $\spd{p}\subset\sym{p}$ denote the subset of positive definite matrices. 
When modelling a time series $S_i\in\spd{p}$, $i=1,\ldots,n$, it is natural to consider representing the $(i+1)$th matrix as
\begin{equation}\label{equ:additivemodel}
S_{i+1} = S_i + V_i + \epsilon_i
\end{equation}
where $V_i\in\sym{p}$ is a deterministic perturbation which might depend on preceding matrices in the time series or some covariates, and where $\epsilon_i$ represents noise. 
For example, a very simple model is defined by $V_i = S_{i}-S_{i-1}$, so that $V_i$ is an estimate of the direction of travel of the time series at the $i$th time point, and by assuming that the elements of $\epsilon_i$ are distributed according to some appropriate multivariate normal distribution. 
However, without additional constraints on $V_i$ and $\epsilon_i$, the right-hand side of Equation~$\eqref{equ:additivemodel}$ will not generally be positive definite, and so the model is ill-defined in the sense that it gives probability mass to regions of the sample space that cannot contain data. 
The additive model in Equation~$\eqref{equ:additivemodel}$ therefore needs modifying in order to take into account the non-Euclidean nature of the space of positive definite $p\times p$ matrices (see Figure~\ref{fig:manifold}). 
This can be achieved canonically by regarding $\spd{p}$ as a Riemannian manifold, and working instrinsically within the Riemannian geometry.  
The notion of perturbing a matrix $S_i\in\spd{p}$ in some direction $V_i\in\sym{p}$, expressed in Equation~$\eqref{equ:additivemodel}$ by matrix addition, is instead achieved by the Riemannian exponential map which operates intrinsically within $\spd{p}$ (see Section~\ref{sec:Riemannian} for background on Riemannian geometry). 
As a key part of this study we analyse time series $\{S_i\}$ with the standard Euclidean geometry on $\sym{p}$, and compare the results to those obtained when the model is adapted to work intrinsically within $\spd{p}$ with the structure of a Riemannian manifold. 

In addition to assessing the effect of the underlying geometry, we propose a model which is capable of capturing different dynamics within the time series $S_i$, with the aim of providing an interpretable set of parameters for each data set. 
Specifically, the tangent (or direction) vector $V_i$ is modelled as the sum of three terms: 
\begin{enumerate} 
\item An autoregressive term which combines direction vectors at previous lagged time points. 
For certain parameter values, this term can, for example, give rise to approximate movement along a geodesic in the manifold if this terms dominates the other two terms.
In the Euclidean setting, this would correspond to drift in an approximately constant direction.
\item A mean reverting term. 
It is plausible that the brain has some `mean' state
about which it varies, and that observed EEG data should also have a tendancy
to return to a certain point or region of the state space.
In the absence of the autogressive term, the mean reverting and noise term could be considered as a discrete-time analog of an Ornstein-Uhlenbeck process. 
\item A noise term. This is a stochastic term with a multivariate normal distribution on the tangent space, which gives rise to the so-called wrapped Gaussian distribution on the manifold (see Section~\ref{sec:statsonmfds}). 
If this term dominates, then the time series will resemble a type of random walk over the manifold.
\end{enumerate}
By combining these terms, the model is able to capture various possible dynamics for the evolution of EEG data, from smooth flow along geodesics to a noisy mean reverting random walk on the underlying manifold, for example. 
More details are given in Section~\ref{sec:generalmodel}.
Although we apply the model to EEG data in $\spd{p}$, it is formulated in a general way and could be applied to data on other Riemannian manifolds.

The remainder of the paper is structured as follows. 
In Section~\ref{sec:background} we review essential elements of Riemannian geometry and statistics on manifolds and describe the affine invariant geometry on $\spd{p}$. 
In Section~\ref{sec:data} we describe dimensional reduction methods and perform an initial analysis of the seizure and interictal time series using multidimensional scaling. 
Section~\ref{sec:model} defines our model for manifold-valued time series and explains parameter inference. 
The model is fitted to the EEG times series and the results are given in Section~\ref{sec:results}. 
Finally, in Section~\ref{sec:conclusion} we discuss general findings, limitations of the model and potential for further research.

\section{Background}\label{sec:background}

\subsection{Riemannian geometry}\label{sec:Riemannian}

This section briefly describes the elements of Riemannian geometry required for our analysis. 
We bypass many of the details, but these can be found in standard textbooks such as those by \cite{do_carmo_riemannian_1992} and \cite{lang_fundamentals_1999}. 
We present this material for a general Riemannian manifold, and then in Section~\ref{sec:geometries} we give explicit formulae for the affine invariant geometry on $\spd{p}$. 

\subsubsection{Definitions}

Smooth manifolds are spaces which locally resemble $\R^m$ for fixed dimension $m$, but which are more general than vector spaces in order to incorporate curvature. 
Each smooth manifold $\mfd$ is equipped with charts; a chart is a map $\varphi:U\subset\mfd\rightarrow\varphi(U)\subset\R^m$. 
The union of the chart domains covers the manifold $\mfd$, and the charts satisfy a compatibility condition: for every pair of charts $(\varphi,U)$, $(\psi,V)$ the transition map $\psi\circ\varphi^{-1}:\varphi(U\cap V)\rightarrow \psi(U\cap V)$ is bijective, smooth and with smooth inverse. 
A path $\gamma:[a,b]\rightarrow\mfd$ is then said to be smooth if its composition with every chart $\varphi\circ\gamma:[a,b]\rightarrow\R^m$ is a smooth map. 
The tangent space $\tgt{x}$ at $x\in\mfd$ can be thought of as the vector space of direction vectors of smooth paths through $x$.  

A Riemannian metric $g$ on $\mfd$ is an assignment of an inner product $g_x$ on $\tgt{x}$ which varies smoothly with $x\in\mfd$. 
Given a smooth path $\gamma:[a,b]\rightarrow\mfd$, the norm of its velocity is $\{g_{\gamma(t)}(\dot{\gamma}(t),\dot{\gamma}(t))\}^{1/2}$ at $t\in[a,b]$, and integrating this gives the length of $\gamma$. 
Given two points $x,y\in\mfd$, let $d_g(x,y)$ denote the infimum of path length over the set of smooth paths connecting $x$ to $y$. 
It can be shown that $d_g$ is a metric on $\mfd$. 
Any path between $x$ and $y$ which realizes the infimum is called a geodesic; it is a shortest path between $x$ and $y$. 

The Riemannian exponential map at $x$, $\Expx:\tgt{x}\rightarrow \mfd$ is defined for all tangent vectors with sufficiently small norm, and can be thought of as the result of firing a geodesic from $x$ in the direction of a particular tangent vector. 
Specifically, given $v\in \tgt{x}$, the path $\Expx(tv)$ defined on $t\in[0,1]$ is a geodesic from $x$ to $\Expx(v)$. 
The inverse of the the exponential map at $x$, called the Riemannian logarithm map, is denoted $\Logx:\mfd\rightarrow \tgt{x}$. 
In general, it is defined on some neighbourhood of $x$. 
For two points $x,y\in\mfd$, the logarithm map $\Logx(y)$ can be thought of as the tangent vector at $x$ to the geodesic from $x$ to $y$. 
For the manifolds we consider, the exponential map is in fact defined on the whole of each tangent space $\tgt{x}$ for all $x\in\mfd$, and the logarithm map $\Logx(y)$ is similarly defined for all $x,y\in\mfd$. 

Given a curve $\gamma$ between two points $x,y\in\mfd$, the Riemannian geometry yields a natural isomorphism from $\tgt{x}$ to $\tgt{y}$ called \emph{parallel translation} along $\gamma$. 
Throughout we will assume that $\gamma$ is the geodesic from $x$ to $y$, and for the manifolds we consider this gives a well-defined parallel translation map defined for all $x,y$ and denoted $\ptrans{x}{y}:\tgt{x}\rightarrow \tgt{y}$. 
It is an isometry in the sense that it preserves the Riemannian inner product $g$. 
It follows that given an orthonormal basis $\{v_i : i=1,\ldots,m\}$ of $\tgt{x}$, then for all $y\in\mfd$ the vectors $w_i=\ptrans{x}{y}(v_i)$ form an orthonormal basis of $\tgt{y}$. 

A Riemannian manifold which admits a globally defined smooth parallel orthonormal basis of vectors fields is called a \emph{parallelizable} manifold. 
A manifold which is parallelizable is not necessarily diffeomorphic to Euclidean space: examples of parallelizable manifolds include certain spheres $S^m$ and $\spd{p}$ with the affine invariant geometry.

\subsubsection{Statistics on Riemannian manifolds}\label{sec:statsonmfds}

To complete this section, we require some concepts for doing statistics on Riemannian manifolds. 
There are several different ways to define analogs of Gaussian distributions on manifolds; we will adopt an approach first developed in the context of directional data \citep{jupp2009directional}, and use \emph{wrapped} Gaussian distributions. 
Suppose $\{v_i : i=1,\ldots,m\}$ is an orthonormal basis of $\tgt{x}$ and $\Sigma$ is a $m\times m$ positive semi-definite matrix. 
Then a random point $X\in\mfd$ has the \emph{wrapped Gaussian distribution} centred at $x\in\mfd$ with covariance $\Sigma$ in the basis $\{v_i\}$ if
\begin{align}
Y &=(Y_1,\ldots,Y_m) \sim N(0,\Sigma),\label{equ:zeromeangaussian}\\
U &= \sum_i Y_iv_i\in\tgt{x},\quad\text{and}\\
X &= \Expx\left(U\right).
\end{align}
Here $N(0,\Sigma)$ is the zero-mean multivariate Gaussian distribution on $\R^m$ with covariance $\Sigma$. 
When $\Sigma = \sigma^2 \Id{m}$ for some $\sigma>0$, then the distribution is defined independently from the choice of orthonormal basis.
The wrapped Gaussian distribution represents a bump of density around the point $x$ with the degree of spread and shape of the density contours determined by the matrix $\Sigma$. 
In fact we will work with a related set of distributions obtained by adding a non-zero mean $w\in\R^m$ to the Gaussian in Equation~$\eqref{equ:zeromeangaussian}$. 
In this case we write 
\begin{equation}\label{equ:mvntangent}
U\sim N_{T_x\mfd}(w,\Sigma)
\end{equation}
to denote the distribution of the random tangent vector $U$. 
The resulting distribution of the random point $X= \Expx\left(U\right)$ closely approximates the wrapped Gaussian with zero mean centred at the point $\Expx(\sum w_iv_i)$ when the vector $w$ is sufficiently small. 

Since the addition operation is not defined for points on a general manifold, the definition of a sample mean needs to be adapted. 
Given a set of data points $x_i\in\mfd$, $i=1,\ldots,n$, define the \emph{Fr\'echet function} $\mathcal{F}$ by
\begin{equation}\label{equ:frechetfunction}
\mathcal{F}(y; \{x_i\}) = \frac{1}{n}\sum_{i=1}^n d_g(y,x_i)^2
\end{equation}
where $d_g$ is the geodesic distance. 
A point $x=\argmin_{y}\mathcal{F}(y; \{x_i\})$ is called a \emph{Fr\'echet sample mean} of the data $\{x_i\}$. 
The value of $\mathcal{F}$ at the Fr\'echet sample mean is called the \emph{Fr\'echet sample variance}. 
Fr\'echet sample means do not necessarily exist for all data sets on a general manifold, nor are they generally unique. 
However, the Fr\'echet sample mean exists and is unique on $\spd{p}$ with the affine invariant geometry \cite{sturm_probability_2003}. 
Of course, on $\R^p$ with the Euclidean metric, the Fr\'echet sample mean and standard sample mean coincide.

\subsection{Geometry on $\sym{p}$ and $\spd{p}$}\label{sec:geometries}

Each time series $\{S_i:i=1,\ldots,n\}\in\spd{p}$ of covariance matrices can be analysed within $\sym{p}$ (giving a standard Euclidean model) or intrinsically within $\spd{p}$ viewed as a Riemannian manifold. 
In fact, $\sym{p}$ with its standard Euclidean geometry can be specified as a Riemannian manifold, and it is convenient to present it in that way for comparison with the affine invariant geometry on $\spd{p}$ -- this forms Section~\ref{sec:eucgeom} below.  
Then in Section~\ref{sec:aigeom} we give closed formulae for Riemannian operations in the affine invariant geometry on $\spd{p}$. 
The nesting of the spaces $\spd{p}\subset\sym{p}$ represents an increasing refinement of analysis, and is associated with increasing computational cost. 

\subsubsection{The Euclidean geometry on $\sym{p}$}\label{sec:eucgeom}

When $\mfd=\sym{p}$ it is straightforward to show that every tangent space is a copy of $\sym{p}$. 
We define a Riemannian metric at $S\in\sym{p}$ by
\begin{equation*}
\geuc{V}{W}{S} = \trace \left(VW\right)
\end{equation*}
for $V,W\in\tgt{S}=\sym{p}$ 
so that the inner product is in fact independent of the base point $S$. 
The corresponding Riemannian geometry is exactly that of the symmetric matrices equipped with the standard Euclidean (or Frobenius) distance. 
Specifically, it is straightforward to show that for all $S,S_1,S_2,V\in\sym{p}$ :
\begin{align*}
\deuc{S_1}{S_2} &= \norm{S_1-S_2}\\
\intertext{where $\|\cdot\|$ is the Frobenius norm;}
\Exp_S(V) &= S+V\\
\Log_{S_1}(S_2) &= S_2-S_1;\\
\intertext{and}
\ptrans{S_1}{S_2}(V) &= V.
\end{align*}
Substituting $tV$ for the tangent vector in the exponential map for any $t\geq 0$ gives the geodesic from $S$ in direction $V$ as $S+tV$, i.e.~the straight line segmet in direction $V$. 

If $e_1, \ldots,e_p$ denotes the standard basis of $\R^p$, then the matrices
\begin{equation}\label{equ:EucONbasis}
E_{qr} = \begin{cases}
\frac{\sqrt{2}}{2}(e_qe_r^T+e_re_q^T) & \text{when $q\neq r$, and}\\
e_qe_q^T & \text{when $q=r$}
\end{cases}
\end{equation}
for $1\leq q \leq r \leq p$
define an orthonormal basis of $T_S\sym{p}$ for all $S\in\sym{p}$. 

The Fr\'echet sample mean of matrices $S_1,\ldots,S_n\in\sym{p}$ is the standard Euclidean mean $\frac{1}{n}\sum S_i$. 
Since positive definiteness is a convex property, if $S_i\in\spd{p}$ for all $i$ it follows that the Euclidean mean is also positive definite.

\subsubsection{The affine invariant geometry on $\spd{p}$}\label{sec:aigeom}

The affine invariant, or Fisher-Rao, geometry on $\mfd=\spd{p}$ arises naturally from the information geometry of multivariate normal distributions with zero mean \citep{skovgaard1984riemannian,lenglet2006}. 
The metric satisfies the property 
\begin{equation}\label{equ:aiproperty}
\daff{AS_1A^T}{AS_2A^T}=\daff{S_1}{S_2}
\end{equation} 
for all $S_1,S_2\in\spd{p}$ and $A\in GL(p)$, so in other words, it is invariant under changes of basis in $\R^p$ when the $S_i$ are covariance matrices.  
This is particularly relevant when analysing covariance matrices from EEG recordings: the overall scale of measurements is affected by
instrument sensitivity and positioning which between electrodes and so is largely artefactual. 
The metric is invariant under re-scaling of the signal from each electrode, which corresponds to the the action of a diagonal matrix $A$ with positive entries in Equation~$\eqref{equ:aiproperty}$. 
The affine invariant geometry offers additional advantages over the Euclidean geometry, for example, in that the determinant of matrices along geodesics is better behaved. (In the Euclidean geometry the determinant can become inflated relative to its value at the end points of the geodesic.)

Suppose $S,S_1,S_2\in\spd{p}$. 
It is straightforward to show that $T_S\spd{p}=\sym{p}$ for all $S$. 
The Riemannian inner product on $T_S\spd{p}$ is defined by
\begin{equation}\label{equ:AIinnerprod}
\gaff{V}{W}{S} = \trace \left(S^{-1}VS^{-1}W\right)
\end{equation}
where $V,W\in T_S\spd{p}=\sym{p}$. 
It follows that
\begin{equation*}
\daff{S_1}{S_2} = \left\| S_1^{-\frac{1}{2}}S_2S_1^{-\frac{1}{2}} \right\|
= \left({\sum_{r=1}^{p}\log^2\lambda_i}\right)^\frac{1}{2}
\end{equation*}
where  $\lambda_r, r =1,...,p$ are the eigenvalues of $S_1^{-1/2}S_2S_1^{-1/2}$; 
the exponential map is
\begin{equation}\label{equ:aiexp}
\Exp_S(V) = S^{\frac{1}{2}}\exp (S^{-\frac{1}{2}}VS^{-\frac{1}{2}})S^{\frac{1}{2}}
\end{equation}
where $\exp$ denotes the matrix exponential; and the log map is 
\begin{equation*}
\Log_{S_1}\left({S_2}\right) = S_1^\frac{1}{2}\log\left( S_1^{-\frac{1}{2}}S_2S_1^{-\frac{1}{2}} \right)S_1^\frac{1}{2}
\end{equation*}
where $\log$ denotes the matrix logarithm. 
The parallel transport of $V\in T_{S_1}\spd{p}$ to $T_{S_2}\spd{p}$ is
\begin{equation}\label{equ:AIparatrans}
\ptrans{S_1}{S_2}(V) = (S_2S_1^{-1})^{\frac{1}{2}}V(S_2S_1^{-1})^{\frac{1}{2}}.
\end{equation}
More details are given by \cite{skovgaard1984riemannian} and \cite{lenglet2006}; \cite{yair2019parallel} give details of the calculation of the parallel transport. 

An orthormal basis of $T_{S}\spd{p}$ for each $S\in\spd{p}$ can be defined via parallel translation of a choice of orthonormal basis of the tangent space at the identity matrix $\Id{p}\in\spd{p}$. 
The matrices $E_{qr}$, $1\leq q \leq r \leq p$, defined in Equation~$\eqref{equ:EucONbasis}$ determine an orthormal basis of the tangent space at $\Id{p}$ with respect to the Riemannian inner product~$\eqref{equ:AIinnerprod}$. 
Using parallel translation from $\Id{p}$ to a general point $S\in\spd{p}$, (Equation~$\eqref{equ:AIparatrans}$), we obtain an orthormal basis of $T_{S}\spd{p}$ defined by
\begin{equation}\label{equ:parallelbasis}
\hat{E}_{qr}(S) = S^{\frac{1}{2}}E_{qr}S^{\frac{1}{2}}.
\end{equation}
This basis can be thought of as a global smooth section of the frame bundle of the Riemannian manifold $(\spd{p},g^\mathrm{aff})$. 
By definition, it follows that $(\spd{p},g^\mathrm{aff})$ is parallelizable. 
The basis in~$\eqref{equ:parallelbasis}$ will be indexed via $i=1,\ldots,m=\tfrac{1}{2}p(p+1)$ with $i=\tfrac{1}{2}r(r-1)+p$, and we define $\omega_i(S)=\hat{E}_{qr}(S)$.
 
The Fr\'echet sample mean of any sample $S_1,\ldots,S_n\in\spd{p}$ exists and is unique since $(\spd{p},g^\mathrm{aff})$ has non-positive sectional curvatures \citep{lenglet2006}. 
The Fr\'echet function, defined by Equation~$\eqref{equ:frechetfunction}$ on a general metric space, can be differentiated on $\spd{p}$ to obtain a gradient function. 
Gradient descent can then be used to compute the Fr\'echet mean, and we use the algorithm specified by \cite{lenglet2006} in our analysis.

\section{Data reduction and preliminary analysis}\label{sec:data}

\subsection{EEG data set}

In this study we analyse an open-source dataset of patients with drug-resistant focal epilepsy available from \url{http://ieeg-swez.ethz.ch/}, which was collected at the Sleep-Wake-Epilepsy-Center (SWEC) of the University Department of Neurology at Inselspital Bern. 
The full dataset includes 116 seizures captured during 2656 hours of long-term intracranial electroencephalography (EEG) recordings from 18 patients  \citep{burrello2019laelaps}. 
The Supplementary Material provides further information, such as the duration of recordings, number of electrodes and seizures, and seizure duration for each patient.
The full data set is very large, with data recorded hundreds of times a second for thousands of hours across tens of electrodes in each patient. 
An EEG recording for a single patient, after some pre-processing, consists of a time series $z_i\in\R^q$ $i=1,\ldots,n_z$,  where $i$ is the index across time.
The coordinates $j=1,\ldots,q$ are called channels, with each channel recording activity at a specific electrode. 
Each time series contains seizures, together with the inter-seizure or \emph{interictal} periods. 
 
We model each seizure independently within patients and between patients. 
In order to better understand differences in seizure evolution, for each seizure we identified a comparator interictal period exactly 2 hours before the seizure, and lasting for the same duration as the seizure. 
This comparator period was unavailable for three seizures out of the 116, and so these were removed from consideration. 
Additionally, one seizure with duration less than 10 seconds was also removed, giving a remaining set of 112 seizures and associated interictal periods.

\subsection{Dimensional reduction and covariance matrices}\label{sec:dimred}

Rather than modelling the raw time series $z_i$ of EEG signal, researchers often analyse covariance between channels as a way of compressing information about signals in all recording channels into a meaningful measure. The principal scientific interest is in dependencies between channels, and it is believed that this dependency reflects interactions between brain regions \citep{siddiqui2020review,bastos2016tutorial,chiarion2023connectivity}.
As a result, a sliding window method is typically used whereby the measurements $z_i$ lying in each window are used to calculate a sample covariance matrix for that window. Successive windows can then be understood as a time series, or trajectory in the space of covariance matrices, reflecting evolving `functional connectivity' \citep{schroeder2020seizure}.

In addition, some form of dimensional reduction is often performed over $\R^q$. 
This serves two purposes. 
First the computation time of analysis typically increases polynomially in the number of channels $q$, but $q$ can be large so dimensional reduction is required for computational feasibility. 
Secondly, certain linear combinations of channels can have very low variance. 
These represent a redundancy in the data set, and removing these speeds up analysis, as well as ensuring the covariance matrices obtained are strictly positive definite. 
The end result of these steps is a time series $S_i$, $i=1,\ldots,n$ of $p\times p$ positive definite correlation matrices where $p\leq q$, and $i$ now indices the sequence of windows. 

Dimensional reduction was performed by finding linear combinations of channels which maximized variation in the mean sample covariance matrix as follows. 
We assume that there are $f$ EEG measurements per second, and that we use non-overlapping sliding windows of exactly 1 second duration to compute covariance matrices, so that $n_z = f\times n$ where $n_z$ is the length of the time series $z_i$ and $n$ is the length of the time series $S_i$.
For $i=1,\ldots,n$ we compute the full $q\times q$ sample covariance matrices
\begin{equation*}
S_i' = \mathrm{Cov}\left( \left\{ z_{(i-1)f+1},\ldots,z_{if} \right\} \right).
\end{equation*}
We then seek linear combinations of channels $u\in\R^q$ which maximize $\tfrac{1}{n}\sum_i u^T S_i'u$ where $\cdot^T$ denotes the transpose. 
In other words, we find the eigenvectors $u_1,\ldots,u_p$ of $\tfrac{1}{n}\sum_i S_i'$ with the largest $p$ eigenvalues. 
The time series of covariance matrices $S_i$ is then defined as the sample covariance matrices of these linear combinations:
\begin{equation*}
S_i = \mathrm{Cov}\left( \left\{ Uz_{(i-1)f+1},\ldots,Uz_{if} \right\} \right) = US_i'U^T
\end{equation*}
where the $p\times q$ matrix $U$ has rows $u_1^T,\ldots,u_p^T$. 
We also tested a method of dimensional reduction that identified sets of channels, rather than linear combinations of channels, but this gave inferior results (see Supplementary Material). 

For the interictal time series, dimensional reduction was performed (i) independently from the corresponding seizure time series and (ii) with the $U$ matrix obtained from dimensional reduction of the seizure data, thereby giving two different comparator interictal time series for each seizure time series. 
The first set of interictal series retains a higher proportion of variance, but parameters estimated from the data, such as the Fr\'echet mean covariance matrix cannot be directly compared to estimates obtained from the seizure data since they are in different coordinates. 
The second set of interictal series retains a lower proportion of variance, but direct comparisons can be made with the corresponding seizure data. 

We chose to use a value of $p=15$ for all patients and seizures as a compromise between the computation burden of model fitting which increases with $p$, and the proportion of variance retained in the reduced data sets. 
For $p=15$ the proportion of variance in the reduced data sets is over $80\%$ for more than $80\%$ of the seizure time series (see Supplementary Material). 
From this point on in the paper, we restrict only to the dimensionally reduced data sets with $p=15$. 
Some analysis was also performed with $p=20$ for a few time series, but the results obtained were very similar indeed to those obtained using $p=15$. 

\subsection{Data exploration}

An initial analysis of the seizure and interictal data was carried out in the Euclidean and affine invariant geometries separately. 
Among other approaches, this included: (i) calculating Fr\'echet means and variances, and (ii) performing multidimensional scaling. 
Due to the large quantity of data, we do not give a full set of results but instead give results for a representative selection of patients. 
Throughout the paper, results will be given for patient 18 in particular, and for patients 6, 7, and 13 in the Supplementary Material. 
These patients were selected on the basis of the number and duration of seizures for the patients being representative. 

\begin{figure}
\begin{center}
\includegraphics[width=0.8\textwidth]{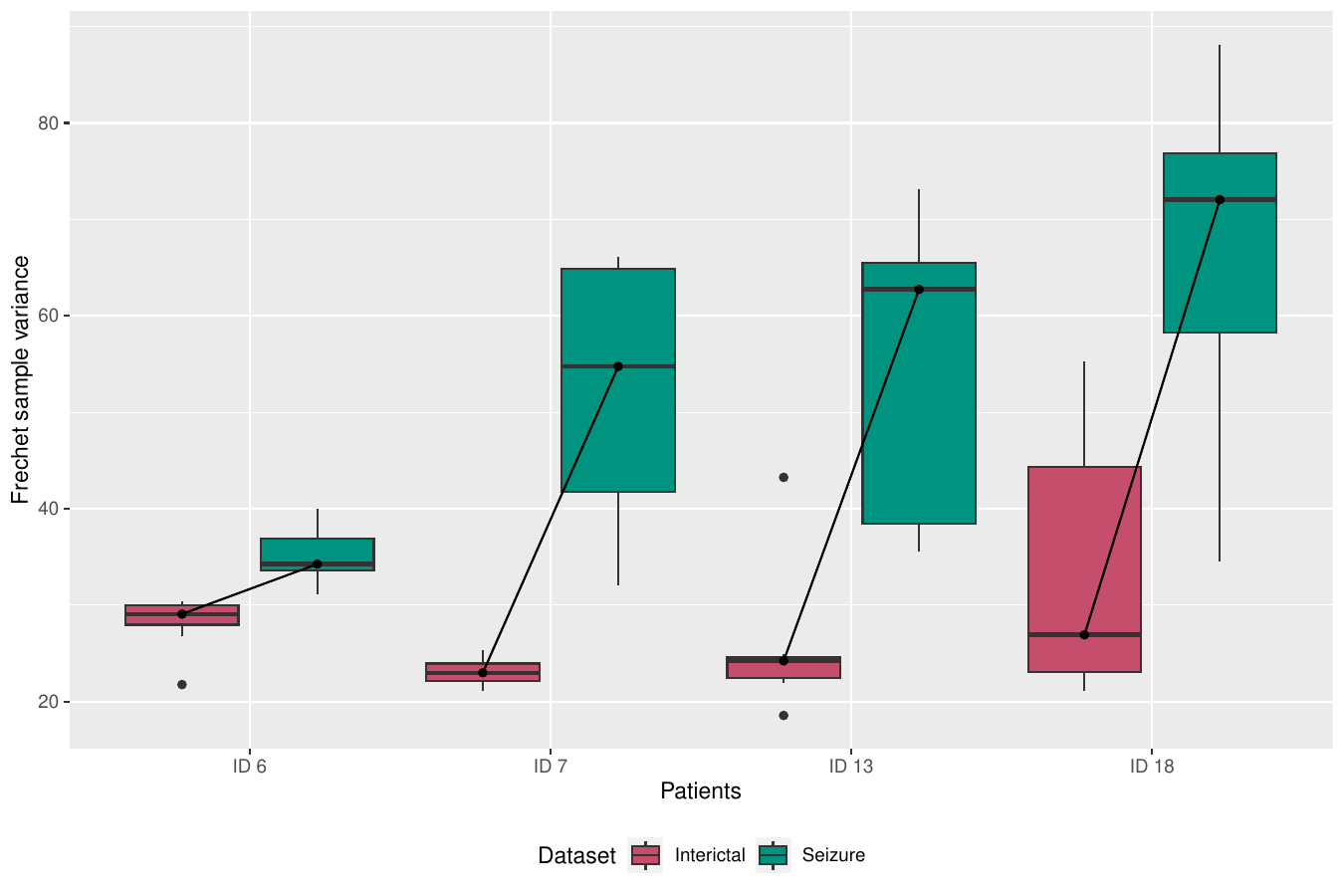}
\begin{caption}{
\label{fig:FSV}
The Fr\'echet sample variance for each seizure time series and the corresponding interictal time series, broken down by patient id (patients 6, 7, 13 \& 18), using the affine invariant metric. 
Dimensional reduction was performed independently for the interictal series. 
The median points of the interictal and seizure distributions for each patient are linked by a line. 
}
\end{caption}
\end{center}
\end{figure}

Figure~\ref{fig:FSV} shows the Fr\'echet sample variance for each seizure time series and the corresponding interictal time series, broken down by patient id for these four patients, using the affine invariant metric. 
(See Supplement for the same plot for the Euclidean geometry.)
The plot shows that there is a substantial increase in variance in the seizure time series compared to interical series. 
However, there is also substantial variation between patients.

\begin{figure}
\begin{center}
\includegraphics[width=0.8\textwidth]{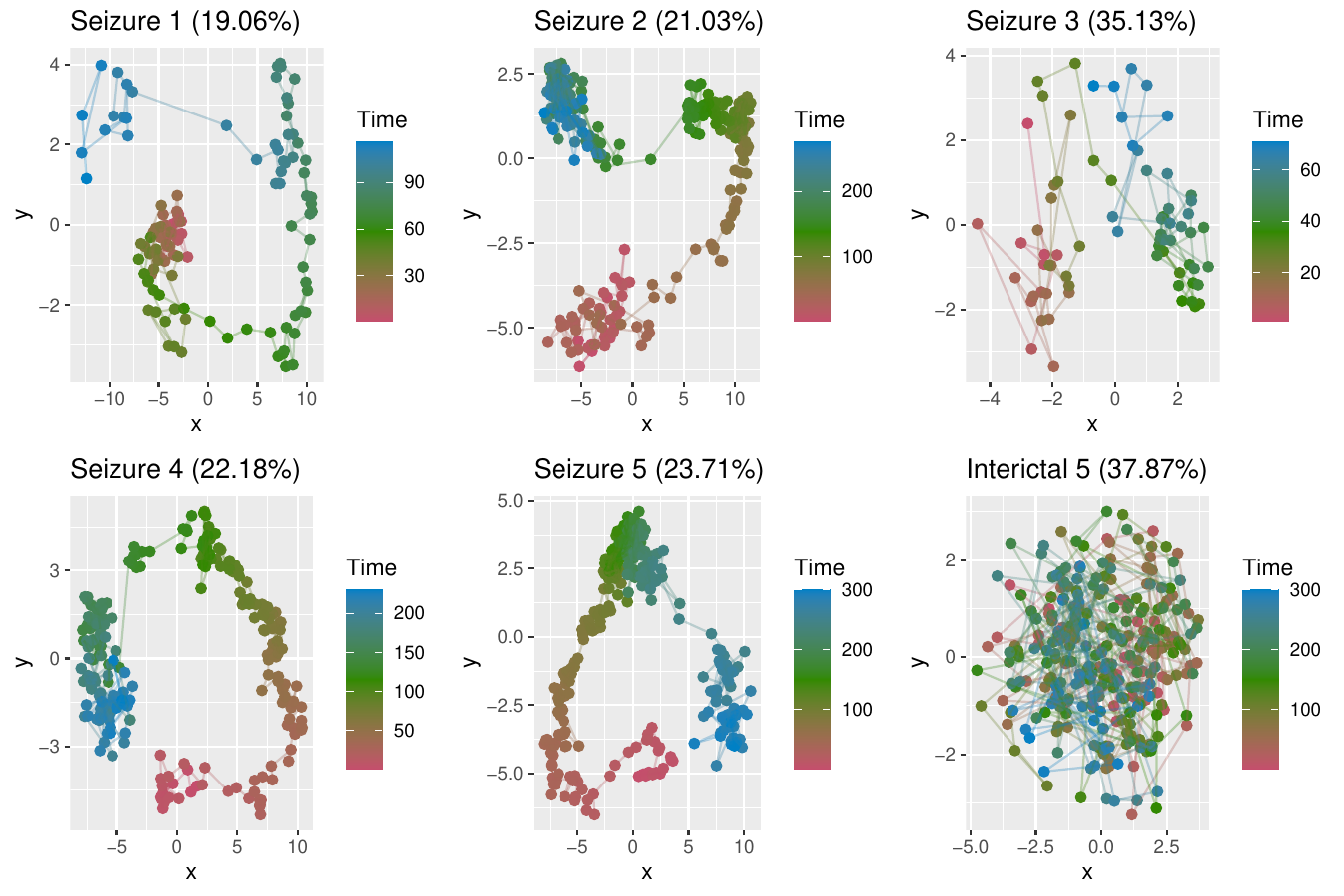}
\begin{caption}{
\label{fig:MDS18AI}
Multidimensional scaling (MDS) of seizure time series for Patient 18 using the affine invariant metric to construct the matrix of distances $\daff{S_i}{S_j}$. 
Patient 18 had 5 seizures (first 5 panels). 
Panel 6 shows MDS results for the interictal period corresponding to seizure 5. 
Plots are coloured to show the development over time. 
Each panel title shows the proportion of variance represented by the 2-dimensional MDS. 
}
\end{caption}
\end{center}
\end{figure}

\begin{figure}
\begin{center}
\includegraphics[width=\textwidth]{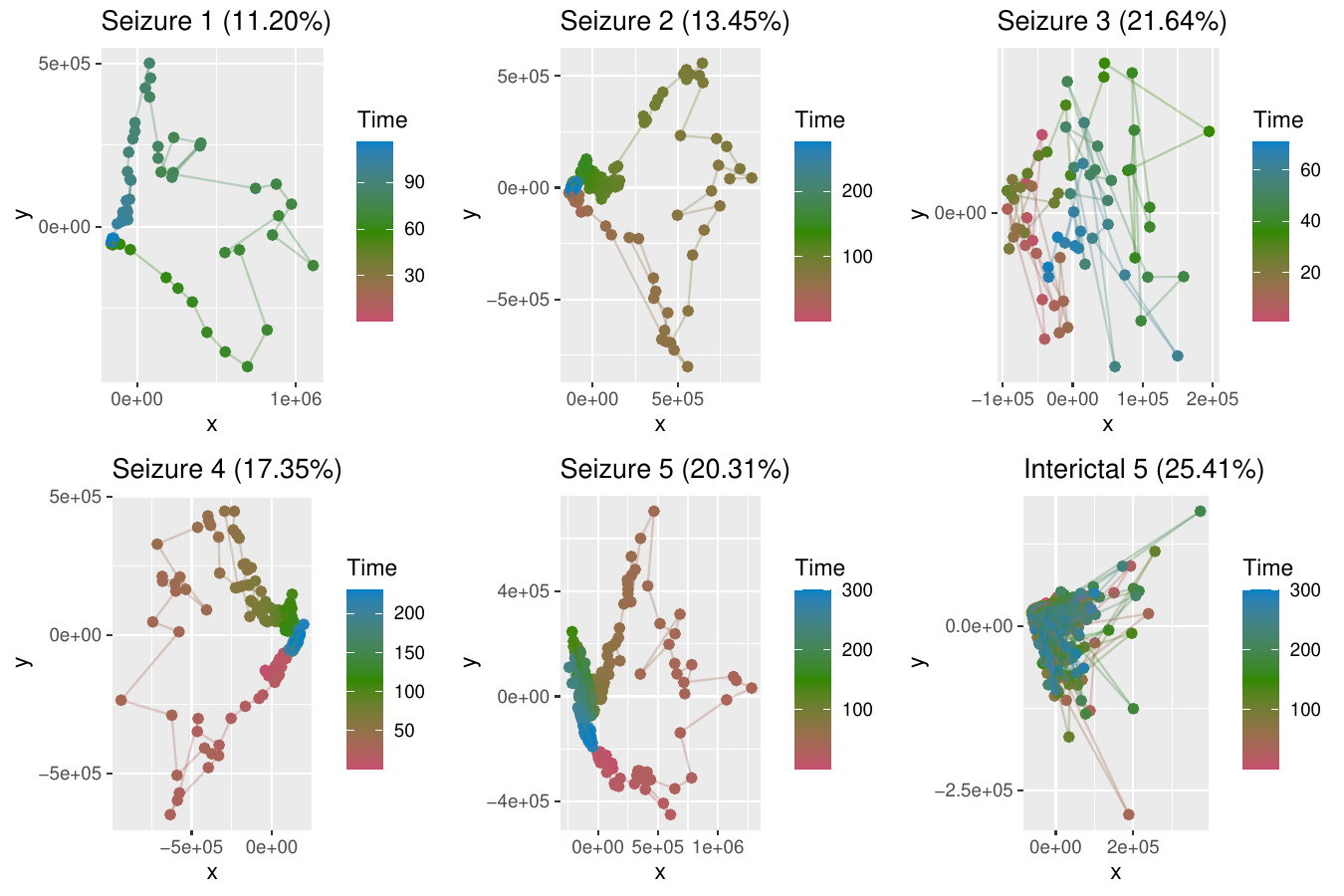}
\begin{caption}{
\label{fig:MDS18Euc}
Multidimensional scaling (MDS) of seizure time series for Patient 18 using the Euclidean metric to construct the matrix of distances $\deuc{S_i}{S_j}$. 
Other details are as for Figure~\ref{fig:MDS18AI}.
}
\end{caption}
\end{center}
\end{figure}

Figures~\ref{fig:MDS18AI} and~\ref{fig:MDS18Euc} show the results of metric multidimensional scaling (MDS) \citep{mardiamultivariate} of the seizure time series and one interictal series for patient 18, using the affine invariant and Euclidean metrics respectively. 
The interictal time series was prepared with dimensional reduction independent from the seizure time series. 
MDS plots with the affine invariant metric for seizures 1, 2, 4, and 5 are quite similar: the plots show evolution in each case along a curving trajectory but with a degree of local noise. 
Seizure 3 is of shorter duration and a curved trajectory is less apparent. 
For the interictal period, there is no apparent trajectory over time, and the plot largely resembles noise. This was also the case for the interictal periods for the other seizures.  
The proportion of variance captured by each 2-dimensional MDS is low, varying from around $20\%$ to $35\%$ between the plots.  
The plots for the Euclidean metric are rather different, since the distances between successive points vary very substantially over time for seizures 1, 2, 4 and 5. 
As a result, certain sub-intervals of time dominate each MDS plot.  
The MDS plot for the interictal period for seizure 5 has a cone-like shape, and seems to be dominated by relatively few distant time points. 

The MDS plots for patient 18 and other patients (see Supplementary Material) suggest that during some seizures the time series of covariance matrices evolves along a directed flow or curve in the space of covariance matrices combined with a degree of local noise, whereas the interictal time series resembles noise around some fixed point.  
Our model is constructed with the aim of capturing and characterising these different modes of evolution, as described in the next section.

\section{Model specification and inference}\label{sec:model}

In this section we propose a model for a time series $S_1,\ldots, S_n\in\mfd$ for a parallelizable Riemannian manifold $(\mfd,g)$. 
A simpler version is also specified for manifolds which are not parallelizable. 
The model is formulated in terms of distributional assumptions for $S_{k+1}$ in terms of the preceding values $S_1, \ldots, S_k$. 
Specifically, we develop a linear model in the tangent space $T_{S_k}\mfd$ for the direction vector $V_k=\Log_{S_k}(S_{k+1})$ of $S_{k+1}$ from $S_k$. 
As described in the Introduction, and motivated by the MDS results above, the expression for $V_k$ has three terms, each of which corresponds to a different possible type of dynamics for time series: a vector autoregressive term, a mean reverting term and a noise term.  

\subsection{General model}\label{sec:generalmodel}

Here we set up notation and specify a general version of our model on a parallelizable Riemannian manifold $(\mfd,g)$.  
Due to the potentially very large number of parameters for the general model, in the next subsection we describe certain lower dimensional reductions of the general model which we fit to data later in the paper. 

Let $V_i = \Log_{S_i}(S_{i+1})\in T_{S_i}\mfd$, $i=1,\ldots,n-1$, be the direction vector at time $i$ and define
\begin{equation*}
V_{i\ell} = \ptrans{S_{i-\ell}}{S_i}\left( V_{i-\ell} \right)
\end{equation*} 
for $\ell = 1,\ldots,L$ and $i>L$. 
Here $L$ is some fixed maximum lag, and $V_{i\ell}$ is the parallel translation of the direction vector $V_{i-\ell}$ at $S_{i-\ell}$ to the point $S_i$ at time $i$. 
The model for $S_{k+1}$ is specified by:
\begin{align}\label{equ:fullmodel}
S_{k+1} &= \Exp_{S_k}\left( V_k \right),\notag\\
V_k &= \sum_{\ell=1}^L A_\ell(S_k) V_{k\ell} + B(S)\, \Log_{S_k}(S^*) +\epsilon_k,\quad\text{and}\\
\epsilon_{k}&\sim N_{T_{S_k}\mfd}\left( 0, \Sigma_{S_k} \right).\notag  
\end{align}
The terms in the model are defined as follows. 
\begin{enumerate}
\item The sum $\sum_\ell A_\ell(S_k) V_{k\ell}$ is a manifold-adapted version of a Euclidean VAR model. 
For each $\ell$, $A_\ell(S)$ is an endomorphism of $T_S\mfd$, which varies smoothly with $S$. 
Each term in the sum is therefore the result of a linear map acting on the lagged direction vector $V_{i-\ell}$ parallel transported to $T_{S_i}\mfd$. 
Given a parallel orthonormal basis $\left\{\omega_i(S):i=1,\ldots,m\right\}$ of $T_S\mfd$, one way to determine the bundle endomorphisms $A_\ell(S)$ is via a constant matrix representation in this basis, and (with a slight abuse of notation) we let $A_\ell$ denote this matrix. 
Each matrix $A_\ell$ is of dimension $m\times m$ where $m$ is the dimension of $\mfd$ (so for example $m=\tfrac{1}{2}p(p+1)$ when $\mfd=\spd{p}$). 
Defined in this general way, the model potentially has a very large number of parameters, and we discuss lower dimensional parametrizations below. 

\item Like the terms $A_\ell(S)$, the term $B(S)$ is an endomorphism of $T_S\mfd$ which varies smoothly with $S$. 
The vector $\Log_{S_k}(S^*)$ is the direction from the current point towards a fixed `attractor' point $S^*\in\mfd$. 
The attractor point $S^*\in\mfd$ can either be fixed or estimated in some way from the data. 
We do not call $S^*$ a \emph{mean} point; it is not necessarily the Fr\'echet mean of a set of data generated according to the model, for example. 
In the same way as for the autoregressive terms, we can use a fixed $m\times m$ matrix $B$ and the parallel basis $\omega_i(S)$ to define $B(S)$ globally. 
In a Euclidean setting, and taking $B(S)$ to be a positive scalar multiple of the identity $B(S)=\beta I_m$, the term $\beta \Log_{S_k}(S^*)$ models a tendency to move towards the attractor point $S^*$, and we refer to this as a \emph{mean reversion} term. 
  
\item The distribution of the random tangent vector $\epsilon_k\in T_{S_k}\mfd$ is determined by a covariance matrix $\Sigma_{S_k}\in\sym{m}$, and this definition implicitly depends on the existence of some orthonormal basis of $T_{S_k}\mfd$. 
Using the parallel orthonormal basis $\omega_i(S)$, we can take $\Sigma_{S_k}$ to be a fixed matrix in this basis, denoted $\Sigma$. 
\end{enumerate}

Without the reversion term $B$, the model reduces to a standard Euclidean VAR model on the tangent space at an arbitrary point $S_0\in\mfd$: the existence of the parallel orthonormal basis enables the tangent vector $\Log_{S_k}(S_{k+1})$ between successive points in the time series to be parallel transported to $T_{S_0}\mfd$ from $S_k\in\mfd$. 
For example, when $\mfd=\spd{p}$ we could conveniently take $S_0= I_p$. 
However, when the reversion term is included, the model does not simplify in this way.

\subsection{Models with reduced parameters}

Due to the potentially very large number of parameters required by the general model in Section~\ref{sec:generalmodel}, in this section we describe two lower dimensional reductions, and these are the models which we fit to data in practice. 

\textbf{Diagonal model.} Working in an orthonormal parallel basis $\omega_i(S)$, $i=1,\ldots,m$, we assume the matrices $A_\ell$, $B$, and $\Sigma$ are all diagonal: 
\begin{align*}
A_\ell &= \diag{a_{\ell 1},\ldots,a_{\ell m}},\\
B &= \diag{b_1,\ldots,b_m}\qquad\textrm{and}\\
\Sigma &= \diag{\sigma_1^2,\ldots,\sigma_m^2}.
\end{align*}
It should be noted that these models are diagonal on the $m\times m$ dimensional tangent space $T\mfd$, so when $\mfd=\spd{p}$ we have $m=\tfrac{1}{2}p(p+1)$, and there is one parameter for every pair of coordinates $1,\ldots,p$.  
The orthonormal parallel basis for the affine invariant geometry is defined in Equation~$\eqref{equ:parallelbasis}$.
For $p=15$ we have $m=120$, so each matrix $A_\ell, B, \Sigma$ involves $120$ scalar parameters. 
The attractor point $S^*\in\spd{p}$ is the remaining model parameter. 

\textbf{Scalar coefficient model.} We assume the matrices $A_\ell$, $B$, and $\Sigma$ are all scalar multiples of the identity on $T_S\mfd$:
\begin{equation*}
A_\ell = \alpha_\ell I_m, \quad B=\beta I_m,\quad \Sigma = \sigma^2 I_m.
\end{equation*}
In fact, with this set of assumptions, the model is defined independently of the choice of orthonormal basis $\omega_i(S)$ at each point $S$ and so it can be applied on a manifold which is not parallelizable. 
As for the diagonal model, the attractor point $S^*\in\spd{p}$ is the remaining model parameter. 

\subsection{Parameter inference}\label{sec:inference}

For fixed $L$ and $S^*$, we can obtain maximum likelihood estimates for the remaining parameters, and estimation of these parameters reduces to a set of linear regressions within each tangent space. 
For the scalar coefficient model, an arbitrary basis of each tangent space can be used to obtain the regression equations, since the model parameters are basis independent. 
For the diagonal model the global parallel basis $\omega_i(S)$, $i=1,\ldots,m$ is required, and so for simplicity of presentation we will work in this basis. 
Working in this basis, a general tangent vector $V\in T_S\mfd$ has coordinate vector $\vec{v}$ such that the $i$th element is $g_S(V, \omega_i(S))$. 
In coordinates, the model for $V_k$ in Equation~\ref{equ:fullmodel} becomes
\begin{equation}\label{equ:modelincoords}
\vec{v}_k = \sum_{l=1}^L A_\ell\vec{v}_{k\ell} +B\vec{v}^*_{k}+\vec{\epsilon}_k, \quad k=L+1,\ldots,n
\end{equation}
where $\vec{v}_k$, $\vec{v}_{k\ell}$ and $\vec{v}^*_{k}$ are the coordinate vectors at $S_k$ for $V_k$, $V_{k\ell}$ and $\Log_{S_k}(S^*)$ respectively, and $\vec{\epsilon}_k\sim N(0,\Sigma)$. 
The Supplementary Material contains derivations of the maximum likelihood estimators for the scalar and diagonal model coefficients under Equation~$\eqref{equ:modelincoords}$ for fixed $L$ and $S^*$. 

Differentiating the likelihood function with respect to $S^*$ is not straightforward, and so joint maximum likelihood estimation of $S^*$ and the other parameters is challenging. 
As a result, when analysing an interictal time series $S_i$, $i=1,\ldots,n$, we fixed $S^*$ to be the Fr\'echet sample mean of the series. 
When analysing a seizure time series we fixed $S^*$ to be the Fr\'echet sample mean of the corresponding interictal series. 
(Note that each paired seizure and interictal series were prepared using the same dimensional reduction to ensure that the Fr\'echet sample mean of the interictal series was in the same basis as the seizure series.)
For comparison, we additionally analysed some seizure series by fixing $S^*$ to be the Fr\'echet sample mean of the seizure sample itself, but this little difference to the results. 
In particular, the $R^2$ values (defined below) were similar for both choices of $S^*$.  

For the scalar coefficient model, the maximum lag $L$ for each seizure or interictal series was selected as follows. 
Models were fitted for increasing values of $L=1,2,\ldots$. 
For each value of $L$, a $95\%$ confidence interval was constructed for $\alpha_L$ to test the hyothesis that $\alpha_L=0$. 
The value of $L$ for each time series was taken to be the largest for which the null hypothesis was not rejected. 
This simplistic approach was adopted since inference of $L$ was not the main focus of this research. 
Recently a more nuanced Bayesian approach to inference of $L$ (though not for manifold-adapted models) has been developed \citep{binks2023} for EEG data. 
In fact, in order to compare seizures we required a fixed value of $L$ to be used across all seizure series -- see the Section~\ref{sec:seizurecomparison}. 

\section{Application to EEG data}\label{sec:results}

\subsection{Scalar coefficient model}

The procedure for selecting the maximum lag $L$ gave a value of $L=0$ for the vast majority of interictal series. 
For seizures, the majority had $L=1$ or $2$.  
These results apply across both geometries. 
We chose to fit all models from this point onwards with $L=4$ for the following reasons. 
First, we needed to use the same value for all time series in order to compare seizures between patients (see below). 
Secondly, the value $L=4$ is greater than or equal to the value for the vast majority of series, though taking $L=4$ gives rise to estimated parameter values close to zero for larger values of $\ell$ for many time series. 

\begin{figure}
\begin{center}
\includegraphics[width=0.9\textwidth]{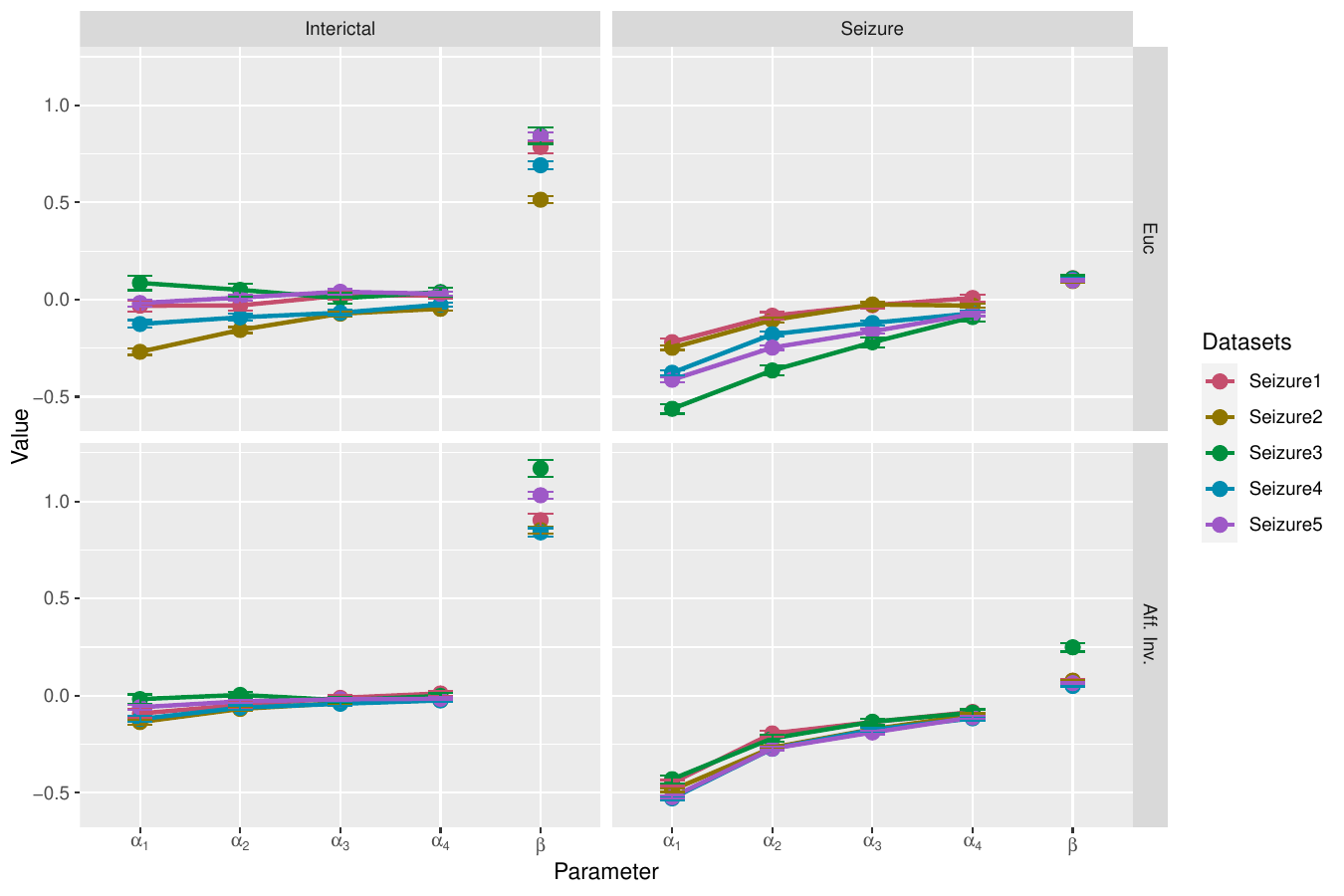}
\end{center}
\begin{caption}{
\label{fig:scalarmodelresults}
Scalar coefficients for patient 18: comparison of estimated values for fives seizures and corresponding interictal series, using Euclidean and affine invariant geometries. 
The value of model coefficients $\alpha_1,\ldots,\alpha_4$ and $\beta$ are shown with $95\%$ confidence intervals marked. 
}
\end{caption}
\end{figure}

Figure~\ref{fig:scalarmodelresults} shows the value of the estimated parameters 
$\alpha_1,\ldots,\alpha_4$ and $\beta$ in the Euclidean and affine invariant geometries for seizures and interictal periods for patient 18. 
We first consider the results for the affine invariant geometry before comparing the geometries below. 
The results under the affine invariant geometry show that for the interictal series the autoregressive terms $\alpha_\ell$ are close to zero, while the mean reversion coefficient $\beta$ is close to $1$. 
In contrast, for the seizure data, the $\alpha_\ell$ are negative, with $95\%$ confidence intervals which do not icnlude zero, and with decreasing absolute value as $\ell$ increases. 
The mean reversion coefficient $\beta$ for the seizures is close to zero apart from the value for seizure 3. 
The inferred values are quite homogeneous between seizures with the exception of seizure 3, which had a noticably different trajectory in the previous MDS analysis (see Figure~\ref{fig:MDS18AI}). 
These results suggest that during interictal periods the brain dynamics follow a mean reverting random walk in $\spd{p}$, while during seizures the mean reversion behaviour is weaker with a significant autogregressive component which gives rise to the trajectories observed in the MDS plots. 
The procedure for selecting the maximum lag $L$ (see results in Supplementary Material) showed that across the entire data set a number of seizures had no significant autoregressive terms ($L=0$). 
These mainly consisted of shorter time series for which the confidence region for $\alpha_1$ was large. 

\begin{table}
\begin{center}
		\scalebox{0.7}{
			\begin{tabular}{l|cc|cc|cc|cc|cc}  
				\hline 
				& \multicolumn{2}{c|}{Dataset 1}& \multicolumn{2}{c|}{Dataset 2} &\multicolumn{2}{c|}{Dataset 3}&\multicolumn{2}{c|}{Dataset 4}&\multicolumn{2}{c}{Dataset 5} \\   
				\hline
				& Seizure & Interictal & Seizure & Interictal & Seizure & Interictal & Seizure & Interictal & Seizure & Interictal \\   
				Full model & 17560 & 12753 & 42016 & 37445 & 11163 & 13462 & 32992 & 28961 & 41434 & 56652 \\   
				$\beta = 0$ & 17816 & 15507 & 42664 & 43996 & 11655 & 15866 & 33356 & 34036 & 42042 & 65909 \\   
				$\alpha = 0$ & 21928 & 13204 & 55541 & 38836 & 13924 & 14298 & 47093 & 29732 & 57787 & 57453 \\    
				\hline
		\end{tabular}}
\end{center}
\begin{caption}
{
\label{tab:AICscalar}
AIC values for scalar models fitted to patient 18. Comparison of seizure / interictal for affine invariant geometry. 
$\alpha=0$ refers to a model with no autoregressive terms $\alpha_\ell$ and $\beta=0$ refers to a model with no mean reversion term.   
} 
\end{caption}
\end{table}

The Akaike information criterion (AIC) was used to compare the scalar model with autoregression and mean reversion (full model) to models without these terms. 
The results for patient 18 are shown in Table~\ref{tab:AICscalar}. 
In all cases the full model has the lowest AIC and is thus preferred. 
However, for seizure series the model with no mean reversion term has a similar AIC to the full model; for interictal series the model with no autoregression terms often has similar AIC to the full model. 

The negative values of $\alpha_\ell$ were investigated further. 
Negative values suggest oscillatory behaviour, and this might be related to the specific window size used to obtain the time series of covariance matrices. 
As a result, data reduction and fitting of the scalar model was repeated for different window sizes (specifically, non-overlapping windows of size $0.5$ seonds, $2$ seconds and $3$ seconds). 
The negative inferred values persisted across these repeated analyses, suggesting oscillatory behaviour at different time scales. 
Next we calculated the quantities $\gaff{V_k}{V_{k,1}}{S_k}/\|V_k\|\|V_{k,1}\|$ $k=2,\ldots,n$ for each seizure for patient 18.
The values were negative, confirming the tendancy for tangent vectors to reverse direction. (See plots in Supplementary Material.) 
Finally we performed a principal component analysis on the tangent vectors $V_k$ for each seizure time series by parallel transporting each vector to the tangent space at the identity matrix $I\in\spd{p}$.  
The two-dimensional PCA plots further also showed this tendancy to reverse. (See Supplementary Material.) 
Our conclusion is that the oscillatory behaviour indicated by the negative autoregressive coefficients is genuine at these time scales, but the autogregressive terms might not be capturing the longer-term evolution of the seizure trajectories observed in the MDS plots. 
More comments are given in Section~\ref{sec:conclusion}.  

\begin{figure}
\begin{center}
\includegraphics[width=\textwidth]{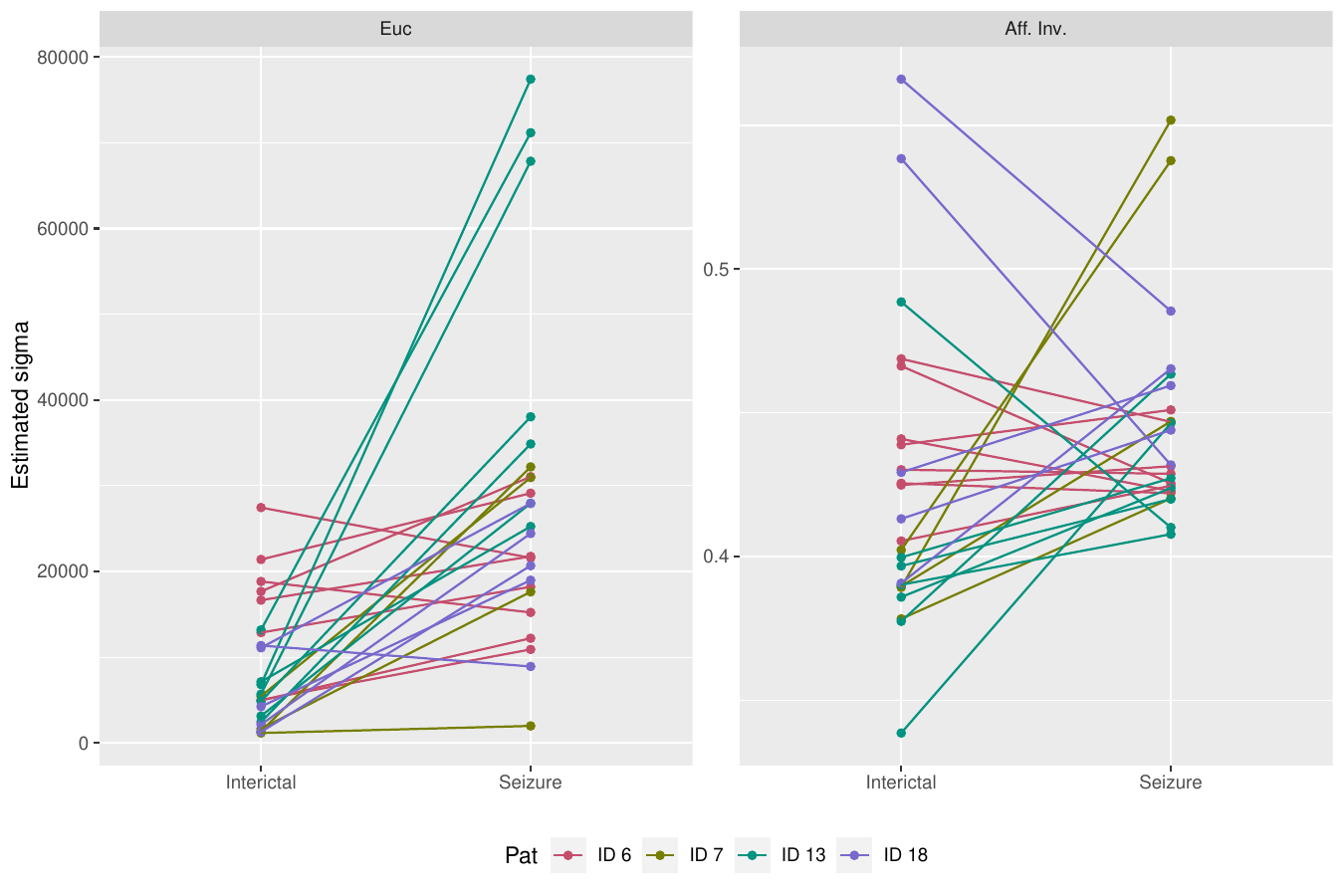}
\end{center}
\begin{caption}{
\label{fig:scalarnoiseparam}
Estimated noise parameter $\hat{\sigma}$ for patients 6, 7, 13 and 18. 
Left: Euclidean geometry. Right: affine invariant geometry. 
The value of $\hat{\sigma}$ for each seizure is connected by a line to the value for the interical series. }
\end{caption}
\end{figure}

Next we compare the results for the affine invariant geometry with those under the Euclidean geometry. 
The values of $\alpha_\ell$ and $\beta$ in Figure~\ref{fig:scalarmodelresults} are similar for the two geometries and show qualitatively similar dynamics for seizures and interictal series as the affine invariant model. 
However, there is greater heterogeneity between series for the estimated parameters under the Euclidean geometry. 
Greater differences between geometries are apparent in the estimated values of $\sigma$ shown in Figure~\ref{fig:scalarnoiseparam}. 
There are very substantial increases in $\sigma$ between interical periods and seizures under the Euclidean geometry (for example by orders of magnitude for patient 13), while the estimated values under the affine invariant geometry are broadly similar between interictal and seizure series. 
Thus under the Euclidean model, much of the variation in seizure data is modelled as noise. 

\begin{table}
\begin{center}
		\scalebox{0.7}{
			\begin{tabular}{l|cc|cc|cc|cc|cc}  
				\hline 
				& \multicolumn{2}{c|}{Dataset 1}& \multicolumn{2}{c|}{Dataset 2} &\multicolumn{2}{c|}{Dataset 3}&\multicolumn{2}{c|}{Dataset 4}&\multicolumn{2}{c}{Dataset 5} \\   
				\hline 
				& Seizure & Interictal & Seizure & Interictal & Seizure & Interictal & Seizure & Interictal & Seizure & Interictal \\   
				Euc. & 0.115 & 0.409 & 0.121 & 0.386 & 0.324 & 0.399 & 0.199 & 0.406 & 0.216 & 0.428 \\   
				Aff. Inv. & 0.330 & 0.541 & 0.426 & 0.563 & 0.501 & 0.651 & 0.450 & 0.535 & 0.442 & 0.613 \\    
				\hline
		\end{tabular}}
\end{center}
\begin{caption}
{
\label{tab:Rsquscalar}
$R^2$ values for scalar model fitted to patient 18. Comparison of seizure / interictal periods and Euclidean / affine invariant geometries.   
} 
\end{caption}
\end{table}

This is further bourne out by the $R^2$ values for patient 18 shown in Table~\ref{tab:Rsquscalar}. 
We define the $R^2$ statistic for each series as
\begin{equation*}
1-\frac{\sum_{i=\ell}^{n-1} \|\epsilon_k\|^2}{\sum_{i=\ell}^{n-1} \|V_k-\bar{V}_k\|^2}
\end{equation*}
where the norms are with the Riemannian metric (either Euclidean or affine invariant) and $\bar{V}_k$ is the mean tangent vector at $S_k$, calculated by parallel transporting all $V_i$, $i=1,\ldots,n-1$, to $S_k$.  
An $R^2$ value close to 1 indicates that a large proportion of the variability in the data is explained by the model. 
Table~\ref{tab:Rsquscalar} shows substantially better model fit for the affine invariant model across all seizures and interictal periods, and this was also the case for other patients. 
This is a result of the intrisic nature of the model in the affine invariant geometry. 
For both geometries the $R^2$ values are higher for interictal series than the corresponding seizures.

\subsection{Diagonal model}

\begin{figure}
\begin{center}
\includegraphics[width=0.8\textwidth]{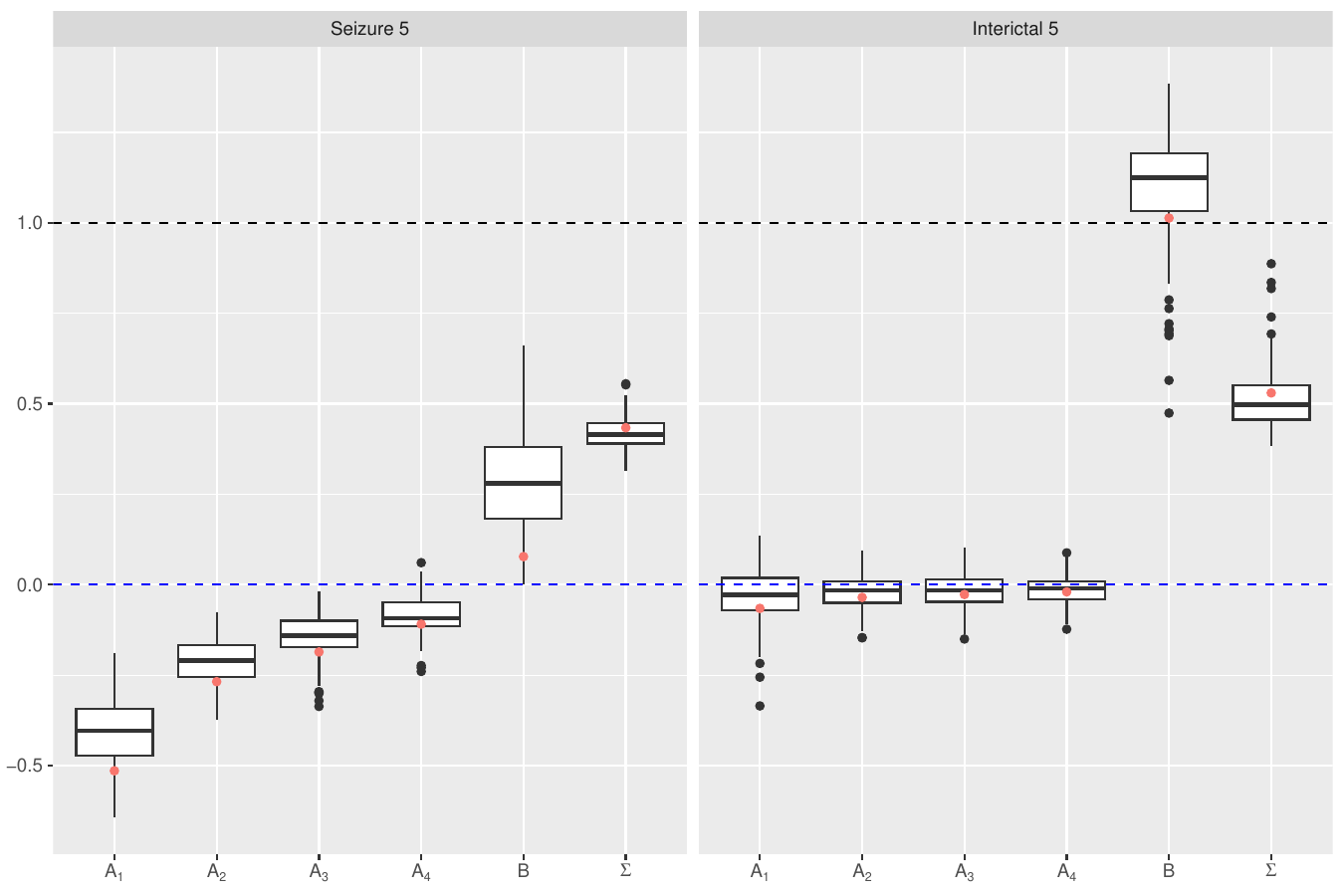}
\end{center}
\begin{caption}
{\label{fig:diagboxplot}
Distribution of fitted diagonal model parameters for patient 18 in the affine invariant geometry. 
The red points show the value of the corresponding coefficients under the scalar model. 
}
\end{caption}
\end{figure}

\begin{figure}
\begin{center}
\includegraphics[width=0.8\textwidth]{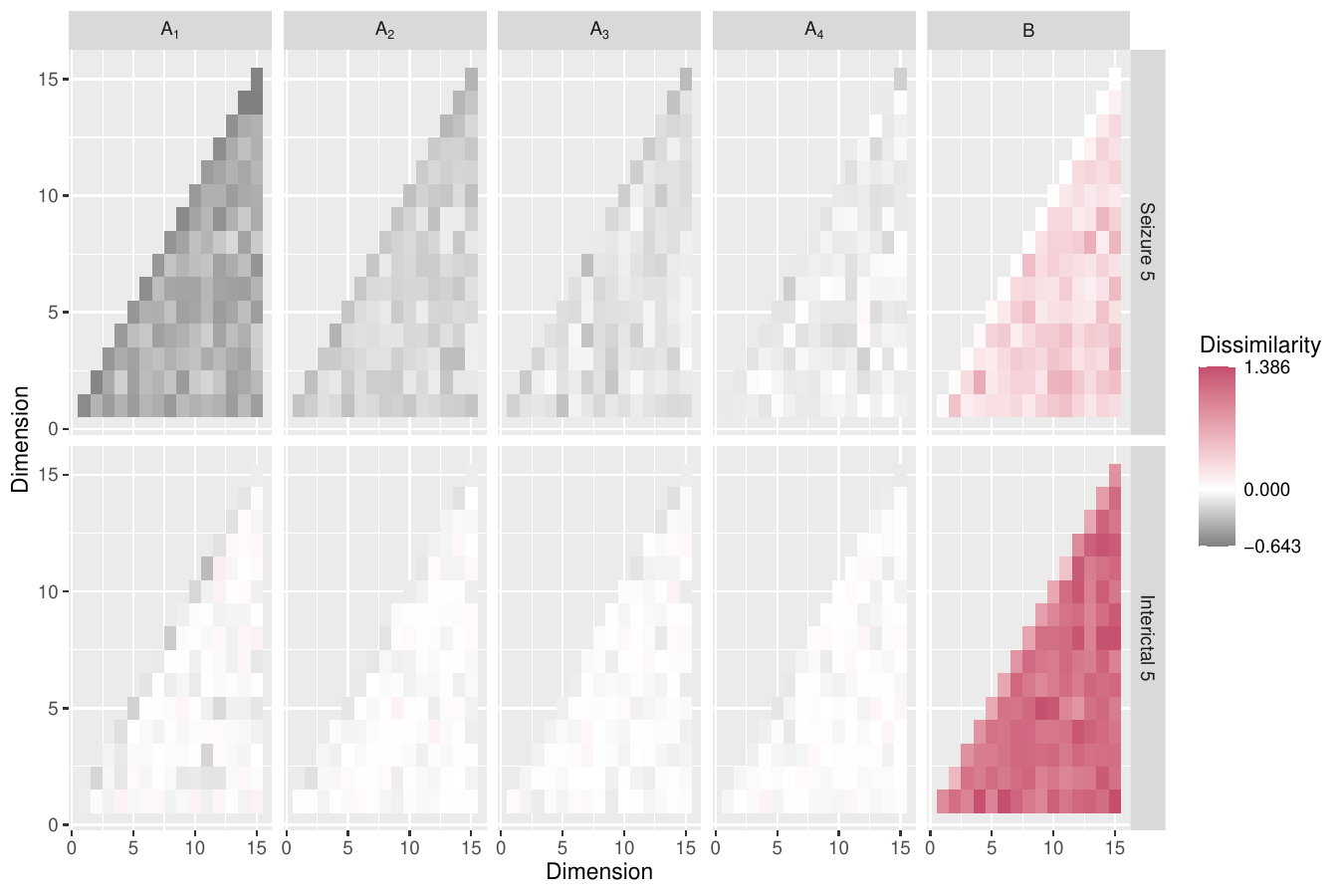}
\end{center}
\begin{caption}
{\label{fig:diagheatmap}
Diagonal model parameters for patient 18. 
Each row / column corresponds to a channel, and the pixel within any given row and column shows the coefficient corresponding to the covariance between those two channels. 
}
\end{caption}
\end{figure}

The diagonal model was fitted to seizures in both the Euclidean and affine invariant geometries, but only results under the affine invariant geometry are shown below due to better model fit (specifically the $R^2$ values). 
Figure~\ref{fig:diagboxplot} shows typical results (for patient 18, seizure 5). 
The figure shows the distribution of the diagonal model parameters $a_{\ell,i}$, $b_{i}$, and $\sigma^2_i$ for $i=1,\ldots,m$, $\ell=1,\ldots,L$ and $L=4$.  
The $a_{\ell,i}$ values are generally close to the value of $\alpha_\ell$ obtained under the scalar model, but the mean reversion coefficients $b_i$ nearly all exceed the value $\beta$ estimated under the scalar model. 
Figure~\ref{fig:diagheatmap} shows a heatmap of the coefficients for the same seizure as Figure~\ref{fig:diagboxplot}. 
For seizures, the diagonal elements for $B$ (i.e. the mean reversion model coefficients corresponding to variances rather than covariances) are zero, while the off-diagonal elements are positive. 
This is not the case for the interictal series, for which the diagonal and off-diagonal elements alike are close to 1. 
Other patterns in the coefficients are harder to discern. 
For example, if a particular channel was instrumental in seizure evolution then it might show up as a vertical or horizontal stripe in the heatmap of coefficients, denoting a departure in value for coefficients corresponding to all the covariances associated with that channel. 
This was not the case, and although there is a degree of heterogeneity in the parameters $a_{\ell,i}$ and $b_{i}$, there is no clear spatial pattern.

\subsection{Seizure comparison}\label{sec:seizurecomparison}

\begin{figure}
\begin{center}
\includegraphics[width=\textwidth]{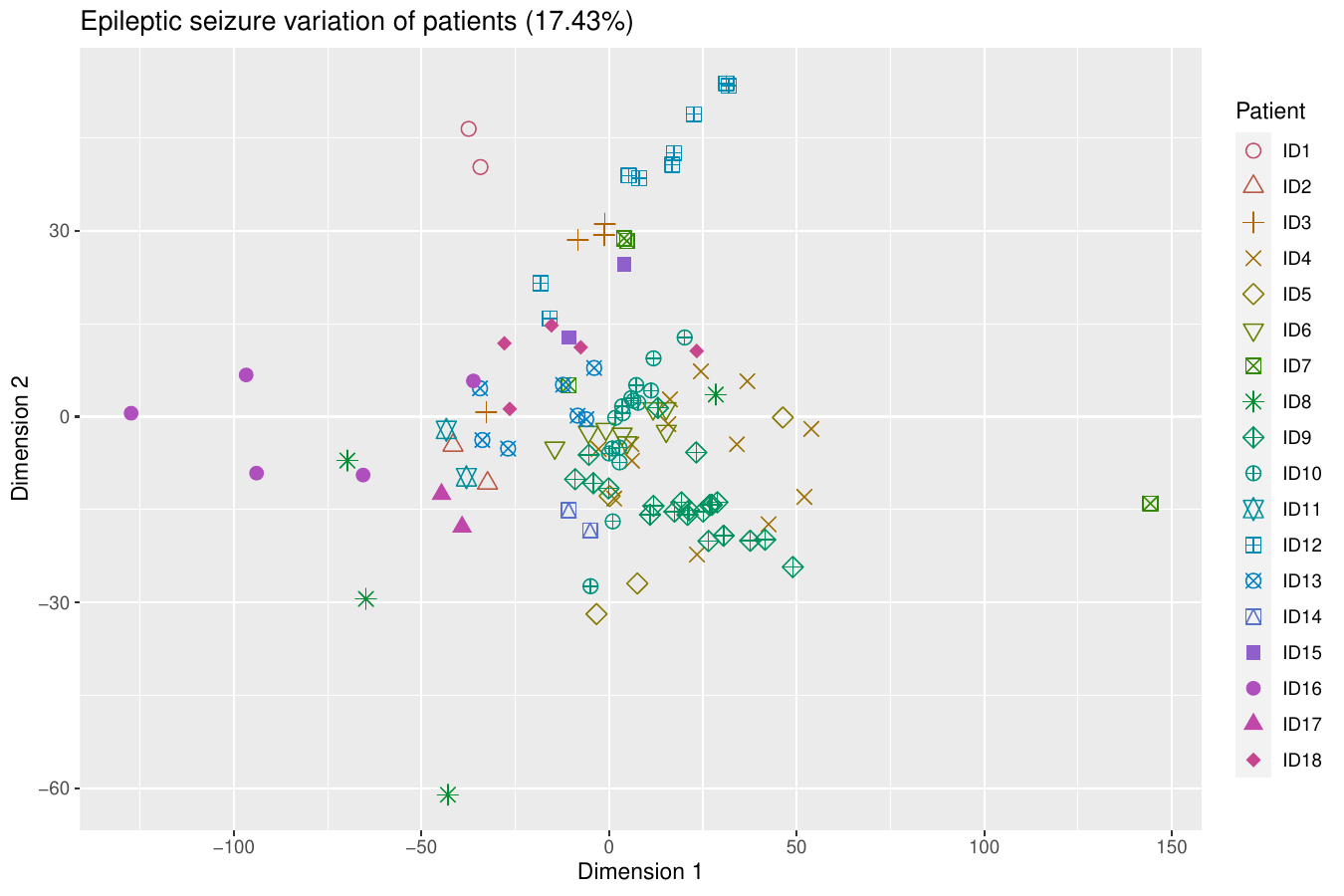}
\end{center}
\begin{caption}
{ 
\label{fig:seizurecomparison}
Multidimensional scaling of Mahalanobis distances between seizures.
Seizures for each patient have the same marker. 
}
\end{caption}
\end{figure}

Comparing seizures in different patients is difficult due to the heterogeneity in electrode number and placement between patients. 
Comparison of fitted model parameters gives a possible method for between-patient seizure comparison. 
We calculated the Mahalanobis distance between vectors of parameters under the scalar model. 
Specifically, using $L=4$, let $\Phi_i=(\hat{\alpha_1},\ldots,\hat{\alpha_L},\hat{\beta},\hat{\sigma})$ denote the estimated parameters for the $i$th seizure, where $i$ ranges across all seizures in all patients. 
The Mahalanobis distance between seizures $i$ and $j$ is defined as
\begin{equation*}
d_{ij} = \frac{1}{2}\left\{ \left[ (\Phi_i-\Phi_j)^T W_i(\Phi_i-\Phi_j) \right]^{1/2}+ \left[ (\Phi_j-\Phi_i)^TW_j(\Phi_j-\Phi_i) \right]^{1/2} \right\}
\end{equation*} 
where $W_i$ and $W_j$ are the estimated asymptotic covariance matrices of estimators. 
Figure~\ref{fig:seizurecomparison} shows a metric multidimensional scaling for the matrix of Mahalanobis distances between seizures.  
The plot shows variation within and between patients, with variation between patients greater than within. 
There are some outlying seizures (from patients 7 and 8), and the seizures from some patients form clusters away from the main group (patients 1, 12 and 16). 
Differences between seizures in patients 7, 8 and 16 appear to drive the first dimension in the MDS.

\section{Conclusion}\label{sec:conclusion}

The aims of this research were (i)  to develop a model for covariance matrix time series with interpretable parameters for different possible modes of EEG dynamics, and (ii) to explore the extent to which modelling results differ between the Euclidean and affine invariant geometries. 
The model we developed, specified for time series on a general Riemannian manifold, effectively captures different dynamics between seizures and interictal periods in the corresponding time series of covariance matrices. 
The time series for interictal periods behave like mean reverting random walks in the space of covariance matrices. 
In contrast, there is a significant autoregressive component in the majority of seizure time series and the mean reversion effect is weaker. 
Although the same modes of dynamics were found under both geometries, model fit was substantially better for the affine invariant geometry. 
In general, possibly due to the invariance properties described in Equation~$\eqref{equ:aiproperty}$, the results obtained under the affine invariant geometry displayed a greater degree of consistency across the analysis in comparison to the Euclidean geometry: for example, seizure trajectories were clearer in the affine invariant MDS plots (Figures~\ref{fig:MDS18AI} and~\ref{fig:MDS18Euc}), and the estimated noise parameter was similar for seizures and interictal periods under the affine invariant geometry, but varied very considerably under the Euclidean geometry (Figure~\ref{fig:scalarnoiseparam}).

In addition to the Euclidean and the affine invariant geometries, a number of other possible Riemannian metrics exist for the space $\spd{p}$ \citep{dryden2009non}. 
Our model requires that any given geometry comes equipped with Riemannian logarithm and parallel transport maps which can be easily computed. 
This reduces the number of geometries which can be used. 
However, the log-Euclidean and log-Cholesky geometries fit into the category of computationally tractable geometries. 
We implemented the model in these geometries and obtained a limited set of results. 
These indicated that model fit was generally inferior to that obtained with the affine invariant geometry, though the estimated parameter values were similar to those obtained under the affine invariant geometry. 
There is potential for more research comparing geometries, possibly analysing time series of correlation matrices as opposed to covariance matrices, by adapting the Riemannian geometry accordingly~\citep{thanwerdas2021geodesics}. 

Under both the scalar and diagonal models, negative values were obtained for the autoregressive coefficients for seizures. 
These results were carefully checked and further analyses were performed to investigate. 
The negative values persisted when the length of sliding windows used to prepare the data sets was varied. 
The inner product between successive tangent vectors was computed and indicated a tendency for reversal of direction which is reflected by the negative estimated coefficient values. 
Together, these results suggest oscillatory behaviour in the covariance time series at the time scale of around 1 second. 
There is potential for future development of alternative models which are better able to capture dynamics at different time scales, in particular the larger time scale trajectories which are apparent in MDS plots of the seizure time series (Figure~\ref{fig:MDS18AI}). 

Although analysis of covariance matrices of EEG data offers advantages over analysis of raw signals, the resulting data sets and models potentially have very high dimension. 
The scalar and diagonal models respectively have $L$ and $\tfrac{1}{2}p(p+1)L$ autoregressive parameters, while in full generality the model would have $O(p^4)$ parameters. 
Despite the large increase in the number of parameters between the scalar and diagonal models, the diagonal model had lower AIC than the scalar model for the seizure time series from the seizures we looked at in detail. 
However, inspection of the estimated autoregressive parameters $A_\ell$ did not yield any easily apparent patterns, for example particular dimensions which were substantially different from others in the seizure covariance matrices. 
On the other hand, the estimated mean reversion matrix $B$ exhibited clear differences between model parameters associated with variances as opposed to covariances for seizure time series. 
Overall, the scalar model is simpler to interpret and easier to implement. 

For simplicity, we took the attractor point $S^*$ for each seizure to be the Fr\'echet mean of the corresponding interictal period. 
However, we also investigated maximum likelihood estimation of $S^*$ jointly with the other parameters using an algorithm which alternated between gradient descent steps which improved the estimate of $S^*$ keeping the other parameters fixed, and steps which kept $S^*$ fixed and estimated the other parameters. 
These results are not reported here due to poor algorithm convergence.

Although widely used to analyse EEG data, use of a sliding window to derive time series of covariance or correlation matrices is problematic in that the choice of the length of the window affects the dynamics which can be revealed by such data. 
A more consistent approach to modelling might be to adopt a continuous time model for the underlying covariance process, like the mean reverting diffusion developed by \cite{bui2021inference}, with the raw EEG signals $z_i$ arising as discrete-time observations from a Gaussian with the underlying covariance process. 
Nonetheless, incorporating autoregressive or other components suitable for modelling seizures into such a continuous time model could be challenging.

\section*{Acknowledgements}
T.D. is funded jointly by the China Scholarship Council and the School of Mathematics, Statistics, and Physics at Newcastle University. Y.W. gratefully acknowledges support by a UKRI Future Leaders Fellowships (MR/V026569/1).

 \newcommand{\noop}[1]{}


\begin{thebibliography}{}

\bibitem[\protect\astroncite{Abreu et~al.}{2020}]{abreu2020pushing}
Abreu, R., Sim{\~o}es, M., and Castelo-Branco, M. (2020).
\newblock Pushing the limits of {EEG}: estimation of large-scale functional
  brain networks and their dynamics validated by simultaneous {fMRI}.
\newblock {\em Front. Neurosci.}, 14:323.

\bibitem[\protect\astroncite{Bastos and Schoffelen}{2016}]{bastos2016tutorial}
Bastos, A.~M. and Schoffelen, J.-M. (2016).
\newblock A tutorial review of functional connectivity analysis methods and
  their interpretational pitfalls.
\newblock {\em Front. Syst. Neurosci.}, 9:175.

\bibitem[\protect\astroncite{Bhattacharya and
  Patrangenaru}{2003}]{bhattacharya2003large}
Bhattacharya, R. and Patrangenaru, V. (2003).
\newblock Large sample theory of intrinsic and extrinsic sample means on
  manifolds {I}.
\newblock {\em Ann. Stat.}, 31:1--29.

\bibitem[\protect\astroncite{Bhattacharya and
  Patrangenaru}{2005}]{bhattacharya2005large}
Bhattacharya, R. and Patrangenaru, V. (2005).
\newblock Large sample theory of intrinsic and extrinsic sample means on
  manifolds {II}.
\newblock {\em Ann. Stat.}, 33:1225--1259.

\bibitem[\protect\astroncite{Binks et~al.}{2023}]{binks2023}
Binks, R.~L., Heaps, S.~E., Panagiotopoulou, M., Wang, Y., and Wilkinson, D.~J.
  (2023).
\newblock Bayesian inference on the order of stationary vector autoregressions.
\newblock {\em arXiv preprint arXiv:2307.05708}.

\bibitem[\protect\astroncite{Boonyakitanont
  et~al.}{2020}]{boonyakitanont2020review}
Boonyakitanont, P., Lek-Uthai, A., Chomtho, K., and Songsiri, J. (2020).
\newblock A review of feature extraction and performance evaluation in
  epileptic seizure detection using {EEG}.
\newblock {\em Biomed. Signal Proces.}, 57:101702.

\bibitem[\protect\astroncite{Bui et~al.}{in press}]{bui2021inference}
Bui, M.~N., Pokern, Y., and Dellaportas, P. (in press).
\newblock Inference for partially observed {R}iemannian {O}rnstein-{U}hlenbeck
  diffusions of covariance matrices.
\newblock {\em Bernoulli}.

\bibitem[\protect\astroncite{Burrello et~al.}{2019}]{burrello2019laelaps}
Burrello, A., Cavigelli, L., Schindler, K., Benini, L., and Rahimi, A. (2019).
\newblock Laelaps: An energy-efficient seizure detection algorithm from
  long-term human i{EEG} recordings without false alarms.
\newblock In {\em 2019 Design, Automation \& Test in Europe Conference \&
  Exhibition}, pages 752--757. IEEE.

\bibitem[\protect\astroncite{Chiarion et~al.}{2023}]{chiarion2023connectivity}
Chiarion, G., Sparacino, L., Antonacci, Y., Faes, L., and Mesin, L. (2023).
\newblock Connectivity analysis in {EEG} data: A tutorial review of the state
  of the art and emerging trends.
\newblock {\em Bioengineering}, 10(3):372.

\bibitem[\protect\astroncite{Costa et~al.}{2017}]{costa2017}
Costa, L., Nichols, T., and Smith, J.~Q. (2017).
\newblock Studying the effective brain connectivity using multiregression
  dynamic models.
\newblock {\em Braz. J. Probab. Stat}, 31(4):765--800.

\bibitem[\protect\astroncite{Do~Carmo}{1992}]{do_carmo_riemannian_1992}
Do~Carmo, M. (1992).
\newblock {\em Riemannian Geometry}.
\newblock Mathematics: {Theory} \& {Applications}. Birkhäuser Basel.

\bibitem[\protect\astroncite{Dryden et~al.}{2009}]{dryden2009non}
Dryden, I.~L., Koloydenko, A., and Zhou, D. (2009).
\newblock Non-{E}uclidean statistics for covariance matrices, with applications
  to diffusion tensor imaging.
\newblock {\em Ann. Appl. Stat.}, 3(3):1102--1123.

\bibitem[\protect\astroncite{Dryden and Mardia}{2016}]{dryden2016shape}
Dryden, I.~L. and Mardia, K.~V. (2016).
\newblock {\em Statistical shape analysis: with applications in R}, volume 995.
\newblock John Wiley \& Sons.

\bibitem[\protect\astroncite{Fletcher et~al.}{2004}]{fletcher2004principal}
Fletcher, P.~T., Lu, C., Pizer, S.~M., and Joshi, S. (2004).
\newblock Principal geodesic analysis for the study of nonlinear statistics of
  shape.
\newblock {\em IEEE T. Med. Imaging}, 23(8):995--1005.

\bibitem[\protect\astroncite{Fletcher}{2013}]{thomas2013geodesic}
Fletcher, T.~P. (2013).
\newblock Geodesic regression and the theory of least squares on {R}iemannian
  manifolds.
\newblock {\em Int. J. Comput. Vision}, 105:171--185.

\bibitem[\protect\astroncite{Fr{\'e}chet}{1948}]{frechet1948elements}
Fr{\'e}chet, M. (1948).
\newblock Les {\'e}l{\'e}ments al{\'e}atoires de nature quelconque dans un
  espace distanci{\'e}.
\newblock In {\em Annales de l'institut Henri Poincar{\'e}}, volume~10, pages
  215--310.

\bibitem[\protect\astroncite{Huckemann and
  Ziezold}{2006}]{huckemann2006principal}
Huckemann, S. and Ziezold, H. (2006).
\newblock Principal component analysis for {R}iemannian manifolds, with an
  application to triangular shape spaces.
\newblock {\em Adv. Appl. Probab.}, 38(2):299--319.

\bibitem[\protect\astroncite{Jung et~al.}{2012}]{jung2012analysis}
Jung, S., Dryden, I.~L., and Marron, J.~S. (2012).
\newblock Analysis of principal nested spheres.
\newblock {\em Biometrika}, 99(3):551--568.

\bibitem[\protect\astroncite{Jupp and Mardia}{2009}]{jupp2009directional}
Jupp, P.~E. and Mardia, K.~V. (2009).
\newblock {\em Directional statistics}.
\newblock John Wiley \& Sons.

\bibitem[\protect\astroncite{Kent et~al.}{1979}]{mardiamultivariate}
Kent, J.~T., Bibby, J., and Mardia, K.~V. (1979).
\newblock {\em Multivariate analysis}.
\newblock Academic Press.

\bibitem[\protect\astroncite{Lang}{1999}]{lang_fundamentals_1999}
Lang, S. (1999).
\newblock {\em Fundamentals of differential geometry}.
\newblock Graduate Texts in Mathematics. Springer.

\bibitem[\protect\astroncite{Lenglet et~al.}{2006}]{lenglet2006}
Lenglet, C., Rousson, M., Deriche, R., and Faugeras, O. (2006).
\newblock Statistics on the manifold of multivariate normal distributions:
  Theory and application to diffusion tensor {MRI} processing.
\newblock {\em J. Math. Imaging Vis.}, 25(3):423--444.

\bibitem[\protect\astroncite{Mallasto and Feragen}{2018}]{mallasto2018wrapped}
Mallasto, A. and Feragen, A. (2018).
\newblock Wrapped {G}aussian process regression on {R}iemannian manifolds.
\newblock In {\em Proceedings of the IEEE Conference on Computer Vision and
  Pattern Recognition}, pages 5580--5588.

\bibitem[\protect\astroncite{Marron and Alonso}{2014}]{marron2014overview}
Marron, J.~S. and Alonso, A.~M. (2014).
\newblock Overview of object oriented data analysis.
\newblock {\em Biometrical J.}, 56(5):732--753.

\bibitem[\protect\astroncite{Pennec et~al.}{2019}]{pennec2019riemannian}
Pennec, X., Sommer, S., and Fletcher, T. (2019).
\newblock {\em Riemannian geometric statistics in medical image analysis}.
\newblock Academic Press.

\bibitem[\protect\astroncite{Schiratti et~al.}{2015}]{NIPS2015_186a157b}
Schiratti, J.-B., Allassonniere, S., Colliot, O., and Durrleman, S. (2015).
\newblock Learning spatiotemporal trajectories from manifold-valued
  longitudinal data.
\newblock In {\em Advances in Neural Information Processing Systems},
  volume~28, pages 2404--2412.

\bibitem[\protect\astroncite{Schroeder et~al.}{2020}]{schroeder2020seizure}
Schroeder, G.~M., Diehl, B., Chowdhury, F.~A., Duncan, J.~S., de~Tisi, J.,
  Trevelyan, A.~J., Forsyth, R., Jackson, A., Taylor, P.~N., and Wang, Y.
  (2020).
\newblock Seizure pathways change on circadian and slower timescales in
  individual patients with focal epilepsy.
\newblock {\em P. Natl. Acad. Sci. USA}, 117(20):11048--11058.

\bibitem[\protect\astroncite{Siddiqui et~al.}{2020}]{siddiqui2020review}
Siddiqui, M.~K., Morales-Menendez, R., Huang, X., and Hussain, N. (2020).
\newblock A review of epileptic seizure detection using machine learning
  classifiers.
\newblock {\em Brain Informatics}, 7(1):1--18.

\bibitem[\protect\astroncite{Skovgaard}{1984}]{skovgaard1984riemannian}
Skovgaard, L.~T. (1984).
\newblock A {R}iemannian geometry of the multivariate normal model.
\newblock {\em Scand. J. Stat.}, pages 211--223.

\bibitem[\protect\astroncite{Sturm}{2003}]{sturm_probability_2003}
Sturm, K.-T. (2003).
\newblock Probability measures on metric spaces of nonpositive curvature.
\newblock In Auscher, P., Coulhon, T., and Grigor’yan, A., editors, {\em Heat
  Kernels and Analysis on Manifolds, Graphs, and Metric Spaces: Lecture Notes
  from a Quarter Program on Heat Kernels, Random Walks, and Analysis on
  Manifolds and Graphs: April 16-July 13, 2002, Emile Borel Centre of the Henri
  Poincar{\'e} Institute, Paris, France}, volume 338 of {\em Contemporary
  Mathematics}. American Mathematical Society.

\bibitem[\protect\astroncite{Thanwerdas and
  Pennec}{2021}]{thanwerdas2021geodesics}
Thanwerdas, Y. and Pennec, X. (2021).
\newblock Geodesics and curvature of the quotient-affine metrics on full-rank
  correlation matrices.
\newblock In {\em Geometric Science of Information: 5th International
  Conference, GSI 2021, Paris, France, July 21--23}, pages 93--102. Springer.

\bibitem[\protect\astroncite{Yair et~al.}{2019}]{yair2019parallel}
Yair, O., Ben-Chen, M., and Talmon, R. (2019).
\newblock Parallel transport on the cone manifold of {SPD} matrices for domain
  adaptation.
\newblock {\em IEEE T. Signal Process.}, 67(7):1797--1811.

\end{thebibliography}
\end{document}


\title[Manifold-valued time series models]{Supplementary material for ``Manifold-valued models for analysis of EEG time series data''}

\author{Tao Ding}
\address[Tao Ding and Tom~M.~W.~Nye]{School of Mathematics, Statistics and Physics\\ Newcastle University\\ Newcastle upon Tyne\\ UK}
\email{t.ding2@newcastle.ac.uk}
\author{Tom~M.~W.~Nye}
\email{tom.nye@ncl.ac.uk}
\author{Yujiang Wang}
\address[Yujiang Wang]{School of Computing\\ Newcastle University\\ Newcastle upon Tyne\\ UK}
\email{yujiang.wang@ncl.ac.uk}

\maketitle

\section{Epilepsy data set}{\label{app:Patinfo}}
The open-source data set analysed in this study is available from \url{http://ieeg-swez.ethz.ch/}. 
The data set consists of EEG recordings for 18 patients (designated ID01 to ID18) who were part of the epilepsy surgery program. 
The number of seizures for each patient varied from 2 to 23, while the total duration of interictal recordings ranged from 41 to 293 hours for each patient. 
Four seizures were removed from consideration: one due to a very short duration (less than 10 seconds) and three due to lack of a comparator interictal period two hours earlier. 
Table \ref{tab:Patinfo} provides an overview of the number of electrodes, total number of seizures, overall recording duration, and seizure duration in seconds for each patient.

\begin{table}[H]
	\centering
	\caption{Patient information.}
	\begin{tabular}{lcccccl}
		\hline
		& Duration & No. of & No. of & \multicolumn{3}{c}{Seizure duration (Seconds)}\\
		\cline{5-7} 
		Subject & (Hours) & electrodes & seizures & \multicolumn{1}{c}{Minimum} & \multicolumn{1}{c}{Maximum} &\multicolumn{1}{c}{Mean} \\
		\hline
		ID01 &293  &88  &2  &589 &613 &601\\
		ID02 &235  &66  &2  &86  &89  &88\\
		ID03 &158  &64  &4  &60  &68  &64\\
		ID04 &41   &32  &13 &31  &68  &44\\
		ID05 &110  &128 &4  &15  &17  &16\\
					{ID06} &{146}  &{32}  &{8}  &{29}  &{126} &{45}\\
					{ID07} &{69}   &{75}  &{4}  &{14}  &{98}  &{69}\\
		ID08 &144  &61  &4  &17 &413 &189\\
		ID09 &41   &48  &21 &22  &136 &40\\
		ID10 &42   &32  &16 &61  &106 &70\\
		ID11 &212  &32  &2  &83  &99  &91\\
		ID12 &191  &56  &9  &106 &194 &146\\
					{ID13} &{104}  &{64}  &{7}  &{40}  &{188} &{103}\\
		ID14 &161  &24  &2  &46  &60  &53\\
		ID15 &196  &98  &2  &69  &119 &94\\
		ID16 &177  &34  &5  &120 &245 &190\\
		ID17 &130  &60  &2  &97  &98  &98\\
					{ID18} &{205}  &{42}  &{5}  &{71}  &{300} &{198}\\
		\hline
		Total  &2656 &    &112  &        &        &\\
		Average&     &58  &     &86      &169     &122\\
		\hline
	\end{tabular}	
	\label{tab:Patinfo}
\end{table}

Data was recorded at a rate of either $512$ or $1024$\textit{Hz}, depending on the experimental subject. 
To ensure the reliability and validity of the EEG recordings, an experienced epileptologist visually inspected all EEG recordings to identify seizure onset and termination and to exclude channels with persistent artifacts. 
Raw EEG signals were preprocessed via protocols described at the web site above. 

\section{Dimensional reduction}

\begin{figure}
	\begin{center}
		\includegraphics[width=0.8\textwidth]{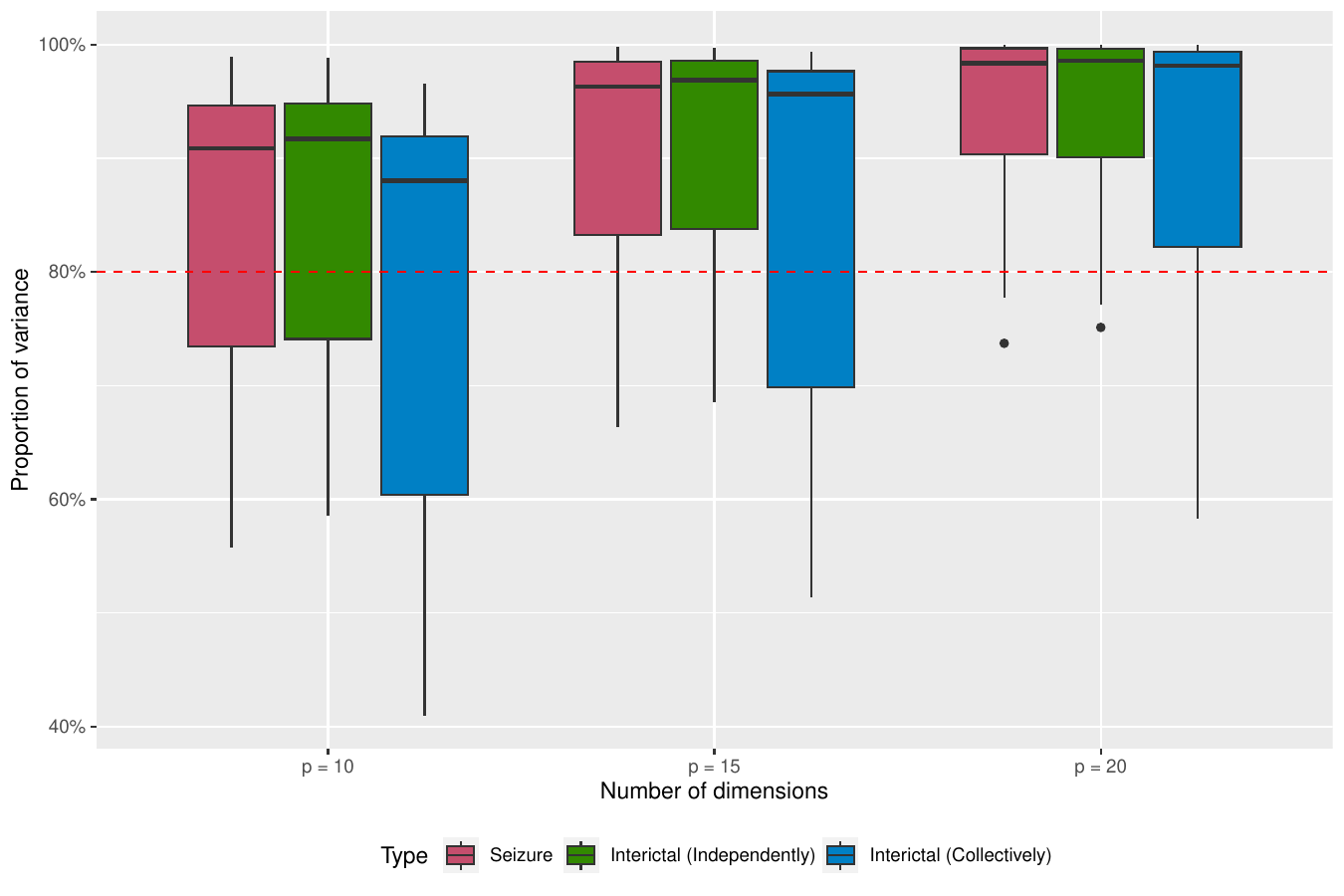}
		\begin{caption}{
				\label{fig:dimred}
				Proportion of variance, as defined by Equation~$\eqref{equ:dimredprop}$, retained in seizure and interictal data after reduction to $p=10, 15, 20$ dimensions. 
				Each boxplot shows the distribution across 112 seizures. 
				Two sets of boxes are shown for the interictal time series: (i) when dimensional reduction was performed independently from reduction of the corresponding seizure data and (ii) when dimensional reduction was performed using the $U$ matrix obtained from the seizure data. 
			}
		\end{caption}
	\end{center}
\end{figure}

Dimensional reduction was performed using the method described in the main text. 
Figure~\ref{fig:dimred} shows the proportion of variance retained in the reduced data sets, as defined by
\begin{equation}\label{equ:dimredprop}
	\frac{\sum_{j=1}^{p}\lambda_j}{\sum_{j=1}^q\lambda_j}
\end{equation}
where $\lambda_1\geq\cdots\geq\lambda_q\geq 0$ are the eigenvalues of $\tfrac{1}{n}\sum_i S_i'$. 

\begin{figure}
	\centering
	\includegraphics[width=0.8\textwidth]{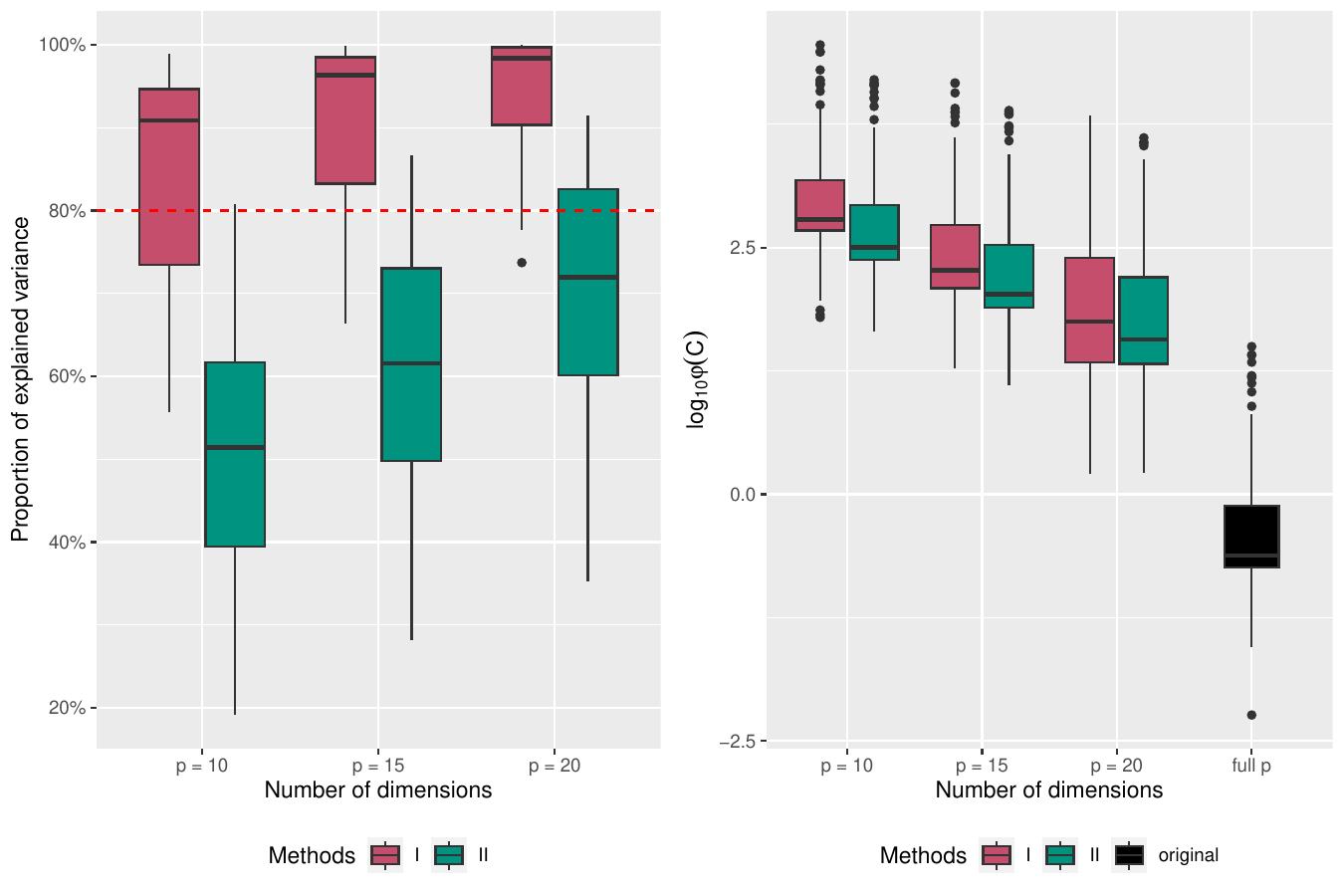}
	\caption{Comparison of reduced datasets across all seizure data. Method I, highlighted in red, aims to capture the maximum variation of the original data. Method II, denoted by green, minimizes redundancy (described in this section). The black box in the right panel represents the original data with full dimensionality. Left: proportion of explained variance. Right: $\log_{10}\varphi(C)$ with reduced dimensions $p = 10, 15$, and $20$.}
	\label{reddim2}
\end{figure}

In addition to the dimensional reduction method described in the main text, we also tested a method that identified sets of channels, rather than linear combinations of channels, thereby potentially giving a more direct interpretation of the reduced data. 
The method aimed at minimizing redundancy in the data by bounding eigenvalues of the reduced covariance matrices away from zero. 
Given a set of channels $C\subset\{1,\ldots,q\}$, let $S'_i(C)$ denote the restriction of the full covariance matrix $S'_i$ to $C$ and define
\begin{equation}\label{equ:dimredomega}
	\varphi(C) = \frac{1}{n}\sum_i \min \sigma\left[ S'_i(C) \right] 
\end{equation}
where $\sigma[A]$ denotes the set of eigenvalues of $A\in\sym{p}$.
A greedy algorithm was used to construct sets $C$ which maximized $\varphi(C)$ for a fixed value of $p$. 
This method of dimensional reduction was inferior to the former method: the proportion of variance in the reduced data (see Equation~$\eqref{equ:dimredprop}$ below) was greater with the former method, but the mean minimum eigenvalue (Equation~$\eqref{equ:dimredomega}$) was similar for both methods as shown in Figure ~\ref{reddim2}. 
Consequently, the former method, detailed in the main text, was chosen for dimensional reduction before further modeling analysis.

\section{Additional exploratory analysis}
\subsection{MDS plots for patient 6,7, and 13}
\begin{figure}
	\centering
	\begin{subfigure}{1\textwidth}
		\centering
		\includegraphics[width=1\textwidth,height = 0.4\textheight]{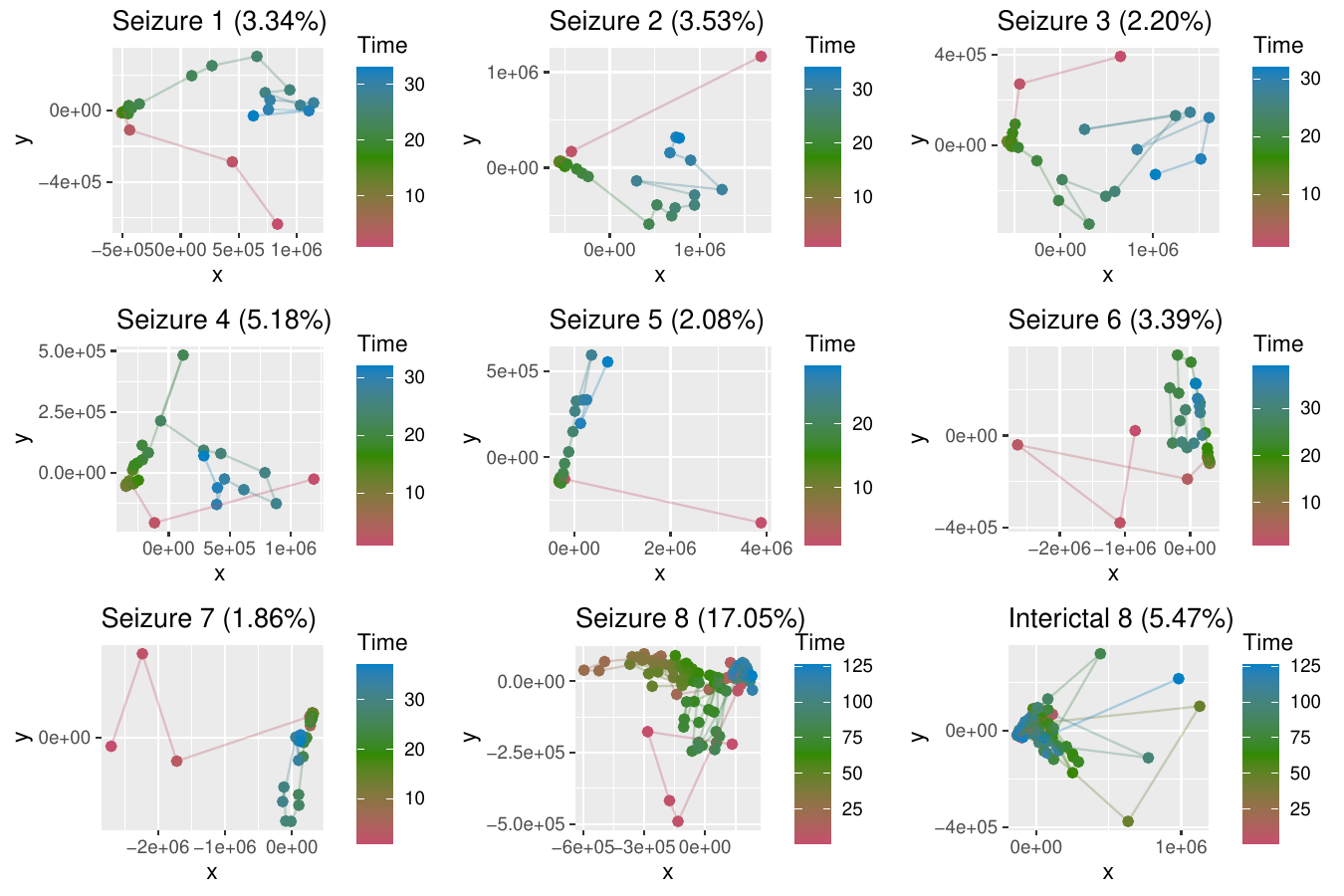}  
		\caption{MDS of seizure time series using the Euclidean metric $\deuc{S_i}{S_{j}}$.}
	\end{subfigure}
	\begin{subfigure}{1\textwidth}
		\centering
		\includegraphics[width=1\textwidth,height = 0.4\textheight]{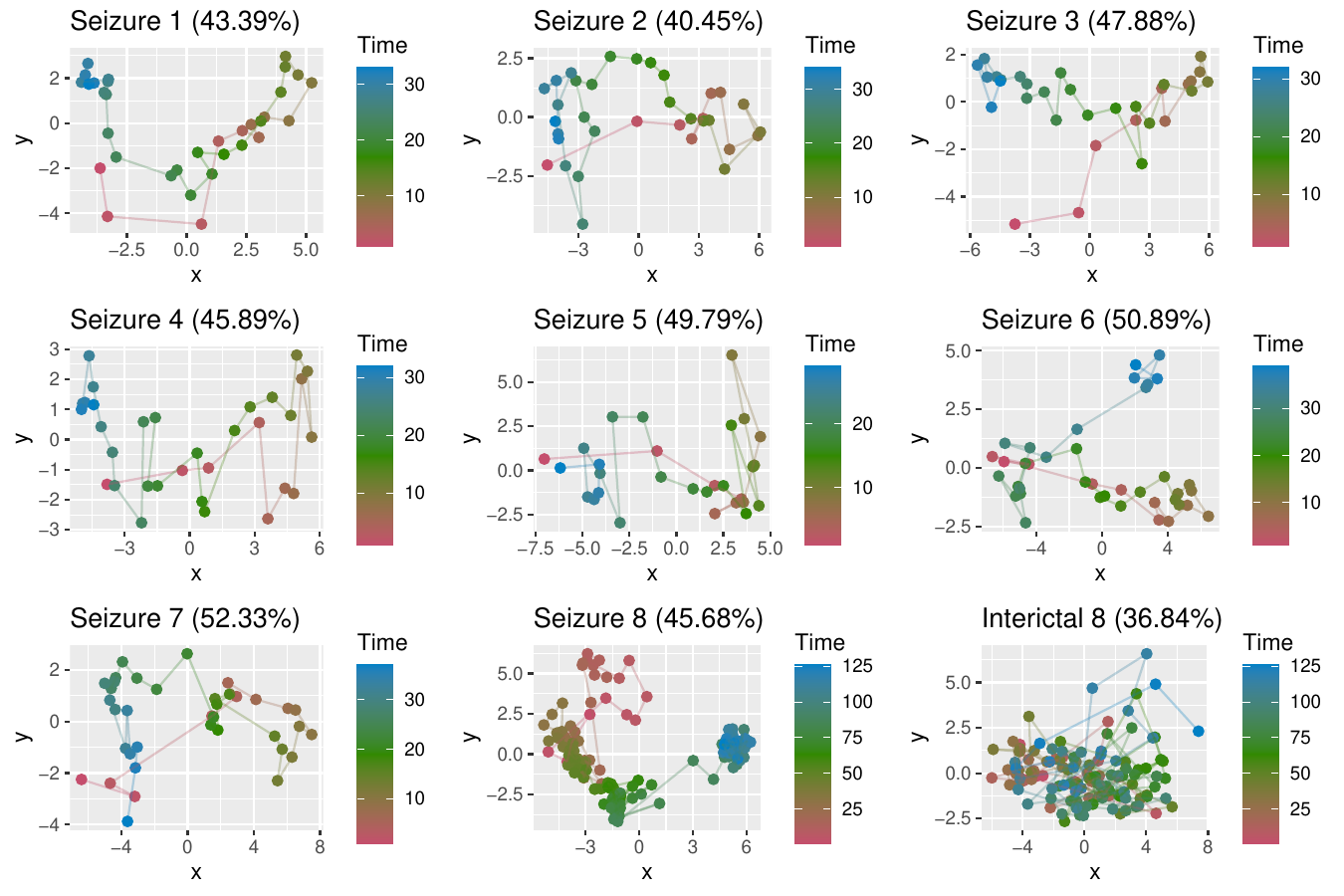}  
		\caption{MDS of seizure time series using the affine invariant metric  $\daff{S_i}{S_{j}}$.}
	\end{subfigure}
	\caption{MDS of seizure time series for Patient 6 using the Euclidean and affine invariant metrics. Patient 6 had 8 seizures (first 8 panels). Panel 9 shows MDS results for the interictal period corresponding to seizure 8. Plots are coloured to show the development over time. Each panel title shows the proportion of stress represented by the 2-dimensional MDS. }
	\label{fig:MDS_6Cov}
\end{figure}

\begin{figure}
	\centering
	\begin{subfigure}{1\textwidth}
		\centering
		\includegraphics[width=1\textwidth,height = 0.4\textheight]{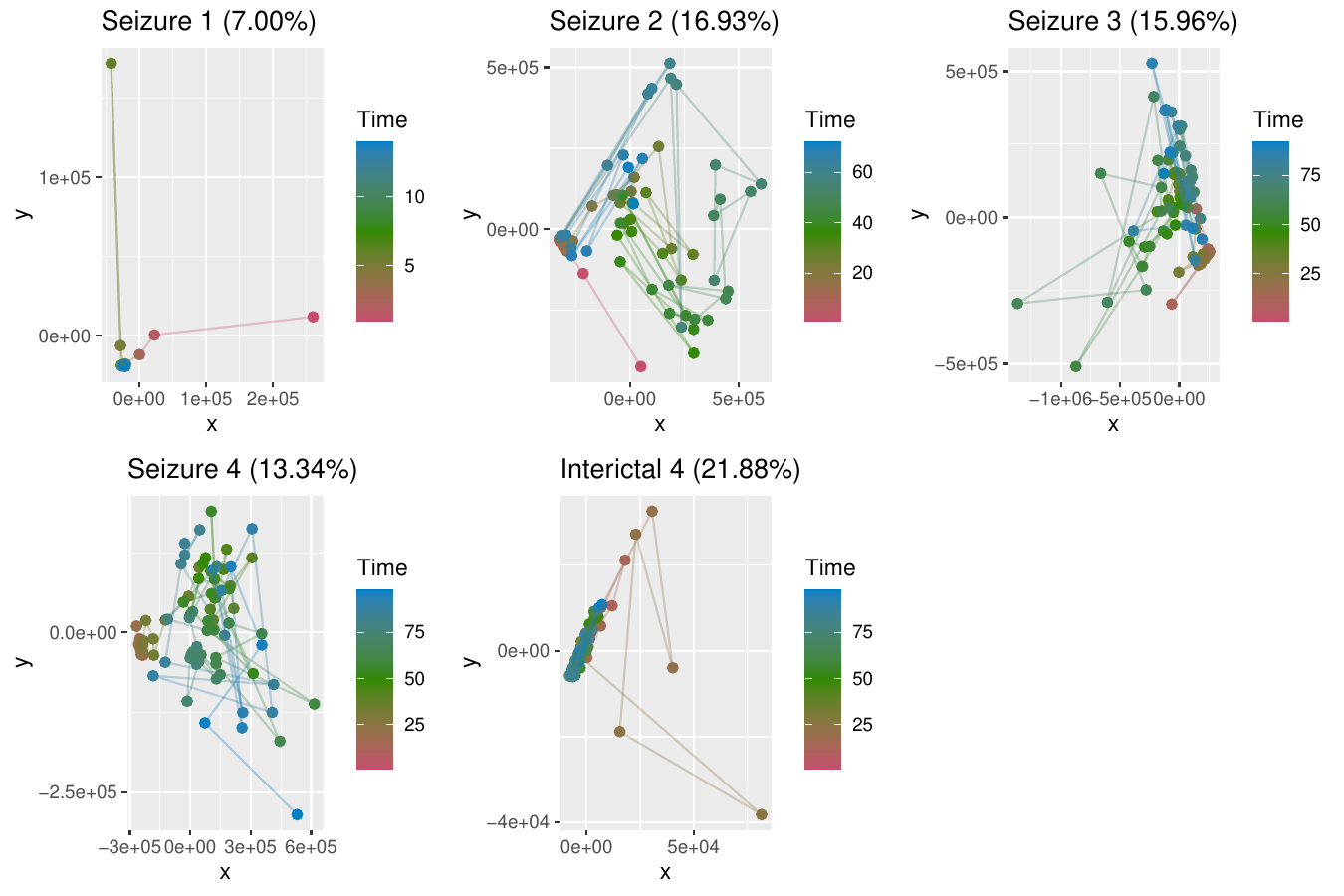}  
		\caption{MDS of seizure time series using the Euclidean metric $\deuc{S_i}{S_{j}}$.}
	\end{subfigure}
	\begin{subfigure}{1\textwidth}
		\centering
		\includegraphics[width=1\textwidth,height = 0.4\textheight]{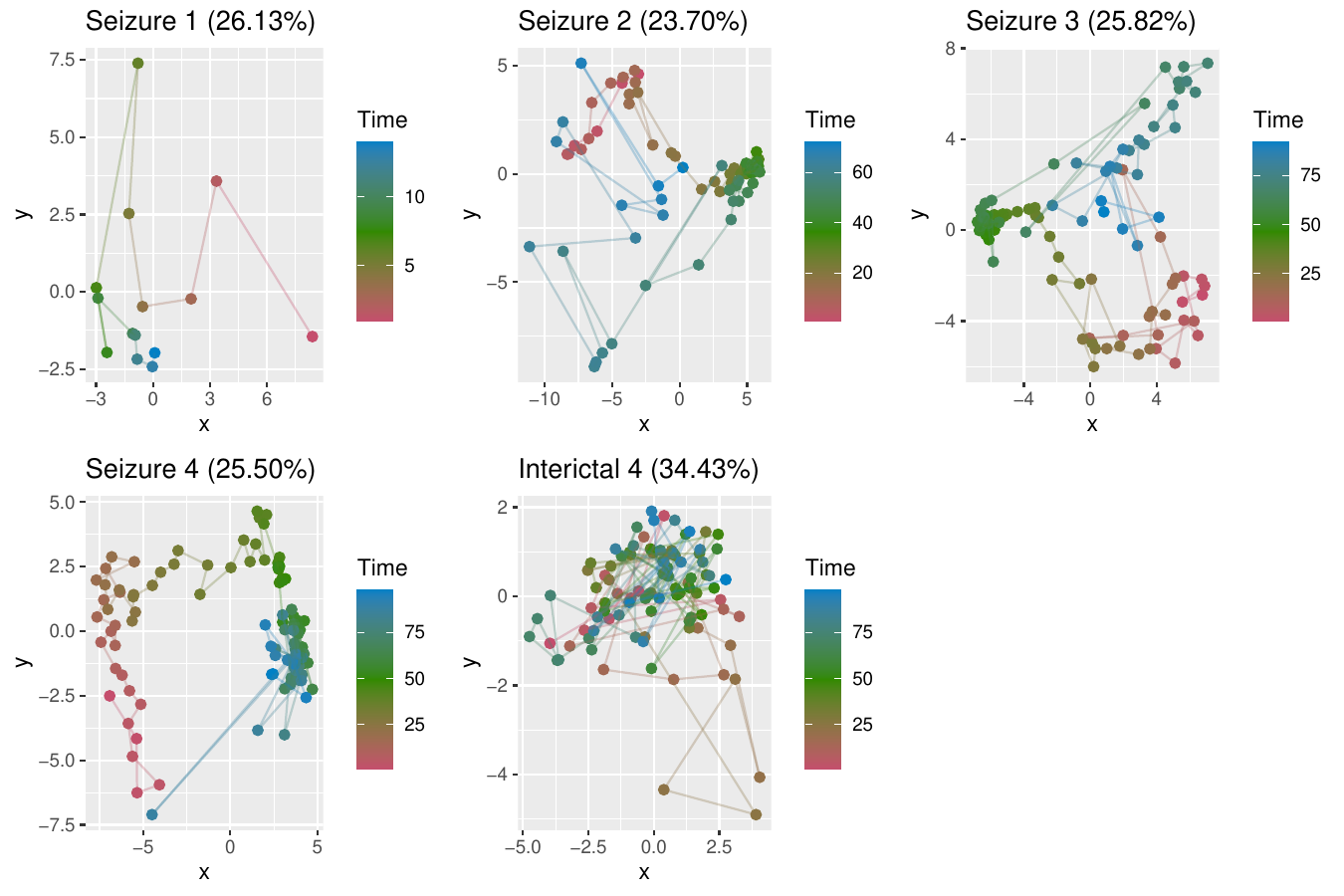}  
		\caption{MDS of seizure time series using the affine invariant metric $\daff{S_i}{S_{j}}$.}
	\end{subfigure}
	\caption{MDS of seizure time series for Patient 7 using the Euclidean and affine invariant metrics. Patient 7 had 4 seizures (first 4 panels). Panel 5 shows MDS results for the interictal period corresponding to seizure 4. Plots are coloured to show the development over time. Each panel title shows the proportion of stress represented by the 2-dimensional MDS. }
	\label{fig:MDS_7Cov}
\end{figure}

\begin{figure}
	\centering
	\begin{subfigure}{1\textwidth}
		\centering
		\includegraphics[width=1\textwidth,height = 0.4\textheight]{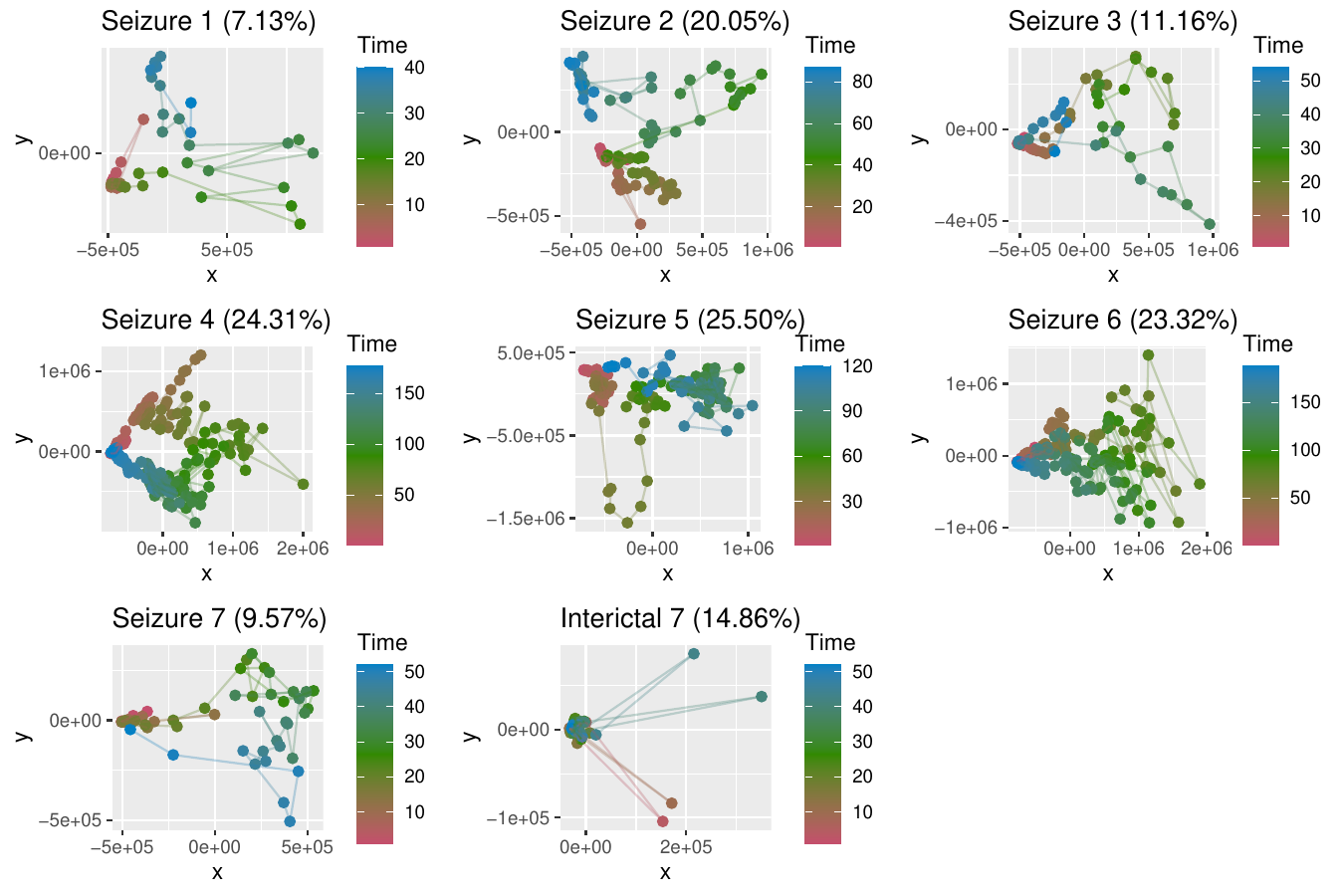}  
		\caption{MDS of seizure time series using the Euclidean metric $\deuc{S_i}{S_{j}}$.}
	\end{subfigure}
	\begin{subfigure}{1\textwidth}
		\centering
		\includegraphics[width=1\textwidth,height = 0.4\textheight]{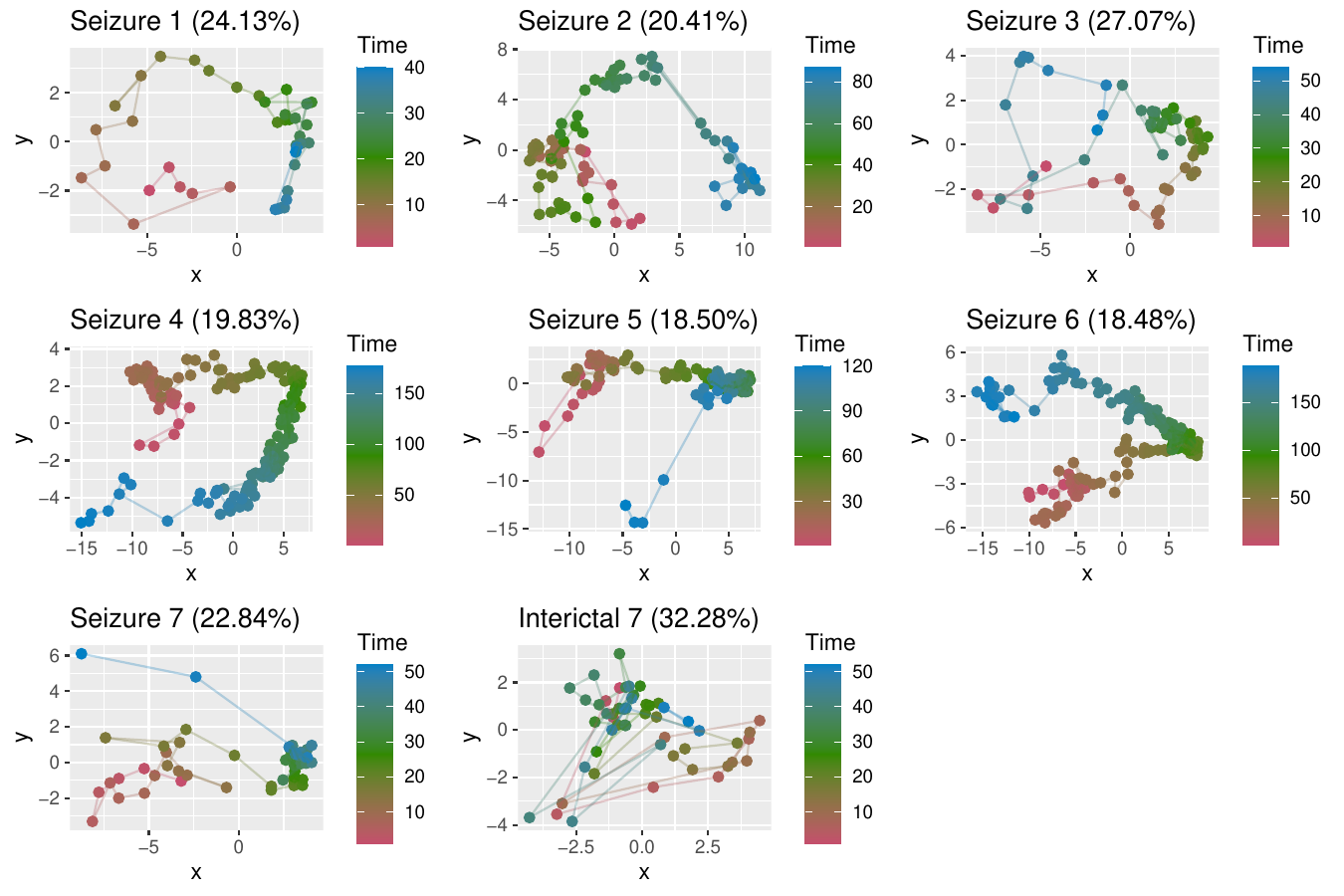}  
		\caption{MDS of seizure time series using the affine invariant metric $\daff{S_i}{S_{j}}$.}
	\end{subfigure}
	\caption{MDS of seizure time series for Patient 13 using the Euclidean and affine invariant metrics. Patient 13 had 7 seizures (first 7 panels). Panel 8 shows MDS results for the interictal period corresponding to seizure 7. Plots are coloured to show the development over time. Each panel title shows the proportion of stress represented by the 2-dimensional MDS. }
	\label{fig:MDS_13Cov}
\end{figure}

MDS plots for patient 6, 7 and 13 are shown in in Figures \ref{fig:MDS_6Cov},  \ref{fig:MDS_7Cov}, and  \ref{fig:MDS_13Cov} using the Euclidean and affine invariant metrics.

\subsection{Fr\'echet sample variance using the Euclidean metric}
\begin{figure}
	\centering
	\includegraphics[width=0.8\textwidth]{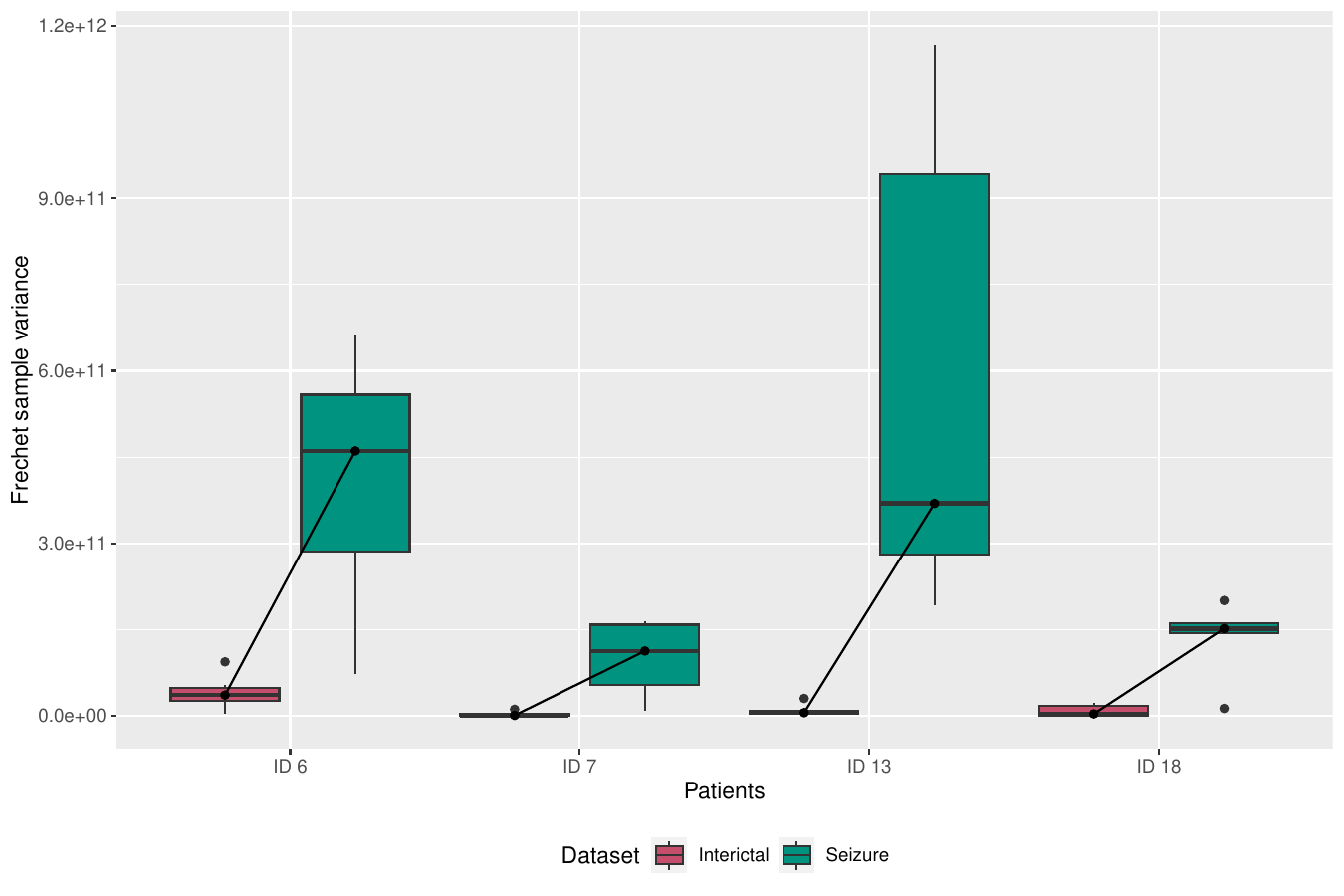}
	\caption{The Fr\'echet sample variance using the Euclidean metric. }
	\label{fig:ESV}
\end{figure}

Similar to Figure 2 in the main text, Figure \ref{fig:ESV} shows the Fr\'echet sample variance using the Euclidean metric for each seizure time series and the corresponding interictal time series, categorized by patient ID for patients 6, 7, 13, and 18. While we can draw similar conclusions to those from Figure  2, it is noteworthy that the variance values in Figure \ref{fig:ESV} change by may orders of magnitude between interictal periods and seizures when the Euclidean geometry is used. 

\section{Parameter inference}

Assuming $\vec{\epsilon}$ is normally distributed, the likelihood function of our manifold-adapted model  is
\begin{equation}{\label{equ:likelihoodfuncs}}
	\begin{split}
		\mathcal{L}(A_1,\ldots,A_L,B,\Sigma|\vec{v}) & = 
		\prod_{k = L+1}^{n}(2\pi)^{-\frac{m}{2}} |\Sigma|^{-\frac{1}{2}} \\
		& \times \exp\left\{ -\frac{1}{2}\left( \vec{v}_k - \sum_{\ell=1}^{L}A_{\ell}\vec{v}_{k\ell}- B \vec{v}^*_{k} \right)^T \Sigma^{-1} \left( \vec{v}_k- \sum_{l=1}^{L}A_{\ell}\vec{v}_{k\ell} - B \vec{v}^*_{k}\right)   \right\}.
	\end{split}
\end{equation}
The following two subsections give a detailed description of MLE for the scalar and diagonal models. 
We then discuss the asymptotic covariance of the estimated parameters.
\subsection{MLE for the scalar coefficient model}
Denote $\vec{\theta} = (\alpha_1,\ldots,\alpha_L, \beta)^T$ and $\Sigma = \sigma^2I_m$. Taking the logarithm of likelihood function $\log(\mathcal{L})$ calculated in Equation (\ref{equ:likelihoodfuncs}) for the  scalar coefficient model, we have
\begin{equation}{\label{equ:logliksca}}
	\log(\mathcal{L}) = -\frac{m(n-L)}{2}\log(2\pi)  - m(n-L)\log(\sigma)
	- \frac{1}{2\sigma^2} \sum_{k = L+1}^{n}\left\| \vec{v}_k - \sum_{l=1}^{L}\alpha_{\ell}\vec{v}_{k\ell}- \beta \vec{v}^*_{k}\right\|^2,
\end{equation}
where $\| \cdot \|$ is the Euclidean norm. 

Subsequently, we can calculate first derivatives with respect to $\alpha_{\ell} ( \ell = 1,\ldots, L)$ and $\beta$ as
\begin{equation*}
	\begin{split}
		\frac{\partial \log(\mathcal{L}) }{\partial \alpha_{\ell}} & = \frac{1}{\sigma^2} \sum_{k = L+1}^{n}\vec{v}_{k\ell}^T(\vec{v}_k- \sum_{\ell=1}^{L}\alpha_{\ell}\vec{v}_{k\ell} - \beta \vec{v}^*_{k}) = 0 \\
		\frac{\partial \log(\mathcal{L})}{\partial \beta} & =  \frac{1}{\sigma^2} \sum_{k = L+1}^{n}{\vec{v}^*_{k}}^T(\vec{v}_k - \sum_{l=1}^{L}\alpha_{\ell}\vec{v}_{k\ell} - \beta \vec{v}^*_{k}) = 0.
	\end{split}
\end{equation*}
Therefore, the MLE parameter vector $\hat{\vec{\theta}} = (\hat{\alpha}_1,...,\hat{\alpha}_L, \hat{\beta})^T$ is obtained by solving
\begin{equation*}
	\sum_{k=L+1}^{n} \begin{bmatrix}
		\|\vec{v}_{k1}\|^2 & \vec{v}_{k1}^T\vec{v}_{k2} & \cdots 
		& \vec{v}_{k1}^T\vec{v}_{kL} &\vec{v}_{k1}^T\vec{v}^*_{k} \\
		\vec{v}_{k2}^T\vec{v}_{k1} &  \|\vec{v}_{k2}\|^2 &\cdots & \vec{v}_{k2}^T\vec{v}_{kL} & \vec{v}_{k2}^T\vec{v}^*_{k}\\
		\vdots &  \vdots &\ddots & \vdots & \vdots\\
		\vec{v}_{kL}^T\vec{v}_{k1} & \vec{v}_{kL}^T\vec{v}_{k2}& \cdots & \|\vec{v}_{kL}\|^2 &
		\vec{v}_{kL}^T\vec{v}^*_{k}\\
		{\vec{v}^*_{k}}^T\vec{v}_{k1} &{\vec{v}^*_{k}}^T\vec{v}_{k2}  &\cdots 
		&{\vec{v}^*_{k}}^T\vec{v}_{kL}&
		\|\vec{v}^*_{k}\|^2
	\end{bmatrix}
	\begin{bmatrix}
		\alpha_1\\
		\alpha_2\\
		\vdots\\
		\alpha_L\\
		\beta
	\end{bmatrix} 
	= 
	\sum_{k=L+1}^{n}\begin{bmatrix}
		\vec{v}_{k1}^T\vec{v}_k\\
		\vec{v}_{k2}^T\vec{v}_k\\
		\vdots\\
		\vec{v}_{kL}^T\vec{v}_k\\
		{\vec{v}^*_{k}}^T\vec{v}_k
	\end{bmatrix} ,
\end{equation*}
so that $\hat{\vec{\theta}} = (Z^TZ)^{-1}Z^TY$, where $Z = \sum_{k=L+1}^{n} \begin{bmatrix}
	\|\vec{v}_{k1}\|^2 &  \cdots 
	&\vec{v}_{k1}^T\vec{v}^*_{k} \\
	\vdots  &\ddots & \vdots \\
	{\vec{v}^*_{k}}^T\vec{v}_{k1}&\cdots 
	& \|\vec{v}^*_{k}\|^2
\end{bmatrix}$ and $Y = \sum_{k=L+1}^{n}\begin{bmatrix}
	\vec{v}_{k1}^T\vec{v}_k\\
	\vdots\\
	{\vec{v}^*_{k}}^T\vec{v}_k
\end{bmatrix} $ for $k = L+1,...,n$. 

Additionally, $\hat{\sigma}$ is found by computing the first derivative of log-likelihood function $\log(\mathcal{L})$ with respect to $\sigma$ and letting it be 0, so that
\begin{equation*}
	\begin{gathered}
		\frac{\partial \log(\mathcal{L})}{\partial \sigma} = -\frac{m(n-L)}{\sigma} + \frac{1}{\sigma^3}\sum_{k = L+1}^n\left\| \vec{v}_k - \hat{\vec{v}}_k \right\|^2  \\
		\hat{\sigma} = \sqrt{\frac{1}{m(n-L)}\sum_{k = L+1}^n\left\| \vec{v}_k -  \hat{\vec{v}}_k \right\|^2 },
	\end{gathered}
\end{equation*}
where $\hat{\vec{v}}_k = \sum_{\ell=1}^{L}\hat{\alpha}_{\ell}\vec{v}_{k\ell}- \hat{\beta} \vec{v}^*_{k}$ is the estimated observation.

\subsection{MLE for the diagonal model}
When $A_{\ell} = \diag{a_{\ell1},\ldots,a_{\ell m}}, B = \diag{b_1,\ldots, b_m}$, and  $\Sigma = \diag{\sigma_1^2,\ldots,\sigma^2_m}$ are diagonal matrices, we can represent model  as
\begin{equation*}
	\begin{split}
		\vec{v}_k & = \sum_{\ell=1}^{L}\begin{bmatrix}
			v_{k\ell,1} & \cdots & 0 \\
			\vdots  & \ddots & \vdots\\
			0& \cdots & v_{k\ell,m}
		\end{bmatrix}\begin{bmatrix}
			a_{\ell 1}\\
			\vdots\\
			a_{\ell m}
		\end{bmatrix} + \begin{bmatrix}
			v^*_{k,1} & \cdots & 0 \\
			\vdots  & \ddots & \vdots\\
			0& \cdots & v^*_{k,m}
		\end{bmatrix}\begin{bmatrix}
			b_{1}\\
			\vdots\\
			b_{m}
		\end{bmatrix} + \begin{bmatrix}
			\epsilon_{k1} \\
			\vdots\\
			\epsilon_{km}
		\end{bmatrix} \\
		& = \sum_{l=1}^{L}\vec{u}_{il}\vec{a}_{\ell}+ \vec{u}^*_{k}\vec{b} + \vec{\epsilon}_{i},
	\end{split}
\end{equation*}
where $\vec{u}$ are diagonal matrices of $\vec{v}$, i.e. $\vec{u} = \diag{v_1,\ldots, v_m}$; $\vec{a}_{\ell} = (a_{\ell1},...,a_{\ell m})^T, \vec{b} =(b_1,...,b_m)^T$ are vectors from diagonal matrix $A_{\ell}, B, \ell = 1, \ldots,L$. Furthermore, $\vec{\epsilon}_i \sim N(0,\Sigma)$ and $\Sigma = \diag{\sigma_1^2,\ldots, \sigma_m^2}$.

The log-likelihood function is then written as
\begin{equation}{\label{equ:loglikdiag}}
	\begin{split}
		\log(\mathcal{L}) & = -\frac{m(n-L)}{2}\log(2\pi) - \frac{n-L}{2}\log(\sigma_1^2\cdots\sigma_m^2)  \\
		& - \frac{1}{2} \sum_{k = L+1}^{n}\left( \vec{v}_k - \sum_{\ell=1}^{L}\vec{u}_{k\ell}\vec{a}_{\ell}-  \vec{u}^*_{k}\vec{b} \right)^T\Sigma^{-1}\left( \vec{v}_k - \sum_{\ell=1}^{L}\vec{u}_{k\ell}\vec{a}_{\ell}- \vec{u}^*_{k}\vec{b} \right).
	\end{split}
\end{equation} 
To calculate the maximum likelihoood estimators, the first derivatives of the log-likelihood with respect to coefficients $A_{\ell}$ and $B$ are calculated and set to 0. Then, the estimated $\hat{\Theta} = (\hat{\vec{a}}_1,...,\hat{\vec{a}}_L,\hat{\vec{b}})^T$  is found using 
\begin{equation*}
	\hat{\Theta} = (Z^TZ)^{-1}Z^TY,
\end{equation*}
where $Z = \sum_{k=L+1}^{n} \begin{bmatrix}
	\vec{u}_{k1}\vec{u}_{k1} & \cdots &\vec{u}_{k1}\vec{u}_{kL}& \vec{u}_{k1}\vec{u}^*_{k}\\
	\vdots & \ddots & \vdots & \vdots \\
	\vec{u}_{kL}\vec{u}_{k1} & \cdots &\vec{u}_{kL}\vec{u}_{kL}& \vec{u}_{kL}\vec{u}^*_{k}\\
	\vec{u}^*_{k}\vec{u}_{k1} & \cdots &\vec{u}^*_{k}\vec{u}_{kL}& \vec{u}^*_{k}\vec{u}^*_{k}\\
\end{bmatrix}$ and $Y = \sum_{k = L+1}^n\begin{bmatrix}
	\vec{u}_{k1}\vec{v}_{k}\\
	\vdots\\
	\vec{u}_{kL}\vec{v}_{k}\\
	\vec{u}^*_{k}\vec{v}_{k}
\end{bmatrix}$.

Similarly, $\hat{\sigma}_r$ is found setting the derivative of the log-likelihood to zero: 
\begin{equation*}
	\begin{split}
		\frac{\partial \log(\mathcal{L}) }{\partial \sigma_r} & = - \frac{n-L}{\sigma_r} + \frac{1}{\sigma_r^3} \sum_{k = L+1}^n \left(\vec{v}_k -  \hat{\vec{v}}_k \right)_r^2 = 0 \\
		\hat{\sigma}_r & = \sqrt{\frac{ \sum_{k = L+1}^n \left(\vec{v}_k -  \hat{\vec{v}}_k \right)_r^2}{n-L}},
	\end{split}
\end{equation*}
where $(\vec{v}_k - \hat{\vec{v}}_k)_r$ is the $r$th component of residual vector $\sum_{k = L+1}^n \left(\vec{v}_k -  \hat{\vec{v}}_k \right)^2$, $r = 1,...,m$. 

\subsection{Asymptotic covariance matrix of estimators} 
The asymptotic covariance matrix of maximum likelihood estimator of $\Phi = (\alpha_1,...,\alpha_L,\beta, \sigma)^T$ in the scalar coefficient model (or $\Phi = (\vec{a}_1,\ldots,\vec{a}_L,\vec{b},\vec{\sigma})^T$  in the diagonal  model)  is used to construct confidence intervals and provides a measurement of the uncertainty associated with the estimated parameters. 

\subsubsection{Fisher information matrix of estimators in the scalar coefficient model.}

In model  with scalar coefficients, we firstly compute the \emph{Hessian matrix}, which is the second derivative of log-likelihood function defined as
\begin{equation*}
	H( \log(\mathcal{L}))  = \begin{bmatrix}
		\frac{\partial^2\log(\mathcal{L})}{\partial \alpha_1^2} & \cdots &\frac{\partial^2\log(\mathcal{L})}{\partial \alpha_1\partial\alpha_L}
		&\frac{\partial^2\log(\mathcal{L})}{\partial \alpha_1\partial\beta}
		&\frac{\partial^2\log(\mathcal{L})}{\partial \alpha_1\partial\sigma}\\
		\vdots & \ddots & \vdots & \vdots & \vdots\\
		\frac{\partial^2\log(\mathcal{L})}{\partial \alpha_1\partial\alpha_L}  & \cdots
		&\frac{\partial^2\log(\mathcal{L})}{\partial \alpha_L^2}
		&\frac{\partial^2\log(\mathcal{L})}{\partial \alpha_L\partial\beta}
		& \frac{\partial^2\log(\mathcal{L})}{\partial \alpha_L\partial\sigma}\\
		
		\frac{\partial^2\log(\mathcal{L})}{\partial \alpha_1\partial\beta}  & \cdots
		&\frac{\partial^2\log(\mathcal{L})}{\partial \alpha_L\partial\beta}
		&\frac{\partial^2\log(\mathcal{L})}{\partial \beta^2}
		& \frac{\partial^2\log(\mathcal{L})}{\partial \beta\partial\sigma}\\
		
		\frac{\partial^2\log(\mathcal{L})}{\partial \alpha_1\partial\sigma}  & \cdots
		&\frac{\partial^2\log(\mathcal{L})}{\partial \alpha_L\partial\sigma}
		&\frac{\partial^2\log(\mathcal{L})}{\partial \beta \partial\sigma}
		& \frac{\partial^2\log(\mathcal{L})}{\partial \sigma^2}
	\end{bmatrix}
\end{equation*}
where the second cross-partial derivative of $\log(\mathcal{L})$ is
\begin{align*}
	\frac{\partial^2 \log(\mathcal{L})}{\partial \alpha_{\ell} \partial \alpha_o} & = -\frac{1}{\sigma^2} \sum_{k=L+1}^{n}\vec{v}_{k\ell}^T\vec{v}^*_{k}\\
	\frac{\partial^2  \log(\mathcal{L})}{\partial \alpha_{\ell} \partial \beta} & = -\frac{1}{\sigma^2} \sum_{k=L+1}^{n}\vec{v}_{k\ell}^T\vec{v}^*_{k}\\
	\frac{\partial^2 \log(\mathcal{L})}{\partial \sigma^2} & = \frac{m(n-L)}{\sigma^2} - \frac{3}{\sigma^4} \sum_{k = L+1}^n\left\| \vec{v}_i - \sum_{\ell=1}^{L}\hat{\alpha}_{\ell}\vec{v}_{k\ell}- \hat{\beta} \vec{v}^*_{k} \right\|^2\\
	\frac{\partial^2  \log(\mathcal{L})}{\partial \alpha_{\ell} \partial\sigma} &= \frac{\partial}{\partial \alpha_{\ell}} \left( -\frac{m(n-L)}{\sigma} + \frac{1}{\sigma^3} \sum_{k = L+1}^n\left\| \vec{v}_k - \sum_{\ell =1}^{L}\hat{\alpha}_{\ell}\vec{v}_{k\ell}- \hat{\beta} \vec{v}^*_{k} \right\|^2  \right)\\
	& = \frac{\partial}{\partial \sigma} \left( \frac{1}{\sigma^2} \sum_{k = L+1}^{n}\vec{v}_{k \ell}^T \left( \vec{v}_k - \sum_{\ell=1}^{L}\hat{\alpha}_{\ell}\vec{v}_{k\ell}- \hat{\beta} \vec{v}^*_{k} \right)  \right) \\
	&= - \frac{2}{\sigma^3} \sum_{k = L+1}^{n}\vec{v}_{k\ell}^T \left( \vec{v}_k - \sum_{\ell =1}^{L}\hat{\alpha}_{\ell}\vec{v}_{k\ell}- \hat{\beta} \vec{v}^*_{k} \right)
\end{align*}	
where $\ell,o = 1,\ldots,L$. 

Recall the model $\vec{v}_k =\sum_{\ell =1}^{L}\hat{\alpha}_{\ell}\vec{v}_{k\ell} + \hat{\beta} \vec{v}^*_{k} + \vec{\epsilon}_i$. We have
\begin{align*}
	\mathbb{E}\left[\vec{v}_k- \sum_{\ell=1}^{L}\alpha_{\ell}\vec{v}_{k\ell}- \beta \vec{v}^*_{k}\right] &= 0 \\
	\mathbb{E}\left[ \left\|\vec{v}_i -  \sum_{\ell =1}^{L}\alpha_{\ell}\vec{v}_{k \ell}- \beta \vec{v}^*_{k}\right\|^2\right] &= m\sigma^2.
\end{align*}
Consequently, the negative expectation of the second partial derivative of $ \log(\mathcal{L})$ are
\begin{align*}
		-\mathbb{E}\left[\frac{\partial^2  \log(\mathcal{L})}{\partial \alpha_{\ell} \partial \alpha_o} \right] & = \frac{1}{\sigma^2}\sum_{k = L+1}^{n}\vec{v}^T_{k\ell}\vec{v}_{ko}\\
		-\mathbb{E}\left[\frac{\partial^2  \log(\mathcal{L})}{\partial \alpha_{\ell} \partial \beta} \right] & =  \frac{1}{\sigma^2}\sum_{k = L+1}^{n}\vec{v}^T_{k\ell}\vec{v}^*_{k}\\
		\begin{split}
			-\mathbb{E}\left[\frac{\partial^2  \log(\mathcal{L})}{\partial \sigma^2} \right]  & = -
			\frac{m(n-L)}{\sigma^2} + \frac{3}{\sigma^4} \sum_{k = L+1}^{n} \mathbb{E}\left[ \left\|\vec{v}_k -  \sum_{\ell =1}^{L}\alpha_{\ell}\vec{v}_{k\ell}- \beta \vec{v}^*_{k}\right\|^2\right]\\
			&= - \frac{m(n-L)}{\sigma^2} + \frac{3}{\sigma^4}\sum_{k = L+1}^{n}m \sigma^2\\
			&= - \frac{m(n-L)}{\sigma^2} + \frac{3m(n-L)\sigma^2}{\sigma^4}\\
			& = \frac{2m(n-L)}{\sigma^2}
		\end{split}\\
		-\mathbb{E}\left[\frac{\partial^2  \log(\mathcal{L})}{\partial \alpha_{\ell} \partial \sigma} \right]   & = 
		-\frac{2}{\sigma^3} \sum_{k = L+1}^{n} \left( \mathbb{E}[\vec{v}_{k\ell}^T]\mathbb{E}\left[\vec{v}_k - \sum_{\ell=1}^{L}\alpha_{\ell}\vec{v}_{k\ell}- \beta \vec{v}^*_{k}\right]\right) = 0
\end{align*}

Therefore, the \emph{Fisher information matrix} is 
\begin{align*}
	I(\Phi) & =  -\mathbb{E}[H(\log(\mathcal{L}))] \\
	&= \frac{1}{\sigma^2}\sum_{k = L+1}^{n} \begin{bmatrix}
		\vec{v}^T_{k1}\vec{v}_{k1} &   \cdots & \vec{v}^T_{k1}\vec{v}_{kL}  & \vec{v}^T_{k1}\vec{v}^*_{k}& 0 \\
		\vdots &  \ddots & \vdots & \vdots &  0\\
		{\vec{v}^*}^T_{k}\vec{v}_{i1} &   \cdots & {\vec{v}^*}^T_{k}\vec{v}_{iL} &{\vec{v}^*}^T_{k}\vec{v}^*_{k} & 0 \\		0 &   \cdots & 0 &   0 &  2m
	\end{bmatrix}
\end{align*}

\subsubsection{Fisher information matrix of estimation in the diagonal model.}

Analogously, we can obtain the Fisher information matrix of maximum likelihood estimator $\Phi = (a_{11},\ldots,a_{1m},a_{21},\ldots,a_{Lm},b_{1},\ldots,b_m,\sigma_1,\ldots,\sigma_m)^T$  from the Hessian matrix. Recalling the log-likelihood function of the manifold-adapted model  with diagonal matrix coefficients and representing the diagonal matrix coefficients as vectors, log-likelihood function in (\ref{equ:loglikdiag}) could be written as
\begin{equation}
	\begin{split}
		\log(\mathcal{L}) = & -\frac{m(n-L)}{2}\log(2\pi) - \frac{n-L}{2}\log(\sigma_1^2\cdots\sigma_m^2) \\
		& - \frac{1}{2} \sum_{k = L+1}^n  \left(   \sum_{r=1}^m \frac{1}{ \sigma_r^{2}} (v_{kr} - \sum_{\ell = 1}^{L} a_{\ell r} v_{k\ell,r} -  v^*_{k,r}b_r)^2
		\right)
	\end{split}
\end{equation}
where $ r = 1,\ldots, m; \ell = 1,\ldots, L; k = L+1,\ldots, n$.  Compute the first derivatives of log-likelihood function with respect to all parameters as
\begin{align*}
	\frac{\partial \log(\mathcal{L})}{\partial a_{\ell r}}  &=  \frac{1}{\sigma_r^2} \sum_{k = L+1}^{n} v_{k\ell,r}(v_{i,r} - \sum_{\ell=1}^{L} v_{k\ell,r}a_{\ell r} - b_r v^*_{k,r}) \\
	\frac{\partial \log(\mathcal{L})}{\partial b_{r}} & =  \frac{1}{\sigma_r^2}  \sum_{k = L+1}^{n} v^*_{k,r}(v_{k,r} -  \sum_{\ell=1}^{L} v_{k\ell,r}a_{\ell r} - b_r v^*_{k,r}) \\
	\frac{\partial \log(\mathcal{L})}{\partial \sigma_{r}} &=  -  \frac{n-L}{\sigma_r}  + \frac{1}{\sigma_r^3}  \sum_{k = L+1}^{n} \left(v_{k,r} - \sum_{\ell = 1}^{L}v_{k\ell,r}a_{\ell r} - b_r v^*_{k,r}\right)^2,
\end{align*}
and their second derivatives are
\begin{align*}
	\frac{\partial^2 \log(\mathcal{L})}{\partial a_{\ell r} \partial \alpha_{po}}  & =  - \frac{1}{\sigma_r^2}\sum_{k = L+1}^{n} v_{k\ell,r}v_{kp,o} \\
	\frac{\partial^2 \log(\mathcal{L})}{\partial a_{po} \partial a_{\ell r}}   &=  - \frac{1}{\sigma_o^2}\sum_{k = L+1}^{n} v_{k\ell,r}v_{kp,o} \\
	\frac{\partial^2 \log(\mathcal{L})}{\partial a_{\ell r}  \partial b_r}  & =  - \frac{1}{\sigma_r^2}\sum_{k = L+1}^{n} v_{k\ell,r}v^*_{k,r}  \\
	\frac{\partial^2 \log(\mathcal{L})}{\partial a_{\ell r} \partial \sigma_r}  & =  - \frac{2}{\sigma_r^3} \sum_{k = L+1}^{n} v_{k\ell,r}(v_{k,r} - \sum_{\ell=1}^L v_{k\ell,r}a_{\ell r} - v^*_{k,r}b_r )\\
	\frac{\partial^2 \log(\mathcal{L})}{\partial b_{r} \partial a_{\ell r}}  & =  - \frac{1}{\sigma_r^2}\sum_{k = L+1}^{n} v^*_{k,r}v_{k\ell,r} \\
	\frac{\partial^2 \log(\mathcal{L})}{\partial b_{r}^2}  & =  - \frac{1}{\sigma_r^2}\sum_{k = L+1}^{n}  {v^*}_{k,r}^2\\
	\frac{\partial^2 \log(\mathcal{L})}{\partial b_{o} \partial a_{\ell r}}  & =  - \frac{1}{\sigma_o^2}\sum_{k = L+1}^{n} v^*_{k,o}v_{k\ell,r} \\
	\frac{\partial^2 \log(\mathcal{L})}{\partial b_{r} \partial \sigma_r}  & =  - \frac{2}{\sigma_r^3} \sum_{k = L+1}^{n} v^*_{k,r}(v_{k,r} - \sum_{\ell = 1}^{L} v_{k\ell,r}a_{\ell r} -  v^*_{k,r}b_r)\\
	\frac{\partial^2 \log(\mathcal{L})}{\partial \sigma_r \partial a_{\ell r}}  & =  - \frac{2}{\sigma_r^3}\sum_{k = L+1}^{n}  v_{k\ell,r}(v_{k,r} - \sum_{\ell = 1}^{L} v_{k\ell,r}a_{\ell r} -  v^*_{k,r}b_r) \\
	\frac{\partial^2 \log(\mathcal{L})}{\partial \sigma_r \partial b_r}  & = -   \frac{2}{\sigma_r^3}\sum_{k = L+1}^{n}  v^*_{k,r}(v_{k,r} - \sum_{\ell = 1}^{L} v_{k\ell,r}a_{\ell r} - v^*_{k,r} b_r) \\
	\frac{\partial^2 \log(\mathcal{L})}{\partial \sigma_r^2}  & =  \frac{n-L}{\sigma_r^2} - \frac{3}{\sigma_r^4}\sum_{k = L+1}^{n}  \left(v_{k,r} - \sum_{\ell = 1}^{L} v_{k\ell,r}a_{\ell r} -v^*_{k,r} b_r \right)^2\\
	\frac{\partial^2 \log(\mathcal{L})}{\partial \sigma_o \partial a_{\ell r}}  & =  \frac{\partial^2  \log(\mathcal{L})}{\partial \sigma_o \partial b_r}=  \frac{\partial^2 \log(\mathcal{L})}{\partial \sigma_o\partial\sigma_r} = 0,
\end{align*}
where $\ell ,p = 1,\ldots,L; r,o = 1,\ldots,m$.

Therefore, we can obtain a large  Hessian matrix of dimension ($(L+2)m \times (L+2)m$)  as
\begin{equation*}
	H(\log(\mathcal{L})) = \begin{bmatrix}
		\frac{\partial^2 \log(\mathcal{L})}{\partial a_{11}^2} & \cdots & \frac{\partial^2 \log(\mathcal{L})}{\partial a_{11}a_{1m}} & \cdots   &    \frac{\partial^2 \log(\mathcal{L})}{\partial a_{11} \partial a_{Lm}} &  \frac{\partial^2 \log(\mathcal{L})}{\partial a_{11}\partial b_1}  & \cdots & \frac{\partial^2 \log(\mathcal{L})}{\partial a_{11}\partial\sigma_1} &  \cdots & \frac{\partial^2 \log(\mathcal{L})}{\partial a_{11}\partial\sigma_m} \\
		& \ddots & & \ddots&  & & \ddots  &   & \ddots   & \\
		\frac{\partial^2 \log(\mathcal{L})}{\partial a_{1m}\partial a_{11}}  & \cdots &  \frac{\partial^2 \log(\mathcal{L})}{\partial a_{1m}^2} &\cdots  & \frac{\partial^2 \log(\mathcal{L})}{\partial a_{1m}\partial a_{Lm}}  & \frac{\partial^2 \log(\mathcal{L})}{\partial a_{1m}\partial b_1} &\cdots &  \frac{\partial^2 \log(\mathcal{L})}{\partial a_{1m}\partial \sigma_1}  & \cdots & \frac{\partial^2 \log(\mathcal{L})}{\partial a_{1m}\partial \sigma_m}  \\
		& \ddots & & \ddots&  & & \ddots  &   & \ddots   & \\
		\frac{\partial^2 \log(\mathcal{L})}{\partial a_{Lm}\partial a_{11}}  & \cdots &  \frac{\partial^2 \log(\mathcal{L})}{\partial a_{Lm}^2} &\cdots  & \frac{\partial^2 \log(\mathcal{L})}{\partial a^2_{Lm}}  & \frac{\partial^2 \log(\mathcal{L})}{\partial a_{Lm}\partial b_1} &\cdots &  \frac{\partial^2 \log(\mathcal{L})}{\partial a_{Lm}\partial\sigma_1}& \cdots &\frac{\partial^2 \log(\mathcal{L})}{\partial a_{Lm}\partial \sigma_m} \\
		\frac{\partial^2 \log(\mathcal{L})}{\partial b_1 \partial a_{11}}  & \cdots  &  \frac{\partial^2 \log(\mathcal{L})}{\partial b_1 a_{Lm}} &\cdots  & \frac{\partial^2 \log(\mathcal{L})}{\partial b_1 \partial a_{Lm}}  & \frac{\partial^2 \log(\mathcal{L})}{\partial b_1^2} &\cdots   & \frac{\partial^2 \log(\mathcal{L})}{\partial b_1 \partial\sigma_1} & \cdots &\frac{\partial^2 \log(\mathcal{L})}{\partial b_1 \partial \sigma_m}  \\
		& \ddots & & \ddots&  & & \ddots  &   & \ddots   & \\
		\frac{\partial^2 \log(\mathcal{L})}{\partial \sigma_1 \partial a_{11}} & \cdots  & \frac{\partial^2 \log(\mathcal{L})}{\partial \sigma_1a_{1m}} & \cdots  &  \frac{\partial^2 \log(\mathcal{L})}{\partial \sigma_1 \partial a_{Lm}} &  \frac{\partial^2 \log(\mathcal{L})}{\partial \sigma_1\partial b_1} &\cdots & \frac{\partial^2 \log(\mathcal{L})}{\partial \sigma_1^2} & \cdots & \frac{\partial^2 \log(\mathcal{L})}{\partial \sigma_1\partial\sigma_m} \\
		& \ddots & & \ddots&  & & \ddots  &   & \ddots   & \\
		\frac{\partial^2 \log(\mathcal{L})}{\partial \sigma_m \partial a_{11}} & \cdots  & \frac{\partial^2 \log(\mathcal{L})}{\partial \sigma_ma_{1m}} &\cdots  &  \frac{\partial^2 \log(\mathcal{L})}{\partial \sigma_m \partial a_{Lm}} &  \frac{\partial^2\ell}{\partial \sigma_m\partial b_1} &\cdots & \frac{\partial^2 \log(\mathcal{L})}{\partial \sigma_m\partial\sigma_1} &\cdots & \frac{\partial^2 \log(\mathcal{L})}{\partial\sigma^2_m} 
	\end{bmatrix}
\end{equation*}

Now, we compute the expectation of each derivative in the above Fisher information matrix which is defined as the negative expectation of Hessian matrix $H( \log(\mathcal{L}))$, i.e., $I(\Phi)  =  -\mathbb{E}[H( \log(\mathcal{L}))] $ . According to the model, it is known that
\begin{align*}
	\E\left[ v_{r}  - \sum_{\ell}v_{\ell,r}a_{\ell r} - b_rv_{0,r}\right]  & = 0 \\
	\E\left[ (v_{r}  - \sum_{\ell} v_{\ell,r}a_{\ell r} - b_rv_{0,r})^2\right]  & =  \sigma_r^2.
\end{align*}
It follows that 
\begin{align*}
	-\E \left[ \frac{\partial^2 \log(\mathcal{L})}{\partial a_{\ell r} \partial \sigma_r } \right]   & =   -\E \left[ \frac{\partial^2 \log(\mathcal{L})}{\partial b_{r} \partial \sigma_r}  \right] = \E\left[ \frac{\partial^2 \log(\mathcal{L})}{\partial \sigma_r \partial a_{\ell r}} \right]  = \E\left[ \frac{\partial^2 \log(\mathcal{L})}{\partial \sigma_r \partial b_r} \right]= 0 \\
	- \E \left[  \frac{\partial^2 \log(\mathcal{L})}{\partial \sigma_r^2}  \right] & = -  \frac{n-L}{\sigma_{r}^2} + \frac{3(n- L)}{\sigma_r^2} = \frac{2(n-L)}{\sigma_r^{2}}.
\end{align*}
The Fisher information matrix is then
\begin{equation*}
	I(\Phi) =
	\begin{bmatrix}
		\frac{\sum_{k}v_{k1,1}^2}{\sigma_1^2}&  \cdots &  \frac{\sum_{k}v_{k1,1}v_{k1,m}}{\sigma_1^2}&\cdots  & 
		\frac{\sum_{k}v_{k1,1}v_{kL,m}}{\sigma_1^2}&   \frac{\sum_{k}v_{k1,1}v^*_{k,1}}{\sigma_1^2} &\cdots &0& \cdots & 0 \\
		& \ddots & & \ddots&  & & \ddots  &   & \ddots   & \\
		\frac{\sum_{k}v_{k1,m}v_{k1,1} }{\sigma_m^2} &\cdots  &  \frac{\sum_{k}v^2_{k1,m} }{\sigma_m^2}&\cdots  &  \frac{\sum_{k}v_{k1,m}v_{kL,m} }{\sigma_m^2}& \frac{\sum_{k}v_{k1,m}v^*_{k,1} }{\sigma_m^2} & \cdots  & 0&\cdots& 0  \\
		& \ddots & & \ddots&  & & \ddots  &   & \ddots   & \\
		\frac{\sum_{k}v_{kL,m}v_{k1,1} }{\sigma_m^2} & \cdots &  \frac{\sum_{k}v_{kL,m}v_{k1,m} }{\sigma_m^2}&  \cdots & \frac{\sum_{k}v_{kL,m}^2 }{\sigma_m^2}& \frac{\sum_{k}v_{kL,m}v^*_{k,1}}{\sigma_m^2} &\cdots  &0& \cdots &0  \\
		\frac{\sum_{k}v^*_{k,1}v_{k1,1}}{\sigma_1^2}& \cdots &   \frac{\sum_{k}v^*_{k,1}v_{k1,m} }{\sigma_1^2} &\cdots  & \frac{\sum_{k}v^*_{k,1}v_{kL,m}}{\sigma_1^2}  & \frac{\sum_{k}{v^*}_{k,1}^2}{\sigma_1^2}&  \cdots &0& \cdots & 0 \\
		& \ddots & & \ddots&  & & \ddots  &   & \ddots   & \\
		0& \cdots & 0& \cdots & 
		0 & 0 &\cdots  &\frac{2(n-L)}{\sigma_1^{2}}& \cdots&  0 \\
		& \ddots & & \ddots&  & & \ddots  &   & \ddots   & \\
		0& \cdots & 0& \cdots & 
		0 & 0 &\cdots  & 0& \cdots &\frac{2(n-L)}{\sigma_m^{2}}
	\end{bmatrix} .
\end{equation*}

\textbf{Asymptotic covariance matrices of maximum likelihood estimators.}
We can establish approximate $95\%$ confidence intervals for the estimated parameters using:
\begin{equation*}
	\hat{\Phi}_j \pm 1.96 I(\Phi)_{jj}^{-1/2}
\end{equation*}
where $j = 1,\ldots,J$ and $J$ is the total number of estimated parameters, e.g. $J = L+2$ in the scalar coefficient model and $J = (L+2)m$ in the diagonal  matrix model.

Under the correct specification of the model, the maximum likelihood regularity condition, and additional technical assumptions, $\sqrt{n}(\hat{\Phi}-\Phi)$ converges to a multivariate normal distribution with zero mean and covariance matrix $I(\Phi)^{-1}$, i.e. 
\begin{equation*}
	\sqrt{n}(\hat{\Phi}-\Phi) \rightarrow N(0, I(\Phi)^{-1})
\end{equation*}
More precisely, for `large' $n$, the distribution of the vector $\hat{\Phi}$ can be approximated by a multivariate normal distribution with mean $\Phi$ and covariance matrix 
\begin{equation*}
	\frac{1}{n}I(\Phi)^{-1}.
\end{equation*}

\section{Additional model results: scalar model}

\subsection{Selecting the maximum lag $L$}
\begin{figure}
	\centering
	\includegraphics[width=0.8\textwidth]{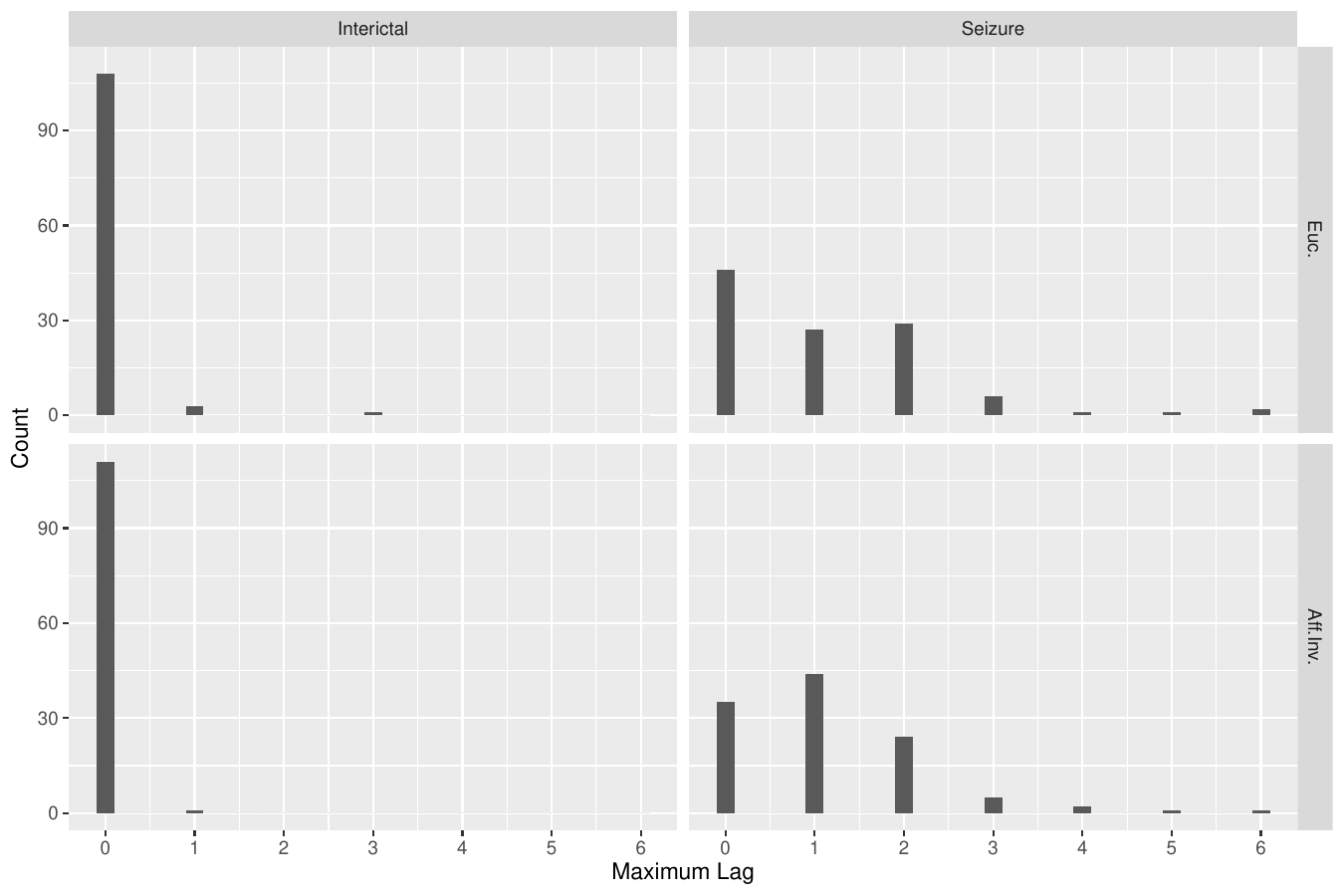}
	\caption{Histograms of the value of $L$ selected for the scalar coefficient model fitted to seizure and interical series in Euclidean and affine invariant geometries.}
	\label{fig:selectlag}
\end{figure}

Figure~\ref{fig:selectlag} displays a histogram of the values of $L$ obtained across all seizures and corresponding interictal periods in both geometries using the procedure given in the main text. 
The plot shows that the majority of seizures have $L\leq 3$ seconds while all but a few interictal series have $L=0$. 

\subsection{Oscillatory behaviour of tangent vectors}
\begin{figure}
	\centering
	\includegraphics[width=0.8\textwidth]{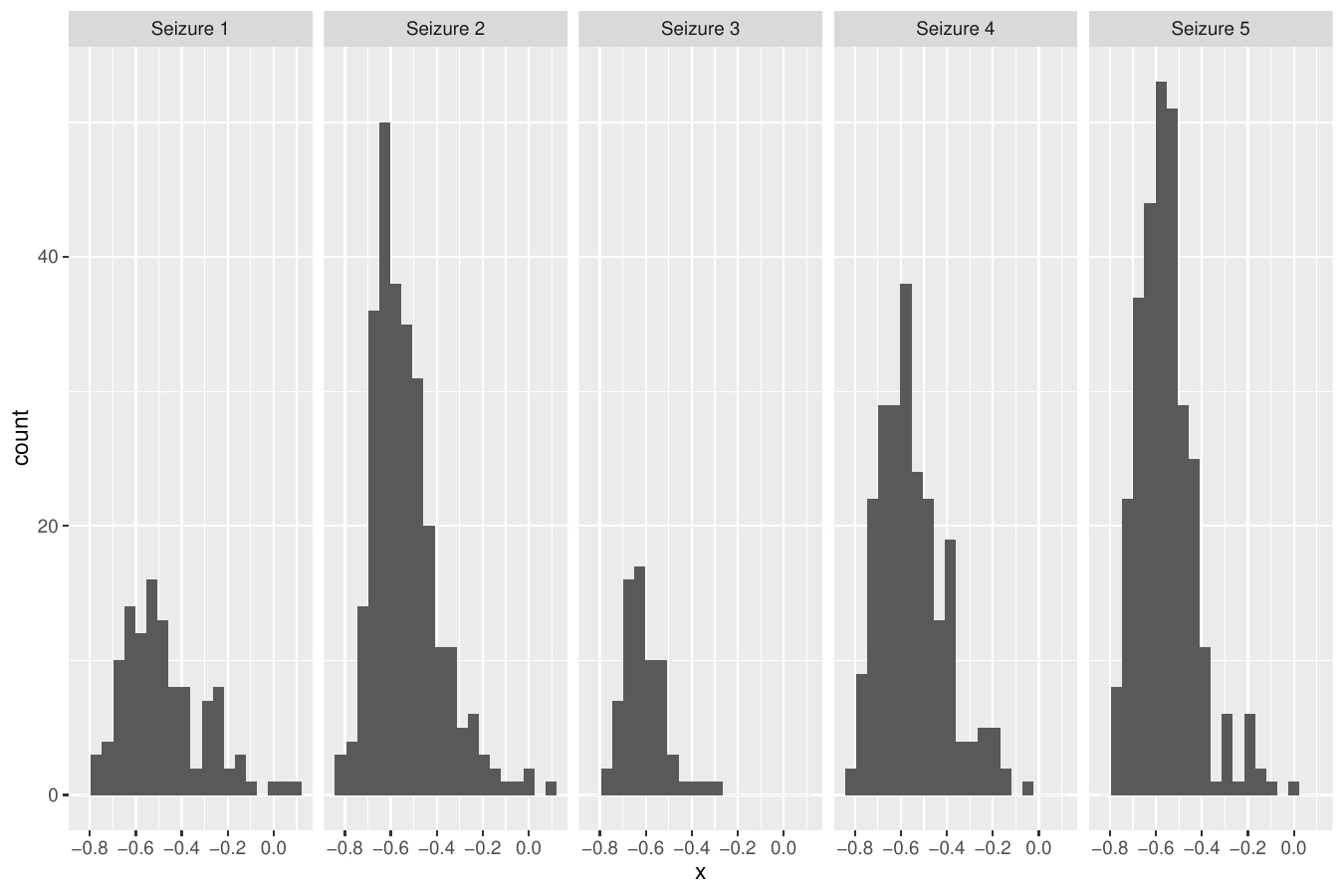}
	\caption{Histograms of  normalised inner products between tangent vectors $V_k$ and $V_{k,1}$ for 5 seizures in patient 18.}
	\label{fig:Histogram_innerpro}
\end{figure}

Negative values for the autoregressive coefficients $\alpha_\ell$ were obtained when the scalar model was fitted to seizures. 
Further analyses were performed to investigate this. 
We calculated the quantities $\gaff{S_k}{V_k}{V_{k,1}}/\|V_k\|\|V_{k,1}\|, k = 2,\ldots,n$ for each seizure for patient 18 as shown in Figure ~\ref{fig:Histogram_innerpro}. 
These values are all negative and values close to $-1$ indicate a complete reversal of direction. 

\begin{figure}[htp]
	\centering
	\includegraphics[width=1\textwidth]{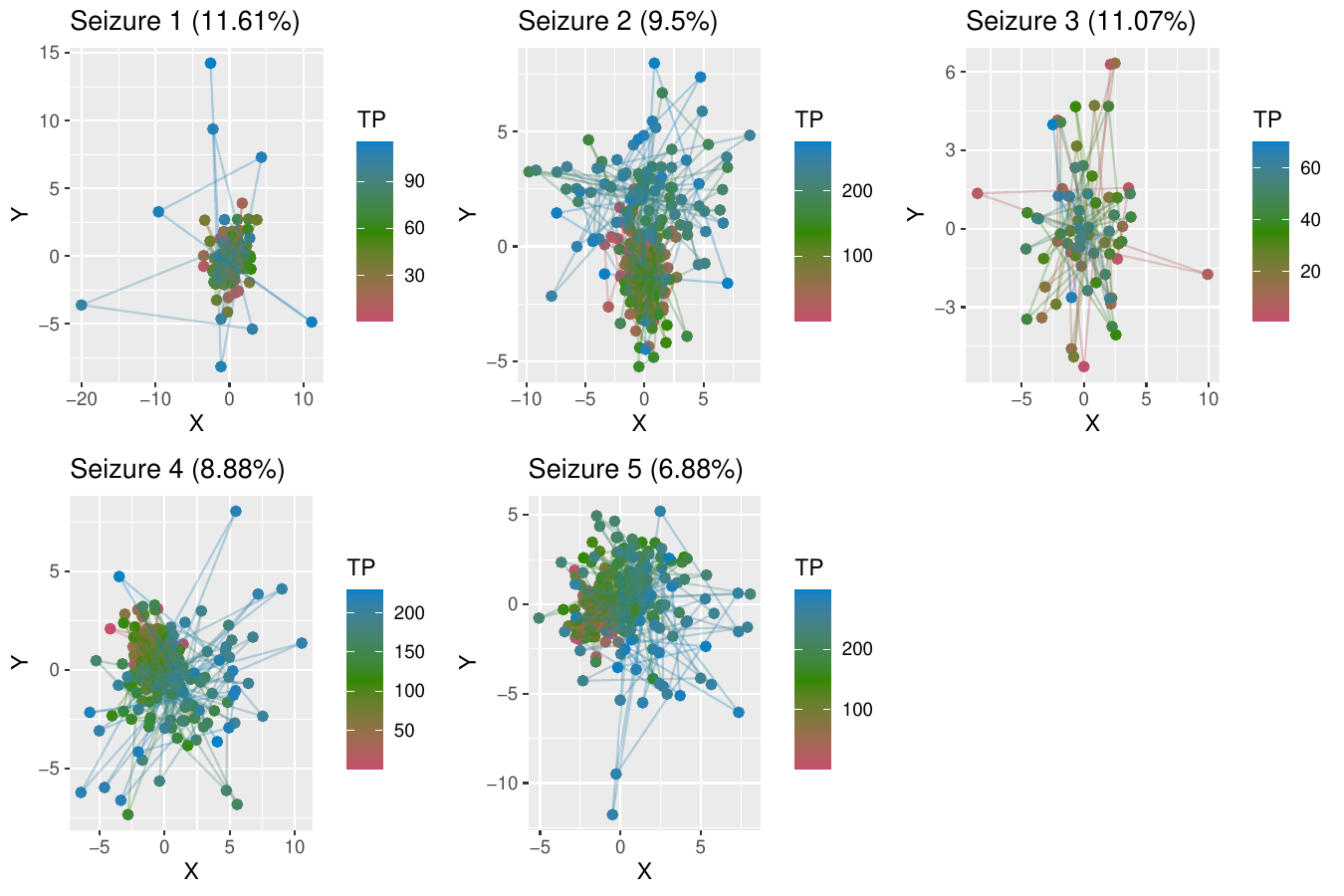}
	\caption{Two-dimensional PCA plots on translated vectors $V_{i,0}$ onto the same tangent space at the identity matrix $I \in \spd{p}$ for 5 seizure  series in patient 18. Plots are coloured to indicate the progression over time. Each panel title indicates the proportion of variance represented by the 2-dimensional eigenvectors with the top 2 eigenvalues.}
	\label{fig:PCA_V}
\end{figure}

Furthermore, we conducted a principal component analysis (PCA) on the tangent vectors $V_{i,0}$ by parallel transporting each vector $V_i$ to the tangent space at the identity matrix $I \in \spd{p}$. Specifically, given a data set ${S_i, i = 1,\ldots,n}$, we computed tangent vectors as $V_i = \Log_{S_i}(S_{i+1})$. Subsequently, we obtained translated vectors $V_{i,0} = \ptrans{S_i}{I}(V_i) \in \tgt{I}$ and performed PCA on these. Figure \ref{fig:PCA_V} displays the two-dimensional PCA plots for patient 18 with 5 seizures, although the proportion of variance captured by two dimensions is low. 
These plots also show a tendancy to reverse direction at each time point. 

\subsection{Squared norms of different terms in the scalar coefficient model}
\begin{figure}
    \centering
    \includegraphics[width=\textwidth]{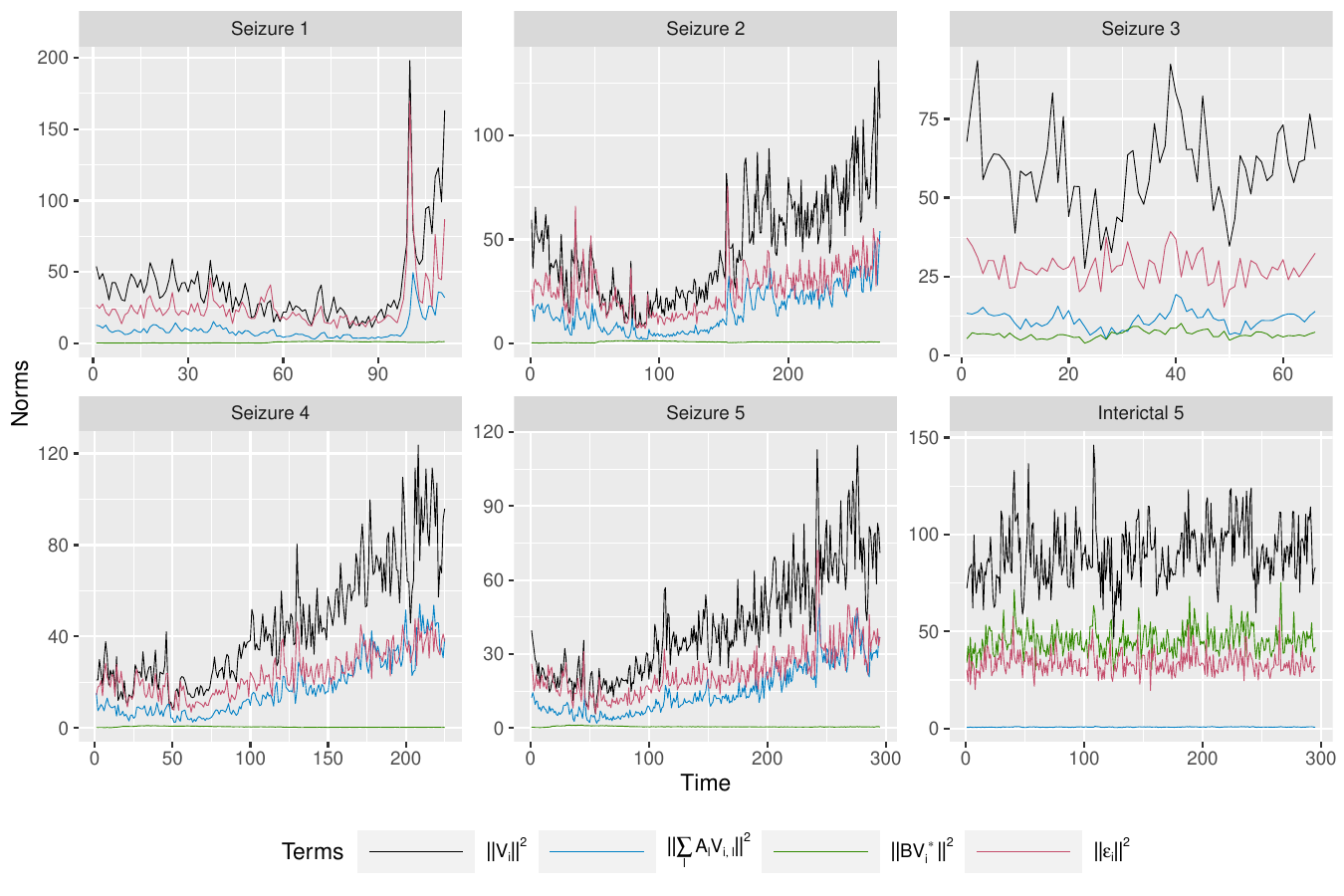}
    \caption{Squared norms of different terms in the scalar coefficient model for patient 18 in the affine invariant geometry. The black line represents the norm of the observed value, the blue line corresponds to the autoregressive term, the green line depicts the mean-reverting term, and the red line denotes the noise term.}
    \label{fig:Norms_terms}
\end{figure}

After fitting the scalar model to patient 18, we computed squared norms for various terms as functions of time $i$: the observed value $\|V_i\|^2$, the autoregressive term $\|\sum_{\ell}^{L}A_{\ell}V_{i\ell}\|^2$, the mean-reverting term $\|BV^*_{i}\|^2$, and the noise term $\|\vec{\epsilon}_i\|^2$. 
The results are shown in Figure \ref{fig:Norms_terms}.
During seizures the norm of the autoregressive term is comparable to the norm of the noise term, and the mean reversion term is almost zero (apart from seizure 3). 
In the interictal period, the autoregressive term has zero norm, and the mena reverting term has norm slightly greater than the noise. 

\subsection{Parameter values for patients 6, 7, and 13}
\begin{figure}
	\centering
	\begin{subfigure}{1\textwidth}
		\centering
		\includegraphics[width=0.8\textwidth]{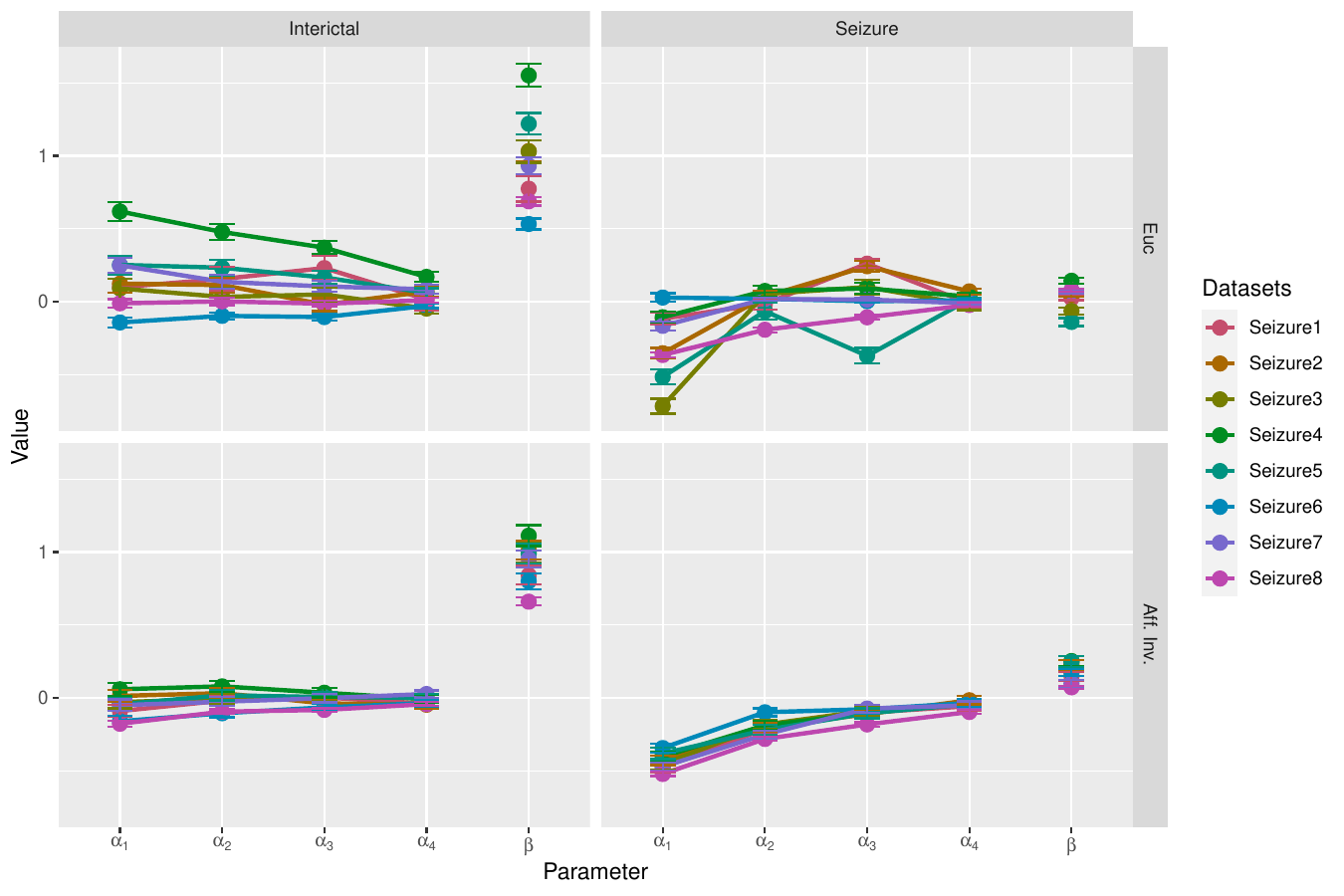}  
		\caption{Scalar coefficients for patient 6.}
	\end{subfigure}
	\begin{subfigure}{1\textwidth}
		\centering
		\includegraphics[width=0.8\textwidth]{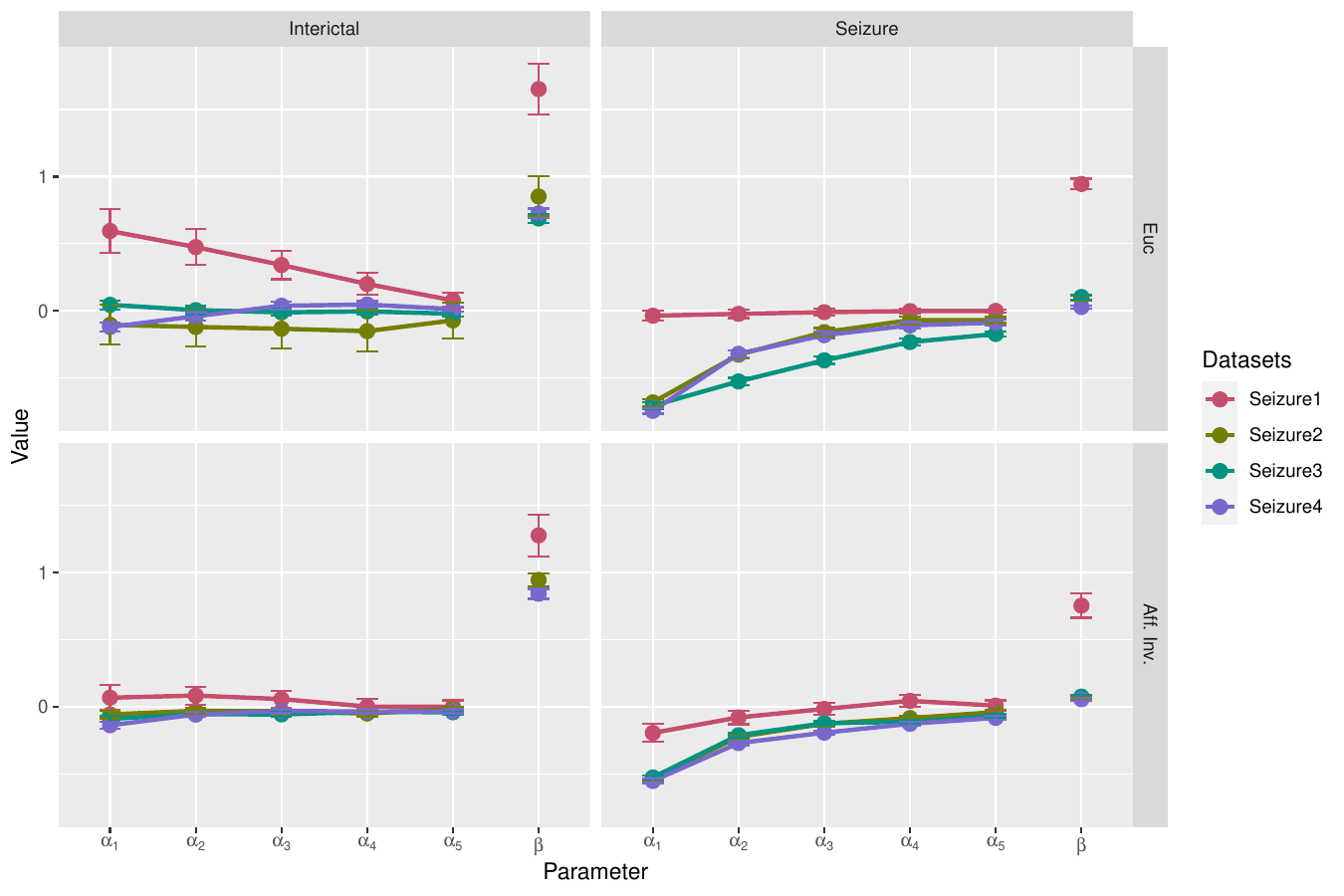}  
		\caption{Scalar coefficients for patient 7.}
	\end{subfigure}

 \begin{subfigure}{1\textwidth}
		\centering
		\includegraphics[width=0.8\textwidth]{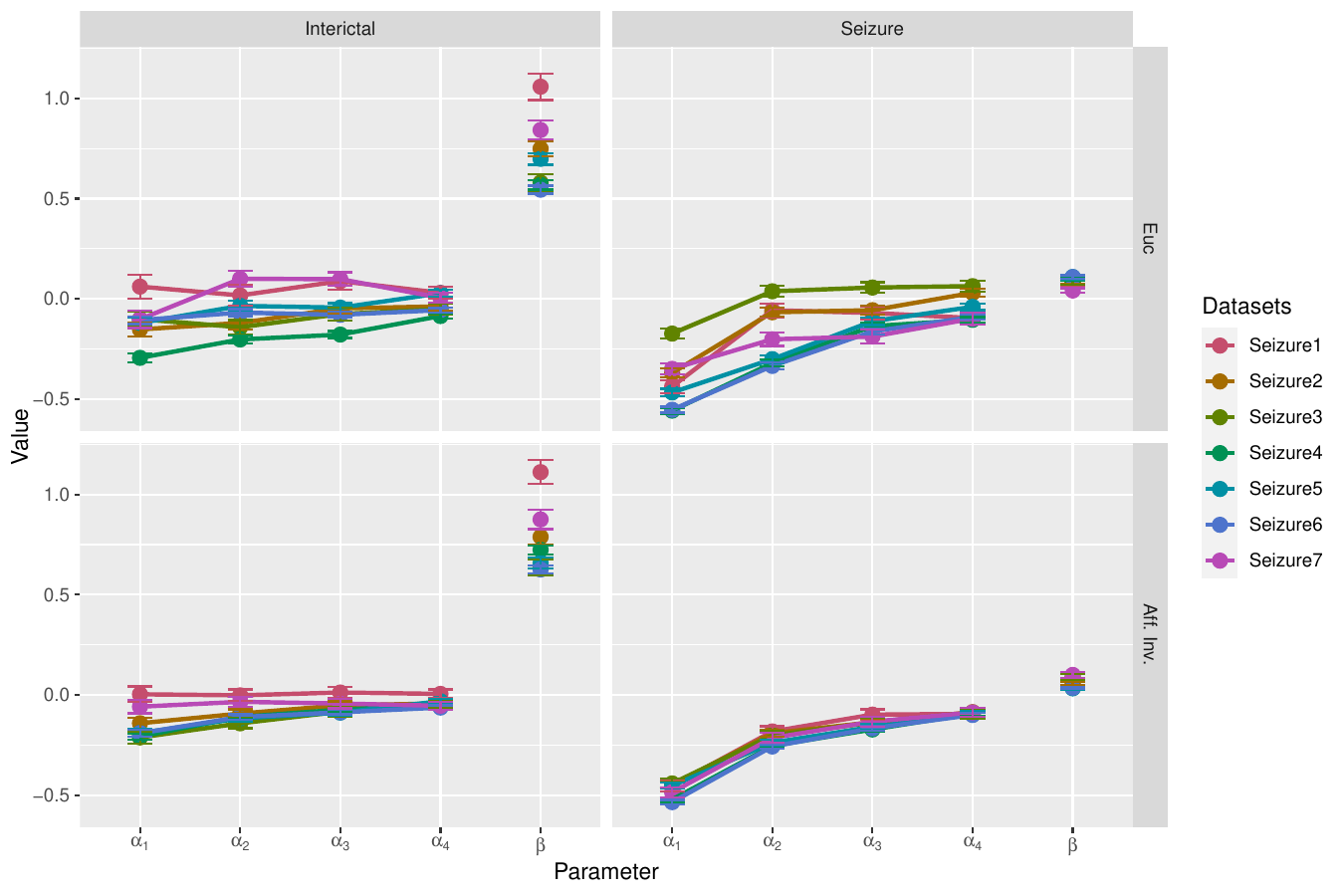}  
		\caption{Scalar coefficients for patient 13.}
	\end{subfigure}
 
	\caption{Scalar coefficients for patients 6, 7, and 13.}
	\label{fig:scalarestpar}
\end{figure}

Similar to the scalar model results depicted in Figure 5 in the main paper, we present the estimated scalar coefficients for patients 6, 7, and 13 in Figure~\ref{fig:scalarestpar}.

\section{Additional model results: diagonal model}

\begin{figure}
    \centering
    \includegraphics[width = 0.8\textwidth]{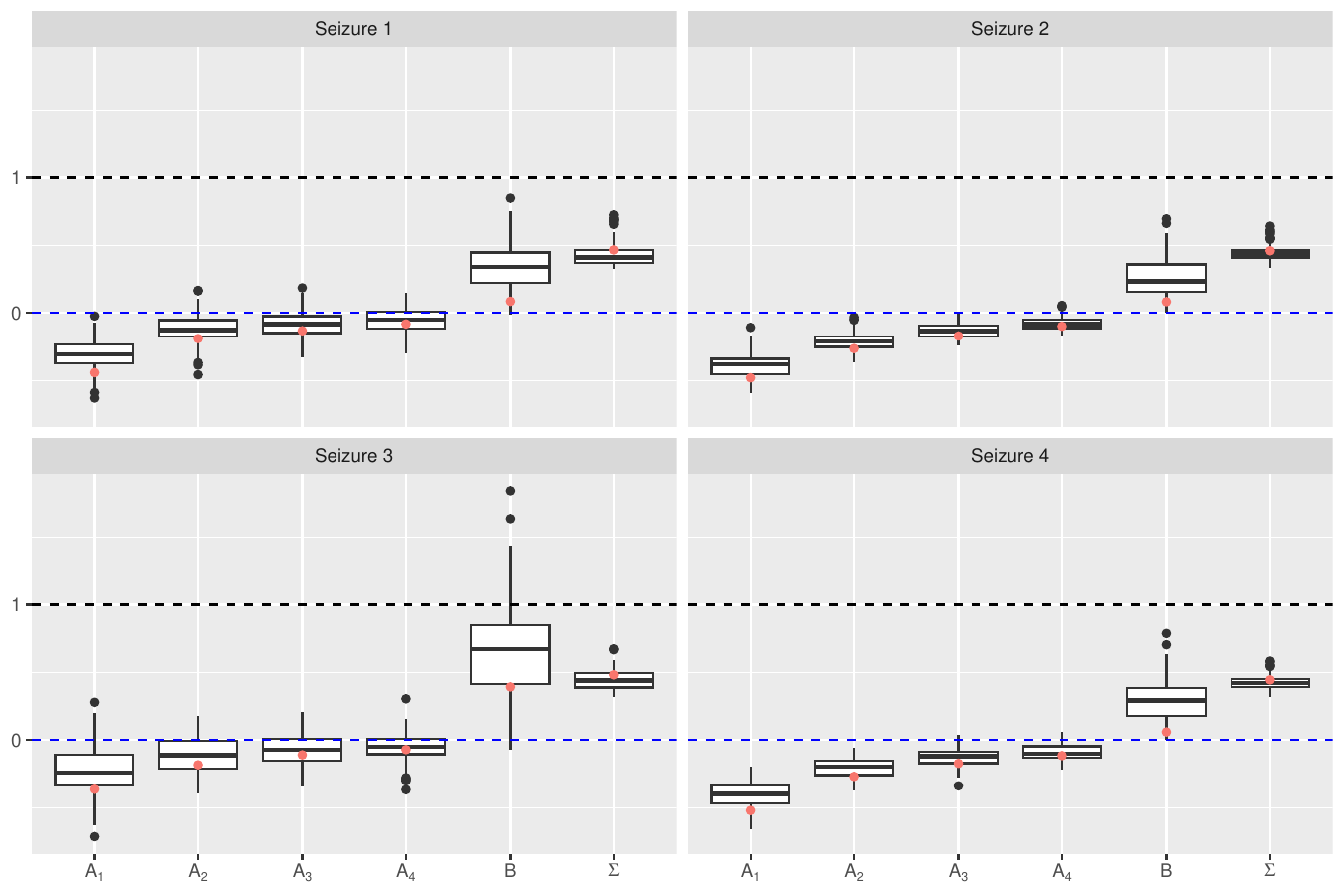}
    \caption{Distribution of fitted diagonal model parameters for
patient 18 (seizure 1-4) in the affine invariant geometry. Other details are as for Figure 7 in the main text.}
    \label{fig:boxplots_18_4}
\end{figure}

\begin{figure}
    \centering
    \includegraphics[width = \textwidth]{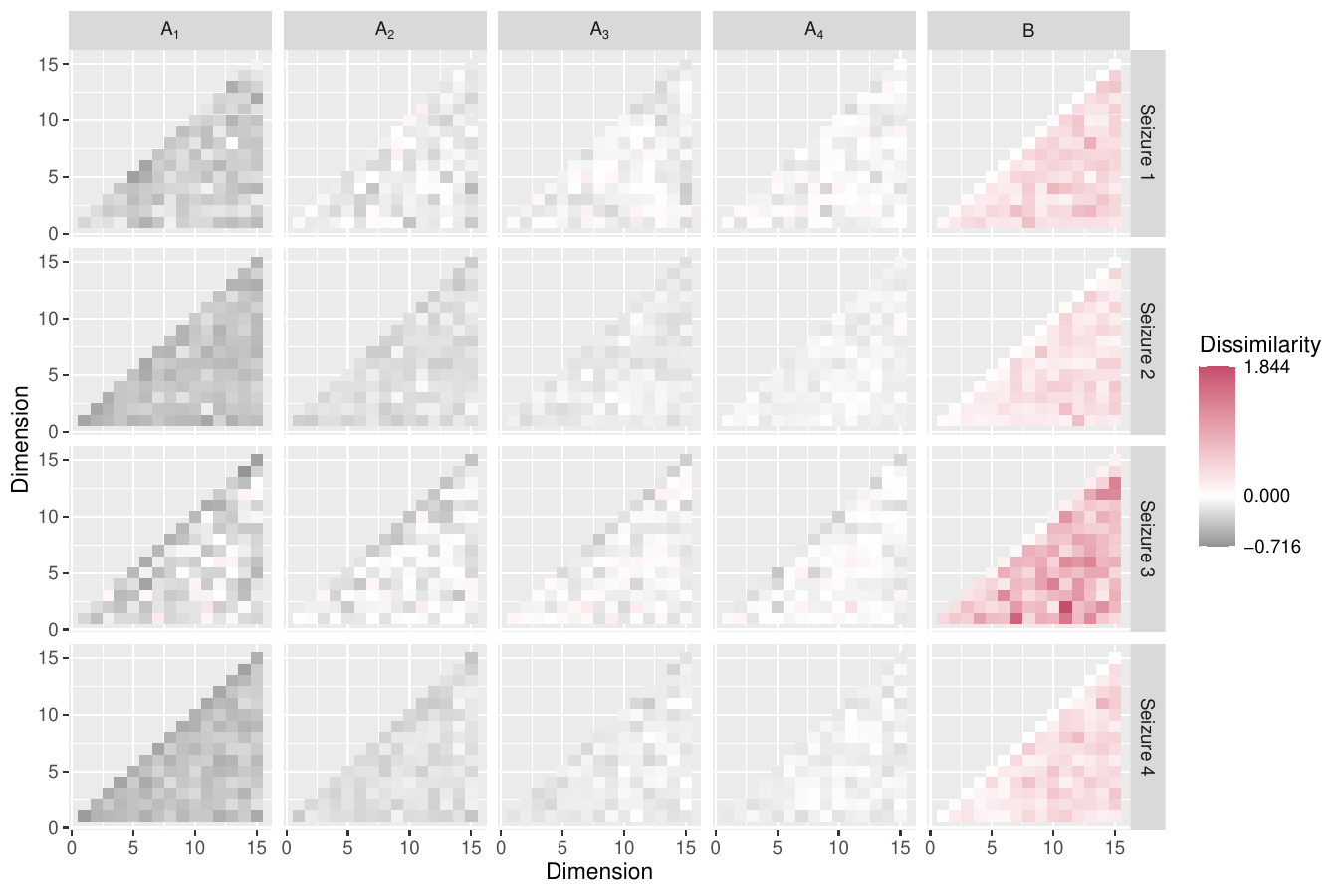}
    \caption{Diagonal model parameters for patient 18 (seizure 1-4) in the affine invariant geometry. Other details are as for Figure 8 in the main text.}
    \label{fig:heatmap_18_4}
\end{figure}

Figures \ref{fig:boxplots_18_4} and \ref{fig:heatmap_18_4} show the diagonal model results for seizures 1 to 4 for patient 18.
(Results for seizure 5 are in the main text.)
Table \ref{tab:diagAIC} gives AIC values for scalar and diagonal models fitted to patient 18 in the affine invariant geometry.
These show the diagonal model is preferred in general (lower values) though the scalar model is preferred for some interictal series. 

\begin{table}[htp]
\centering
		\scalebox{0.7}{
			\begin{tabular}{l|cc|cc|cc|cc|cc}  
				\hline 
				& \multicolumn{2}{c|}{Dataset 1}& \multicolumn{2}{c|}{Dataset 2} &\multicolumn{2}{c|}{Dataset 3}&\multicolumn{2}{c|}{Dataset 4}&\multicolumn{2}{c}{Dataset 5} \\   
				\hline
				& Seizure & Interictal & Seizure & Interictal & Seizure & Interictal & Seizure & Interictal & Seizure & Interictal \\  
sca. & 17560.81 & 12753.47 & 42016.77 & 37445.85 & 11163.30 & 13462.77 & 32992.64 & 28961.46 & 41434.26 & 56652.75 \\ 
diag. & 16427.01 & 13019.79 & 40367.05 & 36557.70 & 10968.24 & 13938.51 & 31583.09 & 27718.58 & 39859.34 & 55863.49 \\ 
   \hline
\end{tabular}}
\caption{AIC values for scalar and diagonal models fitted to patient 18. Comparison of seizure /interictal for affine invariant geometry.}
\label{tab:diagAIC}
\end{table}